\documentclass[fleqn,usenatbib]{mnras}

\usepackage{newtxtext,newtxmath}

\usepackage[T1]{fontenc}

\DeclareRobustCommand{\VAN}[3]{#2}
\let\VANthebibliography\thebibliography
\def\thebibliography{\DeclareRobustCommand{\VAN}[3]{##3}\VANthebibliography}

\usepackage{graphicx}	
\usepackage{amsmath}	

\newcommand{\kms}{\ifmmode{\,\rm{km}\, \rm{s}^{-1}}\else{$\,$km$\,$s$^{-1}$}\fi}
\newcommand{\msun}{\ifmmode{~\rm{M}_{\odot}}\else{M$_{\odot}$}\fi}
\newcommand{\mstar}{\ifmmode{M_{\star}}\else{$M_{\star}$}\fi}
\newcommand{\logm}{\ifmmode{\log(M_{\star}/M_{\odot})}\else{$\log(M_{\star}/M_{\odot})$}\fi}
\newcommand{\loghm}{\ifmmode{\log(M_{\rm{halo}}/M_{\odot})}\else{$\log(M_{\rm{halo}}/M_{\odot})$}\fi}
\newcommand{\mm}{\ifmmode{M_{\star}/M_{\odot}}\else{$M_{\star}/M_{\odot}$}\fi}
\newcommand{\ser}{S\'ersic}
\newcommand{\kinemetry}{\textsc{kinemetry}}

\newcommand{\rmaxs}{\ifmmode{R_{\rm{\sigma}}^{\rm{max}}}\else{$R_{\rm{\sigma}}^{\rm{max}}$}\fi}
\newcommand{\recirc}{\ifmmode{R_{\rm{e,c}}}\else{$R_{\rm{e,c}}$}\fi}
\newcommand{\re}{\ifmmode{R_{\rm{e}}}\else{$R_{\rm{e}}$}\fi}
\newcommand{\ret}{\ifmmode{R_{\rm{e/2}}}\else{$R_{\rm{e/2}}$}\fi}
\newcommand{\retwo}{\ifmmode{R_{\rm{e/2}}}\else{$2R_{\rm{e}}$}\fi}
\newcommand{\aee}{\ifmmode{a_{\rm{e}}}\else{$a_{\rm{e}}$}\fi}
\newcommand{\kpa}{\ifmmode{PA_{\rm{kin}}}\else{$PA_{\rm{kin}}$}\fi}
\newcommand{\eps}{\ifmmode{\varepsilon}\else{$\varepsilon$}\fi}
\newcommand{\ee}{\ifmmode{\varepsilon_{\rm{e}}}\else{$\varepsilon_{\rm{e}}$}\fi}
\newcommand{\epsintr}{\ifmmode{\varepsilon_{\rm{intr}}}\else{$\varepsilon_{\rm{intr}}$}\fi}

\newcommand{\lr}{\ifmmode{\lambda_R}\else{$\lambda_{R}$}\fi}
\newcommand{\lre}{\ifmmode{\lambda_{R_{\rm{e}}}}\else{$\lambda_{R_{\rm{e}}}$}\fi}
\newcommand{\lret}{\ifmmode{\lambda_{R_{\rm{e/2}}}}\else{$\lambda_{R_{\rm{e/2}}}$}\fi}
\newcommand{\lretwo}{\ifmmode{\lambda_{2R_{\rm{e}}}}\else{$\lambda_{2R_{\rm{e}}}$}\fi}
\newcommand{\lreo}{\ifmmode{\lambda_{\,R_{\rm{e}}}^{\rm{\,obs}}}\else{$\lambda_{\,R_{\rm{e}}}^{\rm{\,obs}}$}\fi}
\newcommand{\lrei}{\ifmmode{\lambda_{\,R_{\rm{e}}}^{\rm{\,intr}}}\else{$\lambda_{\,R_{\rm{e}}}^{\rm{\,intr}}$}\fi}
\newcommand{\lreorig}{\ifmmode{\lambda_{\,R_{\rm{e}}}^{\rm{\,orig}}}\else{$\lambda_{\,R_{\rm{e}}}^{\rm{\,orig}}$}\fi}
\newcommand{\lrerep}{\ifmmode{\lambda_{\,R_{\rm{e}}}^{\rm{\,rep}}}\else{$\lambda_{\,R_{\rm{e}}}^{\rm{\,rep}}$}\fi}

\newcommand{\lreeo}{\ifmmode{\lambda_{\,R_{\rm{e}}}^{\rm{\,edge - on}}}\else{$\lambda_{\,R_{\rm{e}}}^{\rm{\,edge - on}}$}\fi}

\newcommand{\vs}{\ifmmode{V / \sigma}\else{$V / \sigma$}\fi}
\newcommand{\vse}{\ifmmode{(V / \sigma)_{\rm{e}}}\else{$(V / \sigma)_{\rm{e}}$}\fi}
\newcommand{\vobs}{\ifmmode{V_{\rm{obs}}}\else{$V_{\rm{obs}}$}\fi}
\newcommand{\vseorig}{\ifmmode{(V / \sigma)_{{\rm{e}}}^{\rm{\,orig}}}\else{$(V / \sigma)_{{\rm{e}}}^{\rm{\,orig}}$}\fi}
\newcommand{\vserep}{\ifmmode{(V / \sigma)_{{\rm{e}}}^{\rm{\,rep}}}\else{$(V / \sigma)_{{\rm{e}}}^{\rm{\,rep}}$}\fi}
\newcommand{\vseo}{\ifmmode{(V / \sigma)_{{\rm{e}}}^{\rm{\,obs}}}\else{$(V / \sigma)_{{\rm{e}}}^{\rm{\,obs}}$}\fi}
\newcommand{\vsei}{\ifmmode{(V / \sigma)_{{\rm{e}}}^{\rm{\,intr}}}\else{$(V / \sigma)_{{\rm{e}}}^{\rm{\,intr}}$}\fi}

\newcommand{\sobs}{\ifmmode{\sigma_{\rm{obs}}}\else{$\sigma_{\rm{obs}}$}\fi}
\newcommand{\se}{\ifmmode{\sigma_{\rm{e}}}\else{$\sigma_{\rm{e}}$}\fi}

\newcommand{\sigpsf}{\ifmmode{\sigma_{\rm{PSF}}}\else{$\sigma_{\rm{PSF}}$}\fi}
\newcommand{\sigpsfre}{\ifmmode{\sigma_{\rm{PSF}}/R_{\rm{e}}}\else{$\sigma_{\rm{PSF}}/R_{\rm{e}}$}\fi}

\newcommand{\smk}{\ifmmode{\langle {k_5 / k_1}\rangle}\else{$\langle{k_5 / k_1}\rangle$}\fi}
\newcommand{\smke}{\ifmmode{{\langle {k_5 / k_1}\rangle-\langle{k_5 / k_1}_{\rm{error}}\rangle}}\else{${\langle{k_5 / k_1}\rangle-\langle{k_5 / k_1}_{\rm{error}}\rangle}$}\fi}

\newcommand{\mk}{\ifmmode{\langle k_{51,\rm{e}} \rangle }\else{$ \langle k_{51,\rm{e}} \rangle $}\fi}

\newcommand{\ea}{\textsc{eagle}}
\newcommand{\eap}{\textsc{eagle$^+$}}
\newcommand{\hy}{\textsc{hydrangea}}
\newcommand{\ha}{\textsc{horizon-agn}}
\newcommand{\ill}{\textsc{illustris}}
\newcommand{\illtng}{\textsc{illustris-tng}}
\newcommand{\ma}{\textsc{magneticum}}
\newcommand{\at}{\ifmmode{\rm{ATLAS}^{\rm{3D}}}\else{ATLAS$^{\rm{3D}}$}\fi}

\graphicspath{{./}{figures/}}

\title[Towards an Optimal Kinematic Classification]{The SAMI Galaxy Survey: A Statistical Approach to an Optimal Classification of Stellar Kinematics in Galaxy Surveys}

\author[Jesse van de Sande]{Jesse van de Sande$^{1,2},$
Sam P. Vaughan$^{1,2}, $
Luca Cortese$^{2,3}, $
Nicholas Scott$^{1,2}, $
Joss Bland-Hawthorn$^{1,2}, $
\newauthor
Scott M. Croom$^{1,2}, $
Claudia D.P. Lagos$^{2,3},$
Sarah Brough$^{2,4}, $
Julia J. Bryant$^{1,2,5}, $
Julien Devriendt$^{6}, $
\newauthor
Yohan Dubois$^{7}, $
Francesco D'Eugenio$^{8},$
Caroline Foster$^{1,2}, $
Amelia Fraser-McKelvie$^{2,3}, $
\newauthor
Katherine E. Harborne$^{2,3}, $
Jon S. Lawrence$^{9}, $
Sree Oh$^{2,10}, $
Matt S. Owers$^{11,12}, $
Adriano Poci$^{12}, $
\newauthor
Rhea-Silvia Remus$^{13}, $
Samuel N. Richards$^{14}, $
Felix Schulze$^{13,15}, $
Sarah M. Sweet$^{2,16}, $ 
Mathew R. Varidel$^{1,2}, $
\newauthor
and Charlotte Welker$^{2,17}$
\\
\\
Affiliations are listed at the end of the paper}
\date{Accepted XXX. Received YYY; in original form ZZZ}

\pubyear{2020}

\begin{document}
\label{firstpage}
\pagerange{\pageref{firstpage}--\pageref{lastpage}}
\maketitle

\begin{abstract}
\noindent Large galaxy samples from multi-object Integral Field Spectroscopic (IFS) surveys now allow for a statistical analysis of the $z\sim0$ galaxy population using resolved kinematic measurements. However, the improvement in number statistics comes at a cost, with multi-object IFS survey more severely impacted by the effect of seeing and lower signal-to-noise. We present an analysis of $\sim1800$ galaxies from the SAMI Galaxy Survey taking into account these effects. We investigate the spread and overlap in the kinematic distributions of the spin parameter proxy \lre\ as a function of stellar mass and ellipticity \ee. For SAMI data, the distributions of galaxies identified as regular and non-regular rotators with \textsc{kinemetry} show considerable overlap in the \lre-\ee\ diagram. In contrast, visually classified galaxies (obvious and non-obvious rotators) are better separated in \lre\ space, with less overlap of both distributions. Then, we use a Bayesian mixture model to analyse the observed \lre-\logm\ distribution. By allowing the mixture probability to vary as a function of mass, we investigate whether the data are best fit with a single kinematic distribution or with two. Below $\log(M_{\star}/M_{\odot})\sim10.5$, a single beta distribution is sufficient to fit the complete \lre\ distribution, whereas a second beta distribution is required above $\log(M_{\star}/M_{\odot})\sim10.5$ to account for a population of low-\lre\ galaxies. While the Bayesian mixture model presents the cleanest separation of the two kinematic populations, we find the unique information provided by visual classification of galaxy kinematic maps should not be disregarded in future studies. Applied to mock-observations from different cosmological simulations, the mixture model also predicts bimodal \lre\ distributions, albeit with different positions of the \lre\ peaks. Our analysis validates the conclusions from previous, smaller IFS surveys, but also demonstrates the importance of using selection criteria for identifying different kinematic classes that are dictated by the quality and resolution of the observed or simulated data.
\end{abstract}

\begin{keywords}
cosmology: observations -- galaxies: evolution -- galaxies: formation -- galaxies: kinematics and dynamics -- galaxies: stellar content -- galaxies: structure
\end{keywords}


\section{Introduction}
\label{sec:intro}

The distribution of ordered to random stellar motions in present-day galaxies provides strong constraints on how  galaxies assembled their mass over cosmic time. Historically, the kinematic properties of spiral galaxies were already known even as the extragalactic nature of these galaxies was still being debated \citep{slipher1914,pease1916}. In contrast, the kinematic variety and complexity of early-type galaxies was revealed at a much later stage \citep[for reviews see][]{dezeeuw1991,cappellari2016}. One of the major discoveries for early-type galaxies was that with increasing luminosity, elliptical galaxies transition from being predominantly rapid to predominantly slow rotators \citep{bertola1975,illingworth1977,davies1983}, and that the flattening of these slowly-rotating ellipticals was due to anisotropy rather than rotation \citep{binney1978,schechter1979}. 

A key remaining question about the kinematic properties of galaxies is whether the distribution of rotation is bimodal with contrasting formation histories or a continuous transition from one type into another. While there are indications for a bimodal distribution between different types of elliptical galaxies or within the early-type population, most of the evidence is circumstantial. The idea of two intrinsically different types of ellipticals originated from the connection between the kinematic properties of ellipticals with boxy and disky isophotes \citep{carter1987,bender1988,kormendy1996}, or with cuspy cores versus inner power-law stellar light profiles \citep{faber1997}, that are thought to be related to the merger history. 

One of the first mentions of a dichotomy in the early-type population, i.e., two physically distinct groups as based on morphological and photometric properties, is by \cite{ferrarese1994} further supported by \citeauthor{lauer1995} (\citeyear{lauer1995}; but see also \citealt{carollo1997}). Subsequently, \citet{kormendy1996} suggest a dichotomy between disky and boxy ellipticals based on the relation between the isophotal boxiness parameter $a_4/a$ and \vs\ (the ratio of the velocity $V$ to the velocity dispersion $\sigma$), however, they mention the possibility that boxy and disky ellipticals form a continuous sequence. In contrast, \citet{ferrarese2006} present evidence in favour of a continuous distribution in the logarithmic inner slopes of early-type galaxies, instead of a bimodality. \cite{kormendy2012} present an extensive list of properties classifying elliptical galaxies into giant ellipticals ($M_V \lesssim -21.5$) versus normal and dwarf true ellipticals ($M_V \gtrsim -21.5$), a continuation of the results from \citet{kormendy1996} and \citet{kormendy2009}. Among other properties, giant ellipticals should have cores, rotate slowly, and have boxy-distorted isophotes. 

With two-dimensional (2D) kinematic measurements from integral field spectroscopy \citep[IFS, e.g., SAURON][]{bacon2001,dezeeuw2002}, \citet{emsellem2007} did not find a clear relation between the spin parameter proxy \lr\ and boxy versus disky early-types, nor between \lr\ and core versus power-law early-type galaxies. Further results from \at\ survey indicate no clear trend between the boxiness parameter $a_4$ and the fast rotator (FR) and slow rotator (SR) classes \citep{emsellem2011}. However, while \citet{krajnovic2013} find no evidence for a bimodal distribution of nuclear slopes of \at\ galaxies, the combination of \se\ and \lre\ is found to be a good predictor for the shape of the inner light profile \citep[see also ][]{krajnovic2020}. 

In parallel, \citet{emsellem2007} identified two rotational types of early-type galaxies from a visual inspection of the $V$ and $\sigma$ maps, quantitatively classified as fast and slow rotators having $\lr\geq0.1$ and $\lr<0.1$, respectively. In subsequent surveys, such as \at, a combination of the \kinemetry\ method, which quantifies the regularity of the velocity field \citep{krajnovic2006,krajnovic2011}, and the spin-parameter proxy \lr\ and ellipticity were used to classify galaxies as fast and slow rotators \citep{emsellem2011}. One of the main results from the \at\ survey is that the vast majority of early-types (86\%) belong to a single family of fast-rotating disk galaxies with ordered rotation and regular velocity fields \citep{krajnovic2011, emsellem2011}. Only a small fraction (14\%) of early-type galaxies are slow rotators with more complex dynamical and morphological (e.g., triaxial) structures.

While the evidence for at least two kinematic populations of ellipticals and of early-type galaxies has been growing \citep[e.g.,][]{cappellari2016,graham2018}, there has been relatively little discussion on the possibility of a continuous distribution or on the overlap of properties between classes. Given the complexity of how massive (\logm>10.5) galaxies assemble their stellar mass over time \citep[e.g., see][]{naab2014}, assigning galaxies to specific classes without expecting considerable overlap might be an unprofitable endeavour. In large volume cosmological simulations (e.g., \ea\ \citealt{schaye2015}, \citealt{crain2015}; \ha\ \citealt{dubois2014}; \ill\ \citealt{Genel2014}, \citealt{vogelsberger2014}; \illtng\ \citealt{springel2018},  \citealt{pillepich2018a}; \ma\ Dolag et al., in prep, \citealt{hirschman2014}) the properties of fast and slow rotators have been studied in detail, but the fast/slow selection methods almost always follow the observational criteria \citep{penoyre2017,choi2018,lagos2018b, schulze2018, walomartin2020, pulsoni2020}. Thus, the question arises as to how much insight we will gain from comparisons to cosmological simulations when quantitatively many fundamental galaxy relations are still poorly matched to observations \citep{vandesande2019}.

With the rise of large multi-object IFS surveys, such as the SAMI Galaxy Survey \citep[Sydney-AAO Multi-object Integral field spectrograph; N $\sim3000$; ][]{croom2012,bryant2015} and the SDSS-IV MaNGA Survey \citep[Sloan Digital Sky Survey Data; Mapping Nearby Galaxies at Apache Point Observatory; N $\sim10000$; ][]{bundy2015}, we are now able to determine the properties of different kinematic populations as a function of stellar mass using a statistical framework that can be similarly applied when studying mock-galaxies extracted from cosmological simulations. However, the observed kinematic measurements of $V$ and $\sigma$ in SAMI and MaNGA are more severely impacted by atmospheric seeing as well as having larger kinematic uncertainties due to the lower signal-to-noise (S/N) as compared to earlier IFS surveys (e.g., SAURON, \citealt{dezeeuw2002}; \at\ \citealt{cappellari2011a}; CALIFA, \citealt{sanchez2012}). Furthermore, these multi-object IFS samples include galaxies of all morphological types and uncertainties in visual morphological classification could introduce additional challenges. Therefore, a new approach is needed in order to separate non-regular or slow-rotator galaxies from the dominant fast or regular rotating population when the S/N and seeing strongly impact the data quality.

Different methods of correcting the measured \lre\ for seeing now exist \citep{graham2018,harborne2020b,chung2020}, although the corrections are less certain for galaxies with irregular velocity fields. It remains to be seen whether the results from statistical samples of recent IFS surveys are significantly impacted by seeing, such that an intrinsically-bimodal galaxy distribution would be observed as unimodal. Thus, we need to reinvestigate and adapt existing fast and slow rotator selection criteria \citep[e.g.,][]{emsellem2007,emsellem2011,cappellari2016} and investigate the amount of overlap between the different distributions using multi-object IFS data combined with mock-observations from simulations with relatively low spatial resolution.

In this paper, we revisit dynamical galaxy demographics in the era of large IFS samples, where the impact of seeing and data quality is more severe than in previous IFS surveys. We use the SAMI Galaxy Survey, which contains $\sim$3000 galaxies across a large range in galaxy stellar mass, morphology, seeing conditions, and data quality. It provides an ideal test set to investigate the challenges set out above. The main goals of this paper are 1) to determine the impact of different seeing correction methods on the kinematic populations, 2) to investigate the scatter or overlap in the fast and slow rotator distributions, 3) to provide updated methods and selection criteria for separating galaxies that belong to different kinematic families in both observations \emph{and} simulations, and 4) to consolidate results from different IFS surveys over a range of sample sizes and data quality.

Given the details required to perform a rigorous statistical investigation to achieve our goals, the paper contains several long sections where we analyse and compare previous methods. These sections are independent and can be skipped without losing the main narrative. We present the data from the observations in Section \ref{sec:data}. The core analysis of observational data is presented in \ref{sec:kinematic_identifiers}, and for large cosmological simulations in Section \ref{sec:simulations_approach}. Readers solely interested in our new classification method can choose to skip Sections \ref{subsec:fast_slow_seeing}-\ref{subsec:or_nor} and \ref{sec:simulations_approach}. Section \ref{sec:discussion} gives our perspective on previous claims of a bimodality (\ref{subsec:dich_vis}-\ref{subsec:dich_sims}) as well as a discussion on the implications of this work (\ref{subsec:best_separate}-\ref{subsec:implications}). A summary and conclusion is given in Section \ref{sec:conclusion}. Throughout the paper we assume a $\Lambda$CDM cosmology with $\Omega_\mathrm{m}$=0.3, $\Omega_{\Lambda}=0.7$, and $H_{0}=70$ km s$^{-1}$ Mpc$^{-1}$. 

\section{Data}
\label{sec:data}

\subsection{SAMI Galaxy Survey}
SAMI is a multi-object IFS mounted at the prime focus of the 3.9m Anglo-Australian Telescope (AAT), with 13 \textit{hexabundles} \citep{blandhawthorn2011,bryant2011,bryant2012a,bryant2014} deployable over a 1 degree diameter field of view. Each hexabundle consists of 61 individual 1\farcs6 fibres, and covers a $\sim15^{\prime\prime}$ diameter region on the sky. The 793 object fibres and 26 individual sky fibres are fed into the AAOmega  spectrograph \citep{saunders2004, smith2004, sharp2006}, with a blue (3750-5750\AA) and red (6300-7400\AA) arm. With the 580V and 1000R grating, the spectral resolution is R$_{\rm{blue}}\sim 1810$ at 4800\AA, and R$_{\rm{red}}\sim4260$ at 6850\AA\ \citep{scott2018}, respectively. In order to cover gaps between fibres and to create data cubes with 0\farcs5 spaxel size, all observations are carried out using a six- to seven-position dither pattern \citep{sharp2015,allen2015}. 

The SAMI Galaxy Survey \citep{croom2012, bryant2015} contains $\sim3000$ galaxies between redshift $0.004<z<0.095$ with a broad range in galaxy stellar mass (M$_* = 10^{8}-10^{12}$\msun) and galaxy environment (field, groups, and clusters). Galaxies were selected from the Galaxy and Mass Assembly \citep[GAMA;][]{driver2011} campaign in the GAMA G09, G12 and G15 regions, in combination with eight high-density cluster regions sampled within radius $R_{200}$ \citep{owers2017}. We use 3072 unique galaxies from internal data release v0.12. Reduced data-cubes and stellar kinematic data products for 1559 galaxies in the GAMA fields are available as part of the first, second, and third SAMI Galaxy Survey data releases \citep{green2018,scott2018, croom2021}.

\subsubsection{Ancillary Data}

For galaxies in the GAMA fields, aperture matched $g-i$ colours were measured from reprocessed SDSS Data Release Seven \citep{york2000, kelvin2012}, by the GAMA survey \citep{hill2011,liske2015}. For the cluster environment, photometry from the SDSS \citep{york2000} and VLT Survey Telescope (VST) ATLAS imaging data are used \citep{shanks2013,owers2017}. Stellar masses are derived from the rest-frame $i$-band absolute magnitude and $g-i$ colour, by employing the colour-mass relation as outlined in \citet{taylor2011}. A \citet{chabrier2003} stellar initial mass function and exponentially declining star formation histories are assumed in deriving the stellar masses. For more details see \citet{bryant2015}.

We use the Multi-Gaussian Expansion \citep[MGE;][]{emsellem1994,cappellari2002} technique, and the code from \citet{scott2013} to derive structural parameters of galaxies from the imaging data from the GAMA-SDSS \citep{driver2011}, SDSS \citep{york2000}, and VST \citep{shanks2013,owers2017}. Those parameters are the effective radius (the half-light radius of the semi-major axis; \re), the ellipticity of the galaxy within one effective radius ($\epsilon_{\rm{e}}$), and position angles. For more details, we refer to D'Eugenio et al. (in prep).

Visual morphological classifications are described in detail in \citet{cortese2016}. The classifications are determined from SDSS and VST {\it gri} colour images and are based on the \mbox{Hubble} type \citep{hubble1926}, following the scheme used by \citet{kelvin2014}. Early- and late-type galaxies are divided according to their shape, the presence of spiral arms and/or signs of star formation. Early-types with disks are then classified as S0s and pure bulges as ellipticals (E). Late-types galaxies with a disk plus bulge component are classified as early-spirals, and galaxies with only a disk component as late-spirals.

\subsubsection{Stellar Kinematics}
\label{subsubsec:stelkin_sami}

The stellar kinematic measurements for the SAMI Galaxy Survey are described in detail in \citet{vandesande2017a}. A short summary is provided below. We use the penalized pixel fitting code \citep[pPXF;][]{cappellari2004,cappellari2017} assuming a Gaussian line-of-sight velocity distribution (LOSVD). Before combining the blue and red spectra, the red spectra are convolved to match the instrumental resolution in the blue. The combined spectra are rebinned onto a logarithmic wavelength scale with constant velocity spacing (57.9 \kms). We derive a set of radially-varying optimal templates from the SAMI annular-binned spectra, using the MILES stellar library \citep{sanchezblazquez2006,falconbarroso2011}. For each individual spaxel, \textsc{pPXF} is given a set of two or three optimal templates from the annular bin in which the spaxel is located as well as the optimal templates from neighbouring annular bins. We estimate the uncertainties on the LOSVD parameters from 150 simulated spectra.

We visually inspect the 3072 SAMI kinematic maps in the GAMA and cluster regions, and 140 galaxies are flagged and excluded due to unreliable kinematic maps caused by nearby objects or mergers that influence the stellar kinematics of the main object. 1025 galaxies are excluded because the radius out to which we can accurately measure the stellar kinematics is less than 2\farcs0 or \re<1\farcs5. We also remove 40 galaxies where the ratio of the point-spread-function to the effective radius of a galaxy is larger than $\sigpsfre>0.6$. We adopt the limit of 0.6 because of the relatively large impact of beam-smearing on \lre\ at these $\sigpsfre$ values \citep[see ][]{harborne2020b}. Lastly, for another 35 galaxies no reliable \lr\ aperture correction out to one \re\ could be derived (see Section \ref{subsec:fast_slow_seeing}). This brings the total sample of galaxies with kinematic measurements to 1832.

\begin{figure}
\includegraphics[width=1.0\linewidth]{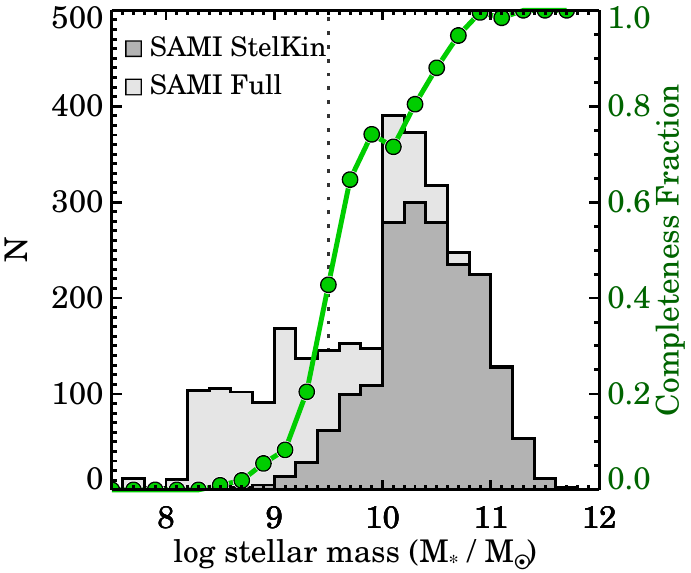}
\caption{Stellar mass distribution of the full SAMI sample (light-grey) and stellar kinematic sample (grey). The green line shows the completeness in bins of stellar mass. The SAMI stellar kinematic sample is biased towards high stellar mass as compared to the full SAMI sample, with 50 percent completeness reached above $\logm \sim 9.55$. 
\label{fig:jvds_mass_completeness}}
\end{figure}

The stellar kinematic completeness as compared to the full SAMI Galaxy Survey sample is presented in Fig.~\ref{fig:jvds_mass_completeness}. The largest fraction of galaxies without kinematic measurements is below stellar mass of $\logm<9.5$. Because the stellar kinematic completeness drops rapidly below 50 percent at low stellar mass (see Fig.~\ref{fig:jvds_mass_completeness}), we do not use the remaining 67 galaxies below $\logm<9.5$ for the core analysis of this paper. 

\begin{figure}
\includegraphics[width=1.0\linewidth]{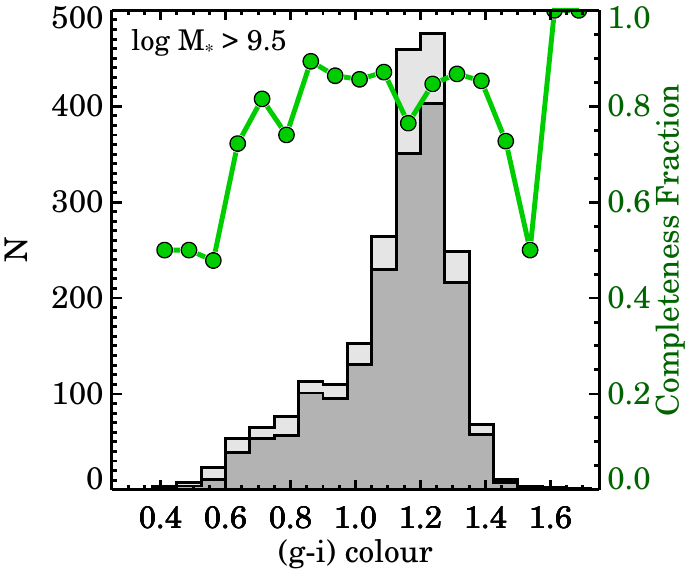}
\caption{Colour distribution of the full SAMI sample (light-grey) and stellar kinematic sample (grey) for all galaxies above a stellar mass of \logm>9.5. The green line shows the completeness in bins of $g-i$ colour. In the region where 95 percent of the data lie ($0.63<g-i<1.36$), we find no colour bias in the stellar kinematic sample as compared to the full SAMI sample. This demonstrates that our kinematic sample is a representative subset of the full galaxy population.
\label{fig:jvds_colour_completeness}}
\end{figure}

We investigate whether the kinematic sample above this mass limit of $\logm=9.5$ is a representative subset of the full SAMI sample by comparing the $g-i$ colour distributions in Fig.\ref{fig:jvds_colour_completeness}. For the vast majority of the sample ($>95$ percent) we find that the colour distribution of the stellar kinematic sample matches that of the full sample, with the exception of the bluest ($g-i<0.6$) and some of the reddest ($1.4<g-i<1.55$) galaxies where the completeness drops below 75 percent. Thus, we conclude that the kinematic sample has no colour bias as compared to the full SAMI sample which was drawn from the volume-limited GAMA survey with high completeness ($\sim90$ percent). The final number of galaxies from the SAMI Galaxy Survey with usable stellar velocity and stellar velocity dispersion maps above a stellar mass of $\logm>9.5$ is 1765; we dub this set of galaxies the "SAMI stellar kinematic sample".

\section{Kinematic Identifiers in Seeing-Impacted Data}
\label{sec:kinematic_identifiers}

\subsection{Fast and Slow Rotators in Seeing Impacted Data}
\label{subsec:fast_slow_seeing}

Fast and slow rotator galaxies are commonly selected from a combination of the spin parameter proxy \lr\ \citep{emsellem2007} and the ellipticity $\varepsilon$. \lr\ quantifies the ratio of the ordered rotation and the random motions in a stellar system, and is given by:

\begin{flushleft}
\begin{equation}
\label{eq:lr}
\lambda_{R} = \frac{\langle R |V| \rangle }{\langle R \sqrt{V^2+\sigma^2} \rangle } = \frac{ \sum_{j=0}^{N_{spx}} F_{j}R_{j}|V_{j}|}{ \sum_{j=0}^{N_{spx}} F_{j}R_{j}\sqrt{V_j^2+\sigma_j^2}}.
\end{equation}
\end{flushleft}

\noindent Here, the subscript $j$ refers to the position of a spaxel within the ellipse, $F_{j}$ the flux of the $j^{th}$ spaxel, $V_{j}$ is the stellar velocity in \kms, $\sigma_{j}$ the velocity dispersion in \kms. $R_{j}$ is the semi-major axis of the ellipse on which spaxel $j$ lies, not the circular projected radius to the centre as is used by e.g.,  \citet{emsellem2007,emsellem2011}. We use the unbinned flux, velocity, and velocity dispersion maps as described in Section \ref{subsubsec:stelkin_sami}. The sum is taken over all spaxels $N_{spx}$ within an ellipse with semi-major axis \re\ and ellipticity \ee\, where the ellipticity is defined from the axis ratio: $\varepsilon=1-b/a$. We use the input galaxy catalogue's R.A. and Dec. and WCS information from the cube headers, to determine a galaxy's centre. The systemic velocity is determined from 9 central spaxels ($1\farcs5 \times 1\farcs5$ box).

We only use spaxels that meet the quality criteria for SAMI Galaxy Survey data as described in \citet{vandesande2017a}: S/N $>3$\AA$^{-1}$, \sobs $>$ FWHM$_{\rm{instr}}/2 \sim 35$\kms\ where the FWHM is the full-width at half-maximum, $V_{\rm{error}}<30$\kms \citep[Q$_1$ from][]{vandesande2017a}, and $\sigma_{\rm{error}} < \sobs *0.1 + 25$\kms\ \citep[Q$_2$ from][]{vandesande2017a}. In practise, as the uncertainties on $V$ and $\sigma$ are strongly correlated with S/N, primarily spaxels towards the galaxy outskirts fail to meet these selection criteria. Kinematic maps with spatially discontinuous $V$ or $\sigma$ measurements (with "holes") are rare, but if the fill factor of good spaxels is less than 85 percent, the galaxy is excluded from the sample. As outlined in \citet{vandesande2017b}, if the fill factor within one effective radius is less than 95 percent, an aperture correction to \lr\ is applied (279 galaxies, 15.8 percent of the stellar kinematic sample).

\begin{figure*}
\includegraphics[width=\linewidth]{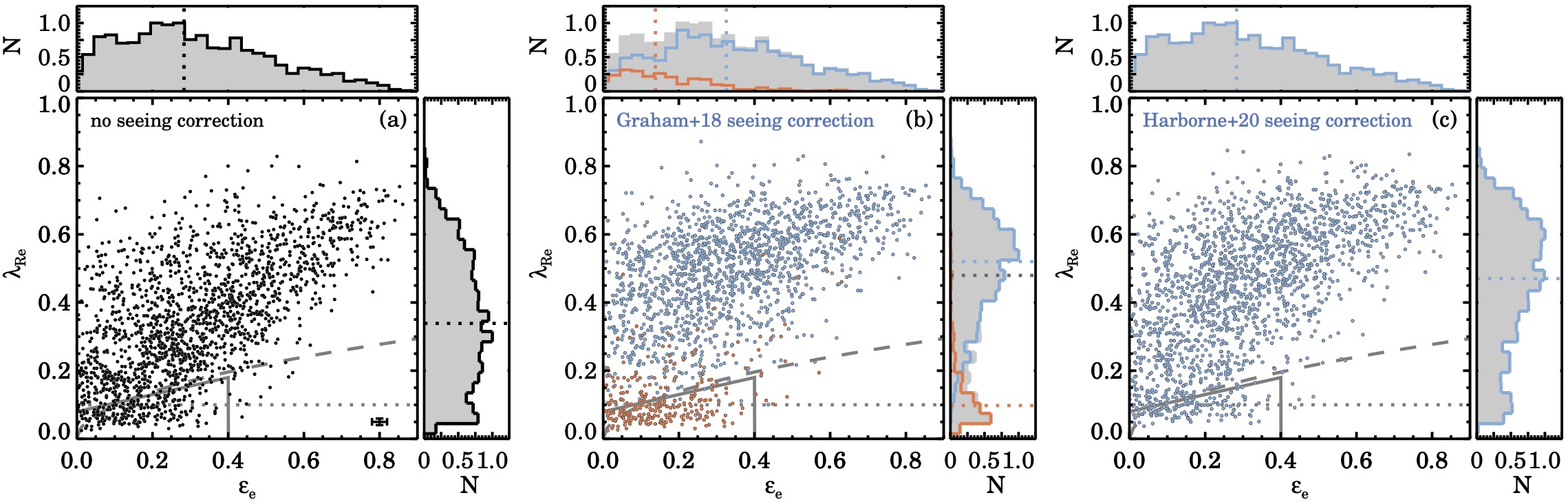}
\caption{Spin parameter proxy \lre\ versus ellipticity \ee\ for the SAMI stellar kinematic sample, without (panel a) and with seeing corrections applied (panels b using the method from \citealt{graham2018} and panel c using the method from \citealt{harborne2020b}). Distributions of \lre\ and \ee\ are shown on the side and on top of each panel. We also show the fast and slow rotator selection criteria from \citet[][dotted line]{emsellem2007}, \citet[][dashed line]{emsellem2011}, and \citet[][solid line]{cappellari2016}. In panel (b) the non-regular rotators defined using \kinemetry\ are colour coded orange, whereas regular rotators are shown in blue. A clear bimodal distribution in \lre\ is only observed when the seeing correction is applied to the regular rotators within the sample (panel b), but we argue this bimodality is artificially enhanced by the seeing correction method (see Section \ref{subsec:fast_slow_seeing}).
\label{fig:jvds_lambdar_eps}}
\end{figure*}

Slow rotators are commonly selected using one of the following criteria, i.e., from \citet[][dotted line in Fig.~\ref{fig:jvds_lambdar_eps}]{emsellem2007}:

\begin{equation*}
\lre < 0.1,
\label{eq:sr_ee07}
\end{equation*}

\noindent or \citet[][dashed-curve in Fig.~\ref{fig:jvds_lambdar_eps}]{emsellem2011}:

\begin{equation*}
\lre < 0.31\times\sqrt{\ee},
\label{eq:sr_ee11}
\end{equation*}

\noindent or with the selection criteria from \citet[][solid line in Fig.~\ref{fig:jvds_lambdar_eps}]{cappellari2016} :
\begin{equation*}
\lre < 0.08+\ee/4 ~~~~~~ \rm{with} ~~~~~~ \ee<0.4.
\label{eq:sr_mc16}
\end{equation*}

\noindent We present the SAMI stellar kinematic sample in Fig.~\ref{fig:jvds_lambdar_eps}(a). The distribution of galaxies in the \lre-\ee\ space is similar to previous studies \citep{emsellem2011,cappellari2016,graham2018,falconbarroso2019}. As noted by \citet{falconbarroso2019}, in contrast to the CALIFA Survey, our sample does not reach values above $\lre>0.8$. This \lr\ ceiling is partially caused by the impact of atmospheric seeing (see next paragraph), but also because of the different radius definition used to calculate \lr\ in Equation \ref{eq:lr}. A similar effect due to seeing is seen in the MaNGA data as presented 
by \citet{frasermckelvie2018}, but not in \citet{graham2018,graham2019b} using the same MaNGA data and \lr\ definition, who find that \lre\ reaches values close to the upper limit of 1.0, with and without a seeing correction applied. In the \lre\ distribution function (Fig.~\ref{fig:jvds_lambdar_eps}a) we see a small peak below $\lre<0.1$, but no clear evidence for two distinct peaks or populations is visible from our seeing-uncorrected data. 

Atmospheric seeing impacts the stellar kinematic measurements by spatially smearing the line-of-sight velocity distribution, which results in a flatter observed velocity gradient but increased overall velocity dispersion. Hence, the seeing-impacted \lre\ values will be lower compared to no-seeing measurements. An analytic correction to account for atmospheric seeing on \lre\ was presented by \citet{graham2018}. This correction was derived by simulating the effect of seeing on kinematic galaxy models constructed with the Jeans Anisotropic MGE modelling method \citep{cappellari2008}, and takes into account the ratio of the seeing to the galaxy effective radius and \ser\ index. 

The accuracy of the analytic \lr\ correction was tested in \citet{harborne2019}. Although the mean correction across a range of morphological types works well, a residual scatter of $\pm0.1$ \lre\ remains as a function of inclination. However, the correction is applicable only for regular rotators \citep{graham2018}. The impact of this limitation is demonstrated in Fig.~\ref{fig:jvds_lambdar_eps}(b). Here, we have seeing-corrected \lre\ for all regular rotating galaxies (blue circles), identified using \kinemetry\ with $\mk<0.07$ \citep[see Section \ref{subsec:kinemetry} and][]{vandesande2017a}, whereas the non-regular rotators are left uncorrected (orange circles). From the \lre\ distribution shown on the side of Panel (b), a clear bimodal distribution appears\footnote{We adopt the definition of bimodality as a distribution with two different modes that appear as distinct peaks in the density distribution.}, although we argue that this separation is artificial enhanced by the seeing correction. 

An alternative seeing correction was presented by \citet{harborne2020b} that has been derived from a suite of hydrodynamical simulations of galaxies with different bulge-to-total ratios. While the method follows the idea of \cite{graham2018}, this new correction includes an inclination term approximated from the observed ellipticity. The residual scatter in \lre\ after applying this correction on a test set of galaxies shows smaller residual scatter as compared to \citet{graham2018}, and also works for all galaxy types within the suite of simulations. Yet, true slow rotators, with complex stellar orbital distributions, kinematically distinct cores, and counter rotating disks, are harder to produce in isolated-galaxy simulations. Instead, galaxies from the \ea\ simulations were used which showed that \lre\ can be seeing-corrected effectively for this type of galaxy with an accuracy of $\Delta\log\lre<0.026$ dex. Furthermore, the absolute impact of the seeing correction on \lre\ for this galaxy type is small. The \citet{harborne2020b} seeing-corrected \lre\ measurements are presented in Fig.~\ref{fig:jvds_lambdar_eps}(c). The low-\lre\ peak that was visible in Fig.~\ref{fig:jvds_lambdar_eps}(b)  \lre\ distribution is no longer as pronounced, and whilst there may be two populations, by eye it is not clear where and how to divide the two possible distributions. 

Including a seeing correction is crucial for recovering an unbiased \lre\ distribution. As galaxies with smaller angular sizes are more severely impacted by seeing, intrinsic differences in the physical sizes of early- and late-type galaxies combined with a redshift-dependent mass selection, can lead to a  morphologically biased \lre\ distribution. Therefore, in what follows we will use the seeing correction from \citet{harborne2020b} as the default. The optimised correction formulae for the SAMI Galaxy Survey data are presented in Appendix \ref{subsec:seeing_corr_analytic}.

However, with the seeing correction applied to all galaxies, it is unclear whether the \cite{emsellem2011} or \citet{cappellari2016} slow rotator selection regions are still valid for our data, or how much overlap there is between the different distributions. As the beam smearing of galaxies with complex inner rotational velocity and dispersion structures behaves differently from regular rotating galaxies, the impact of seeing cannot straightforwardly be predicted with a simple analytic formula; the \lr\ values for complex non-rotating galaxies might be over-corrected. This could explain why it is harder to detect a bimodal distribution in Fig.~\ref{fig:jvds_lambdar_eps}(c). To solve this problem, we will need to include more information than \lre\ and \ee\ alone if we want to determine whether or not we can separate a population of fast and slow rotators in our data.

\subsection{Kinemetry: Regular and Non Regular Rotators}
\label{subsec:kinemetry}

\subsubsection{Description of the \kinemetry\ method}

We first turn to \kinemetry\ for identifying kinematic sub-groups as defined in \citet{krajnovic2011}, because the fast and slow rotator selection regions from \citet{emsellem2011} and \citet{cappellari2016} were designed to best separate regular and non-regular galaxies. In this section we follow a similar approach. The \kinemetry\ method \citep{krajnovic2006,krajnovic2008} provides an estimate of the kinematic asymmetry, under the assumption that the velocity field of a galaxy can be described with a simple cosine law along ellipses: $V(\theta) = V_{\rm{rot}}\cos{\theta}$, where $V_{\rm{rot}}$ is the amplitude of the rotation and $\theta$ is the azimuthal angle. Deviations from this cosine law can then be modelled using Fourier harmonics, where the first order decomposition $k_1$ is equivalent to the rotational velocity and the high-order terms ($k_3$, $k_5$) then describe the kinematic anomalies. The kinematic asymmetry is defined from the amplitudes of the Fourier harmonics $k_5/k_1$ \citep{krajnovic2011}. Our method for measuring the kinematic asymmetry on SAMI Galaxy Survey data is described in detail in \citet{vandesande2017a}. The \kinemetry\ method forms the basis of separating galaxies into regular versus non-regular classes. As was already noted in \citet{vandesande2017a}, the distribution of \smk\ does not show a sharp transition between regular and non-regular rotators. Instead there is a peak in the \smk\ distribution around $\sim0.03$ with a long tail towards high \smk\ values (see also Fig. \ref{fig:jvds_mass_lambdar_kappa_a3d}a).

\begin{figure*}
\includegraphics[width=\linewidth]{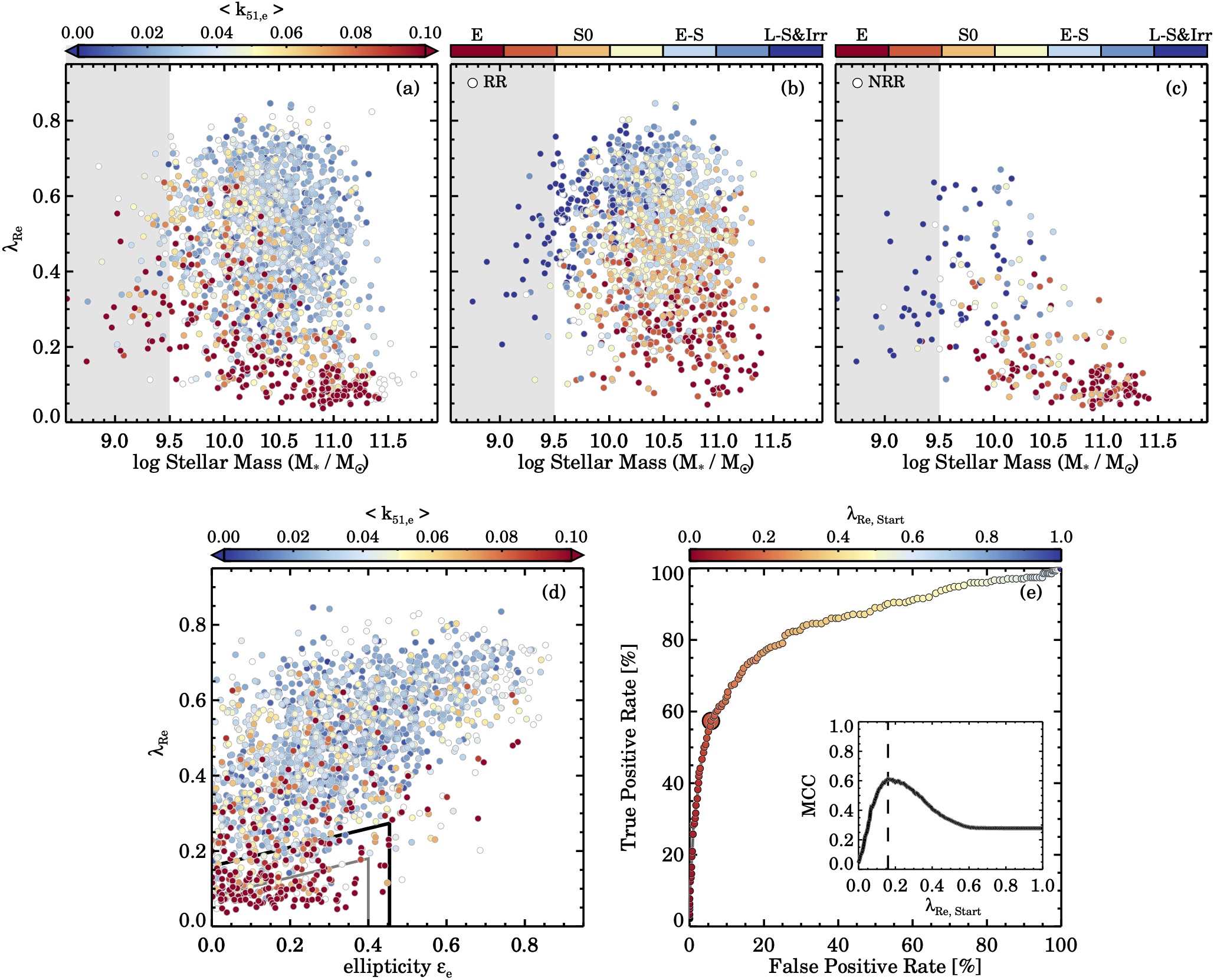}
\caption{Seeing-corrected spin parameter proxy versus stellar mass and ellipticity. Data are colour coded by the kinematic asymmetry parameter \mk\ (panel a) and visual morphological type (panels b and c). Unfilled symbols indicate that \mk\ could not be measured within one \re, or a conclusive visual morphology could not be determined. Galaxies below $\logm<9.5$ are not used in the main analysis, but are shown here for completeness. The overlap of RRs and NRRs increases towards lower stellar mass. The low-mass NRRs also have higher values of \lre\ as compared to high mass NRRs. We show the RRs and NRRs in the \lre-\ee\ space in panel (d) with the optimal selection region (black) derived from panel (e) and the slow rotator selection box from \citet{cappellari2016} in grey. There is considerable overlap of RR and NRR rotators. Panel (e) shows the "Receiver Operating Characteristic Curve" and Matthews correlation coefficient distribution from which we derive the optimal selection region. The most optimal selection region has a True Positive Rate of only 57.4 percent with a False positive rate of 5.7 percent. This suggests that the \lre-\ee\ space is not ideal for distinguishing between regular and non-regular rotators derived from SAMI data.
\label{fig:jvds_mass_lambdar_kinemetry}}
\end{figure*}

Following \citet{emsellem2011} we use the lower limit \smke\ to separate regular and non-regular rotators, taking into account that within uncertainties a galaxy that is classed as non-regular rotator can still be a regular rotator. From here on, we simply refer to \smke\ within an aperture of one effective radius as \mk. The divide 
between Regular Rotators (RR) and Non-Regular Rotators (NRR) was set to 4 percent in \citet{krajnovic2011} based on the peak and error of the distribution, but to 2 percent in \citet{krajnovic2008}. As our data quality is different (median \mk=0.014 for \at\ versus a median \mk=0.029 here), we adjust this limit to \mk=0.07 which corresponds to the 84th percentile of the \mk\ distribution. Note that in \citet{vandesande2017a} we also adopted an intermediate class of Quasi-Regular Rotators, but for the clarity of directly comparing to fast and slow rotators, we do not use the QRR terminology here.

\subsubsection{Identifying Fast and Slow rotators using \kinemetry\ as a prior}

Fig.~\ref{fig:jvds_mass_lambdar_kinemetry}(a) shows the seeing-corrected spin parameter proxy \lre\ using the method from \citet{harborne2020b} versus stellar mass \logm. The data are colour coded by the \mk\ values for the entire sample. There are two clear trends visible. First, at fixed stellar mass, the kinematic asymmetry is higher for low \lre\ values. Secondly, at fixed \lre\ the mean kinematic asymmetry becomes higher towards lower stellar mass, likely to be dominated by a relationship in $k_1$ (rotational velocity) versus stellar mass. To clarify these trends, we show the RRs and NRR separately in Fig.~\ref{fig:jvds_mass_lambdar_kinemetry}(b) and \ref{fig:jvds_mass_lambdar_kinemetry}(c) now colour-coded by visual morphology. As expected, for galaxies with high stellar masses ($\logm>10.75$), NRRs have the lowest values of \lre. However, towards lower stellar mass NRRs demonstrate a large range in \lre, even with our strict definition of non-regularity ($\mk>0.07$). We note that the relatively high-spin NRRs ($\lre>0.4$) roughly fall into two categories: galaxies with late-type spiral morphology and inclination $<45^{\circ}$ with kinematic features in the velocity maps caused by spiral arms or bars, and galaxies with edge-on morphology and low spatial coverage.

The increased scatter in \lre\ towards low stellar masses is caused by a combination of lower $S/N$ and a decrease in the overall rotational velocities ($k_1$) of these galaxies. Because \mk\ is normalised by $k_1$, slower-rotating galaxies that follow a perfect cosine rotation will have higher \mk\ even if uncertainties on $V$ measurements are the same. As galaxies have lower angular momentum towards low stellar mass, higher values of \mk\ are expected. Similarly, as galaxies are inclined from edge-on towards face-on, $k_1$ will become lower, increasing the typical \mk. Galaxies towards low stellar mass and face-on disks with low surface brightness also have lower typical S/N values, causing higher $V$ uncertainties and, therefore, higher \mk. Thus the higher scatter in \mk\ below $\logm<10.5$ is caused by a combination of late-type morphology and observational effects.

The larger scatter between \mk\ and \lre\ leads to considerable overlap between the RR and NRR populations in the \lre-\logm\ diagram. By using a single \lre\ cut-off value to separate regular and non-regular rotators we will not only cause a bias with stellar mass, but also create a large number of \emph{false positives} and \emph{false negatives}, if we assume that \mk\ is the perfect classifier.

The \lre\ versus ellipticity \ee\ diagram, as presented in Fig.~\ref{fig:jvds_mass_lambdar_kinemetry}(d), is now commonly used to separate fast and slow rotators, where the empirical separation between fast and slow rotators is motivated by the location of the regular and non-regular rotators. However, from Fig.~\ref{fig:jvds_mass_lambdar_kinemetry}(d) it is immediately clear that the most-current selection criterion from \citet{cappellari2016} (grey lines) and the previous selection criteria \citep[][not shown]{emsellem2007,emsellem2011}, are unsuccessful in separating RRs and NRRs within our seeing-corrected SAMI sample.

In order to quantify the "success" of the SR selection region for separating RRs and NRRs, we will treat "non-regularity" as a condition that a galaxy can have, while using the \lre-\ee\ diagram as the diagnostic to identify this condition. By adopting this classification, we can calculate statistical measures of performance of this binary test, such as the sensitivity and specificity. To do so, we first construct a confusion matrix (Table \ref{tbl:tbl1}) where we determine the True Positives (TP), True Negatives (TN), False Positives (FP), and False Negatives (FN). A true positive is where a galaxy has the condition of NRR and is also classified (i.e., tested positive) as an SR, whereas a True Negative is an RR that has been classified as an FR.

\begin{table}
	\centering
	\caption{Confusion matrix for the condition of NRR versus RR using the SR versus FR test.}
	\label{tbl:tbl1}
	\begin{tabular}{l c c} 
		\hline
		\hline
		  & NRR & RR \\
		\hline
		SR  & True Positive & False Positive \\
		\hline
		FR  & False Negative & True Negative\\
		\hline
		\hline
	\end{tabular}
\end{table}


There are several statistical measures that quantify the relevance of our statistical test. Here, we will use an "Receiver Operating Characteristic Curve" analysis to quantify how well our test performs \citep[see for example][]{fawcett2006}. Specifically, we will use the sensitivity or true positive rate ($TPR$), the fall-out or False Positive Rate ($FPR$), the Positive Prediction Value ($PPV$), and Matthews correlation coefficient \citep[$MCC$; ][]{matthews1975}:

\begin{equation} 
{TPR} ={\frac {{TP} }{{P} }}={\frac {{TP} }{{TP} +{FN} }}
\label{eq:tpr}
\end{equation}

\begin{equation}
{FPR} ={\frac {{FP} }{{N} }}={\frac {{FP} }{{FP} +{TN} }}
\label{eq:fpr}
\end{equation}

\begin{equation}
{PPV} ={\frac {{TP} }{{TP} +{FP} }}
\label{eq:ppv}
\end{equation}

\begin{equation}
{{MCC}}={\frac {{ {TP}}\times { {TN}}-{ {FP}}\times { {FN}}}{\sqrt {({ {TP}}+{ {FP}})({ {TP}}+{ {FN}})({ {TN}}+{ {FP}})({ {TN}}+{ {FN}})}}}
\label{eq:mcc}
\end{equation}

\noindent Instead of calculating a single set of numbers for the \citet{cappellari2016} FR/SR selection, it will be more insightful to try a variety of selection criteria to determine the optimal selection region. We first explored the full range of selection boxes with different starting and end positions (i.e., with different slopes) in both \lre\ and and \ee\, but the retrieved optimal selection function did not have significantly improved MCC values as compared to the adopted selection function below (there was one exception that we will highlight in Section \ref{subsec:bayesian_approach}). Instead we choose a varying selection region similar to \citet{cappellari2016} as this was well-motivated for higher-S/N and higher-spatial resolution data (e.g., see Appendix \ref{app:kinemetry_a3d}):

\begin{equation}
\lre < \lr_{\rm{start}} + \ee/4, \hspace*{0.1cm} \mathrm{ with } \hspace*{0.1cm} \ee < 0.35 +\frac{ \lr_{\rm{start}} }{1.538}.
\label{eq:sr_vary}
\end{equation}

\noindent We define the optimal selection when the MCC reaches its highest value, which is a trade off between the number of true and false positives and negatives. We note that there are several other optimisation parameters, such as the "Youden's J statistic", the Accuracy, or the F1 score, but they all returned similar results as compared to the MCC.

We show the True Positive Rate versus the False Positive Rate, also known as the "Receiver Operating Characteristic Curve" (ROC-curve), in Fig.~\ref{fig:jvds_mass_lambdar_kinemetry}(e), with an additional inset panel that shows the MCC as a function of $\lr_{\rm{start}}$. A completely random test would result in data residing on the one-to-one line. We test 200 different selection regions, with $\lr_{\rm{start}}$ ranging from 0-1. With increasing values of $\lr_{\rm{start}}$ we find an increase in the TPR but also in the FPR. According to the MCC parameter, the optimal selection region has $\lr_{\rm{start}} = 0.16$ shown as the black line in Fig.~\ref{fig:jvds_mass_lambdar_kinemetry}(d). This value is significantly higher than the $\lr_{\rm{start}}=0.08$ from \citet{cappellari2016}. More importantly, the optimal selection region only has a TPR of 57.4 percent with a FPR of 5.7 percent, and a Positive Prediction Value of 65.8 percent.

Thus, we conclude that using the \lre-\ee\ diagram to separate regular and non-regular rotators is only moderately successful when presented with seeing-dominated data. We emphasise that the \kinemetry\ method was designed for higher quality data than presented here, hence we do not suggest that these results should be interpreted a "failure" of the method. Instead, it is a motivation to explore an alternative kinematic identifier that is better suited for poorer-quality data, which is the goal of the next section.

\subsection{Visual Kinematic Classification: Obvious versus Non-Obvious Rotators}
\label{subsec:or_nor}

\subsubsection{A New Visual Kinematic Classification Scheme}

In the previous section we demonstrated that there is no clean separation of regular and non-regular rotators in the \lre-\ee\ plane or the \lre-\logm\ plane. Nonetheless, when the data quality is good enough, \kinemetry\ provides a quantitative measure of what we visually interpret as kinematic deviations from a regular rotating velocity profile (e.g., see Appendix \ref{app:kinemetry_a3d}). Given that the large variety of the kinematic types as presented by \citet{krajnovic2011} are also easily identified by eye in the \at\ velocity maps, we will now investigate whether a visual kinematic classification of SAMI galaxies offers a clearer separation of galaxies with different kinematic structures.

Visual classification, for example of galaxy morphology, is however subjective from observer to observer and is susceptible to the quality and spatial resolution of the imaging data. Nonetheless, a well-developed framework exists that allows classifiers to determine a galaxy's morphological type with several levels of refinement. Unfortunately, such a clear and well-defined framework does not exist for classifying kinematic maps of galaxies.

\begin{figure*}
\includegraphics[width=1.0\linewidth]{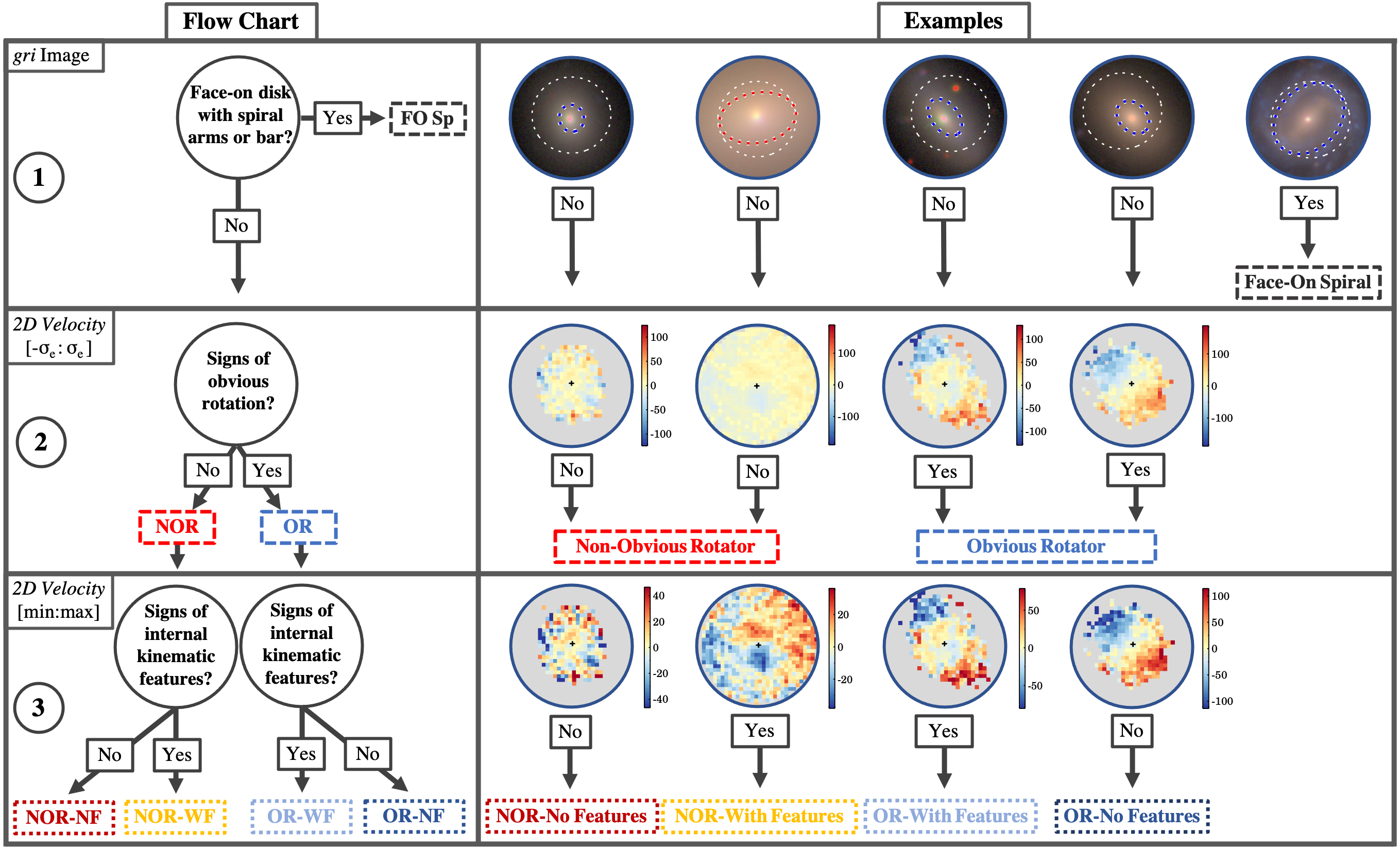}
\caption{Kinematic visual classification flow-chart (left) with example $gri$ colour images and velocity maps as used in the SAMI Galaxy Survey kinematic visual classification. The white-dashed circle on the images from VST-KiDS or Subaru-HSC shows the SAMI field-of-view, whereas the blue- or red-dashed ellipse on the colour images respectively shows 1\re\ or 0.5\re. In the first step, for each galaxy colour images are first used to determine whether the galaxy is Face-On Spiral. Secondly, the galaxy's velocity map is used to classify objects into Non-Obvious Rotators (NORs) and Obvious Rotators (ORs). Here, the velocity scale is derived from the velocity dispersion $\pm\sigma_e$. This integrates the velocity dispersion into the visual classification such that with increasing velocity dispersion the rotational velocity has to become more pronounced in order for a galaxy to be classified as an obvious rotator. The third step employs auto-scaled velocity maps to aid classifiers in identifying kinematic substructures (No-Features versus With-Features).
\label{fig:kin_vis_class}}
\end{figure*}

\citet{krajnovic2011} and \citet{cappellari2016} offer a framework for identifying kinematic features within early-types such a "Kinematically Distinct Core" or "Counter-Rotating Core", yet the classification of when a velocity field is no longer regular rotating is highly subjective. While the origin of this naming convention is closely related to the quantitative \kinemetry\ measurements, comparing flux-weighted measurement within one \re\ and visual classifications are not always straight forward (see Section \ref{subsec:dich_vis} for examples). Furthermore, the different subclasses (e.g., No Features, 2 Maxima, Kinematic Twist) for regular rotators are no longer present in \citet[][fig. 4]{cappellari2016} , who present four classes for galaxies with non-regular velocity fields but only one for regular rotators. However, the main issue with the current kinematic classification scheme is that it is not well adapted for data with different quality. When the S/N decreases and the spatial resolution becomes lower, one would be tempted to classify all galaxies as non-regular rotators if the velocity field appears noisier than the high-quality example galaxies for which the visual classification was designed. 

An initial attempt by three of the authors to classify 50 SAMI galaxies with $\lre<0.35$ into regular versus non-regular rotators led to an identical classification of only 22 galaxies (44 percent). While kinematic features in the core, such as KDCs, are easily classified in nearby galaxies with well-resolved spatial data, they are easily missed in surveys such as SAMI and MaNGA where there is a trade off between multiplexing, spatial resolution, and spatial extent. Furthermore, the regular and non-regular classes that are based on the luminosity weighted \mk\ parameter do not directly translate into a visual classification. As such, we found that the three classifiers had different interpretations of the visual regular versus non-regular classification scheme. This implies that we need to devise a more easily interpretable visual kinematic classification scheme that allows for different levels of data quality.

\begin{figure*}
\includegraphics[width=\linewidth]{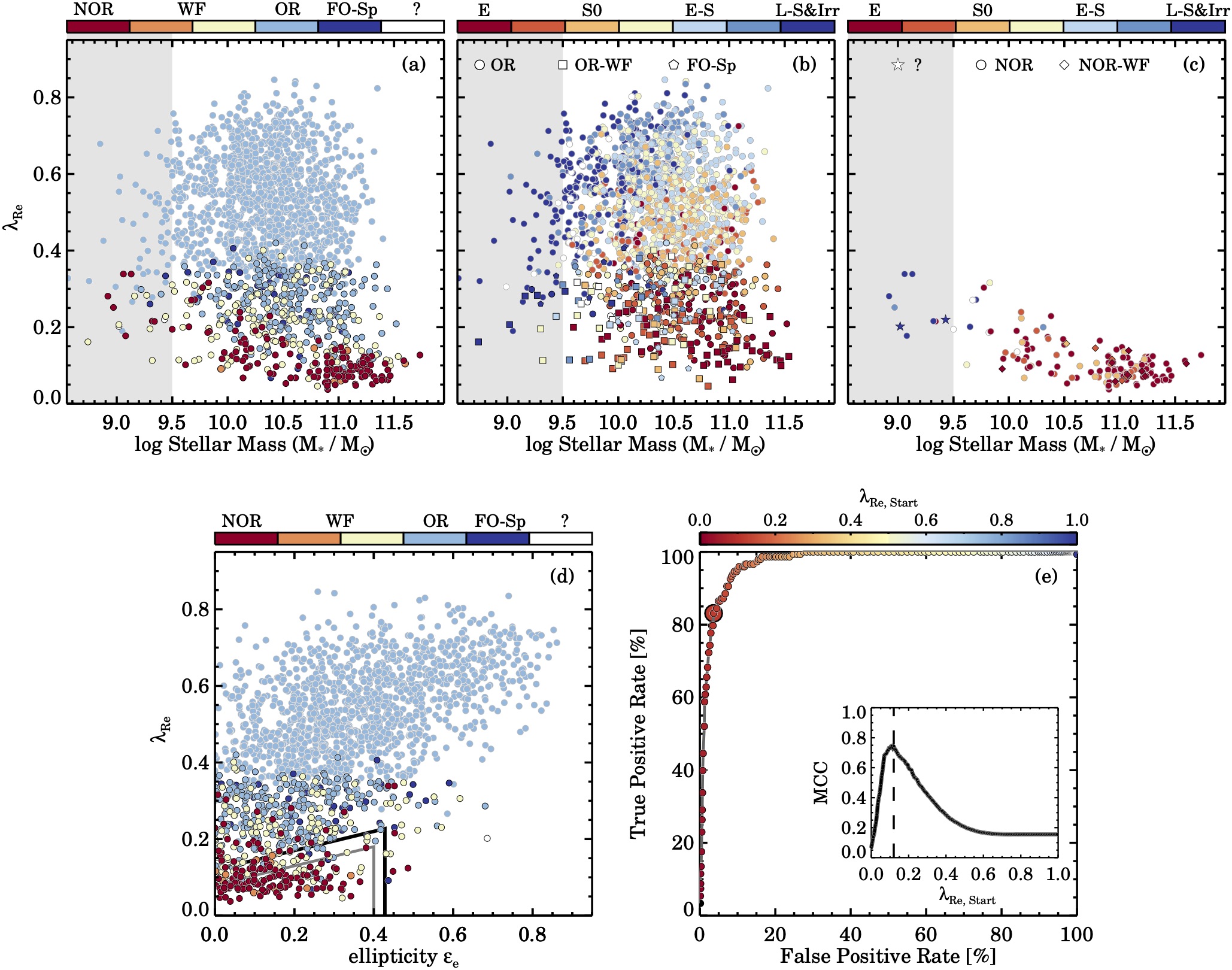}
\caption{Seeing-corrected spin parameter proxy versus stellar mass and ellipticity. Data are colour coded by the kinematic visual classification (panel a) and visual morphological type (panels b and c). Unfilled symbols indicate that a conclusive visual morphology could not be determined. Note that only galaxies with $\lre\lesssim 0.35$ were kinematically visually classified (symbol with black border), all other galaxies are ORs by default (symbols with grey border).
 The overlap between the obvious and non-obvious rotators is considerably less as compared to the results using \kinemetry. With increasing stellar mass the median \lre\ of NORs decreases. We show the ORs and NORs in the \lre-\ee\ space in panel (d) with the optimal selection region (black) and the SR selection box from \citet{cappellari2016} in grey. There is mild overlap of ORs and NORS, but panel (e) indicates a relatively clean selection of NORS can be made using the black selection box.
\label{fig:jvds_mass_lambdar_nor_or}}
\end{figure*}

We propose a kinematic visual classification scheme defined as follows (Fig.~\ref{fig:kin_vis_class}). We begin by defining a specific class for spiral and/or strongly barred galaxies that are close to face-on (FO-Sp), thus showing no obvious rotation. Secondly, we divide the population into "Obvious Rotators" (ORs) and "Non-Obvious Rotators" (NORs). The adopted language is purposely vague to allow for some freedom of interpretation as the classification is qualitative, not quantitative. Whilst the velocity field does not necessarily have to be regular for a galaxy to be classified as an obvious rotator, opposite ends of the velocity field should demonstrate reversed rotation. For the SAMI Galaxy Survey stellar kinematic data, we add one level of refinement. After classifying the kinematic map into OR or NOR, in the third step we check whether the galaxy has an inner kinematic feature ("With Feature"; WF) or not. With improved data quality, this classification scheme can be further refined by adding an extra level to identify the type of kinematic feature \citep[e.g, kinematically distinct core, 2M, etc., from][]{krajnovic2011}.

A flow-chart and five example maps of the different visual kinematic types are presented in Fig.~\ref{fig:kin_vis_class}. For each galaxy we show  the best-available $gri$ colour image derived from VST-KiDS \citep{dejong2017} or Subaru-Hyper Suprime Camera DR1 imaging \citep{aihara2018}, a velocity map with a range set by the average velocity dispersion, as well as a velocity map with auto-scaling. The first velocity map with $\se$-scaling was used to classify galaxies into NORs or ORs, whereas the auto-scaled velocity map is better adjusted for identifying inner kinematic features. The choice for using a velocity range set by the velocity dispersion was motivated by the dependence of the maximum rotational velocity as a function of stellar mass, i.e., the Baryonic Tully-Fisher relation. We also wanted to incorporate the velocity dispersion into the visual classification such that with increasing velocity dispersion the rotational velocity has to become more pronounced in order for a galaxy to be classified as an obvious rotator.

Using similar maps as shown in Fig.~\ref{fig:kin_vis_class}, seven members of the SAMI Galaxy Survey team  visually classified $\sim$600 kinematic maps of galaxies with $\lre\lesssim 0.35$. We chose to visually classify only a selected sample of galaxies, because no NORs were identified at $\lre\gtrsim 0.35$ in a test set of 147 galaxies (10 percent of non-classified galaxies). And because kinematic visual classification is time consuming process we only selected galaxies in the \lre\ region where a mix of ORs and NORs was expected, in order to reduce the total number of galaxies.

Following a similar approach as outlined in \citet{cortese2016}, after all votes were combined, the kinematic type with at least 5/7 votes were chosen (66.7 percent, 399/598). When no absolute majority was found, ORs and ORs-WF were combined into an intermediate type, as well as NORs and NORs-WF. If 5/7 votes then agreed, the galaxy was classified as the intermediate type (23.2 percent, 139/598). For the remaining cases ($\sim10$ percent, 58/598) the classifications of the two most average classifiers were compared and if those agreed that type was chosen (6.2 percent, 37/598). Otherwise, we checked whether a weak majority (4/7) was reached (3.5 percent, 21/598). Only two galaxies in our sample remained unclassified under this scheme.

The results of the kinematic visual classification are presented in the \lre-\logm\ plane (Fig.~\ref{fig:jvds_mass_lambdar_nor_or}a-c). Interestingly, we find that the distribution of ORs and ORs-WF extend to very low values of \lre. While this is expected for face-on spirals, we also find ORs-WFs galaxies with low \lre\ that are classified morphologically as Elliptical and S0s. For NORs, as expected there is an increase in their fraction towards higher stellar mass. The average \lre\ values of NORs also decrease with increasing stellar mass. We find that low-mass ($\logm<10$) NORs are nearly all morphologically classified as late-spiral or irregular. In Fig.~\ref{fig:jvds_mass_lambdar_nor_or}(d) we investigate where ORs and NORs reside in the \lre-\ee\ plane. The NORs have mostly low ellipticity values, and beyond $\ee>0.4$ we only find a handful of NORs. A large fraction of the ORs at low spin-parameter (\lre<0.2) are classified having kinematic features, and similarly for $\lre<0.4$ and $\ee>0.4$, which strengthens the argument for setting an ellipticity limit of $\ee<0.4$ to select galaxies with actual slow rotation rather than low \lre\ values due to counter-rotating disks. Note however that face-on galaxies without a strong bar ($i<10^{\circ}$, $\ee<0.4$) do not have a measurable rotation and will therefore end up in this SR selection region; these galaxies can only be identified as disk-galaxies from a visual morphological classification.

\subsubsection{Identifying Fast and Slow Rotators using Visual Kinematic Morphology as a Prior}

Similar to the test we did for \kinemetry, we will now treat "Non-Obvious Rotation" as a condition that a galaxy can have, while again using the \lre-\ee\ diagram as the diagnostic to identify this condition. The confusion matrix is given in Table \ref{tbl:tbl2}. We then use Equation \ref{eq:sr_vary} to select SRs and FRs for an ensemble of selecting regions and calculate the TPR (Equation \ref{eq:tpr}), the FPR (Equation \ref{eq:fpr}), and the MCC (Equation \ref{eq:mcc}). The optimal selection is defined by the highest MCC value. 

\begin{table}
	\centering
	\caption{Confusion matrix for the condition of NOR versus OR using the SR versus FR test.}
	\label{tbl:tbl2}
	\begin{tabular}{l c c} 
		\hline
		\hline
		  & NOR & OR \\
		\hline
		SR  & True Positive & False Positive \\
		\hline
		FR  & False Negative & True Negative\\
		\hline
		\hline
	\end{tabular}
\end{table}


\noindent The TPR versus FPR, and the MCC distribution are shown in Fig.~\ref{fig:jvds_mass_lambdar_nor_or}(e). According to the MCC parameter, the optimal selection region has $\lr_{\rm{start}} = 0.12$ shown as the black line in Fig.~\ref{fig:jvds_mass_lambdar_nor_or}(d), which is close to the $\lr_{\rm{start}}=0.08$ from \citet{cappellari2016}. The optimal selection region only has a TPR of 83.1 percent with a small FPR of 3.6 percent, and a PPV of 67.6 percent. If we were to accept a higher FPR of 20 percent, which is reached at $\lr_{\rm{start}}=0.27$, then we obtain an impressive TPR of 98.7 percent, but with an unacceptably low PPV of 30.9 percent. Overall, we conclude that there is a good success of selecting Non-Obvious Rotators and Obvious Rotators using the \lre-\ee\ diagram. Nonetheless, as the \lre-\ee\ diagram only shows the average rotational properties within \re\, it cannot replace the spatial information obtained through the process of visual classification.

\subsection{Using Bayesian Mixture Models for Identifying Different Kinematic Families}
\label{subsec:bayesian_approach}

\subsubsection{Description of the Bayesian Mixture Model}

Up to this point, we have been working with the assumption that multiple kinematic populations of galaxies exist. Using \kinemetry\, we separated galaxies into regular and non-regular and for the visual kinematic classification we split galaxies into obvious and non-obvious rotation. Both analyses indicate that the various kinematic classes exist across the full range in stellar masses, with an increased fraction of NRRs and NORs towards high stellar mass. Nonetheless, the question of whether or not a bimodal distribution with two distinct peaks exists has not been answered by this analysis. The \mk\ distribution from \kinemetry\ only reveals a highly skewed distribution, whereas the visual kinematic classification could be tracing two ends of a continuous distribution.

Here, we are interested in analysing the \lre\ distribution as a function of stellar mass without forcing two distinct populations, or assuming where these populations should reside in the \lre-\logm\ plane. To do so, we analyse our data using a Bayesian mixture modelling framework\footnote{Inspired by \citet{taylor2015} who analyse "blue" and "red" galaxies as two naturally overlapping populations using an MCMC analysis.}. The main assumption we make is that the \lre\ distribution of galaxies can be well approximated by a beta distribution, where the probability density function (PDF) is given by:

\begin{equation}
\label{eq:def_beta_function}
f(x, \alpha,\beta) = \frac{x^{\alpha-1}(1-x)^{\beta-1}}{\rm{B}(\alpha,\beta)}
\end{equation}

\noindent with $\rm{B}(\alpha,\beta)$ defined using the Gamma function $\Gamma$:

\begin{equation}
\label{eq:B_function}
\rm {B} (\alpha,\beta )={\frac{\Gamma(\alpha) \Gamma(\beta )}{\Gamma (\alpha +\beta )}}
\end{equation}

\noindent The beta distribution has the property of only being defined on the unit interval, which makes it ideal to describe values of \lr{} that are also constrained to lie between 0 and 1. However, as the maxima and minima of the observed distributions are not perfectly 0 and 1, we rescale the \lr\ values in the following way:

\begin{equation}
\lr_{,\,\rm{rescaled}} = \frac{\lr - \rm{min}(\lr)}{\rm{max}(\lr)-\rm{min}(\lr)}.
\end{equation}

\noindent To model the locations of galaxies in the \lre-\logm{} plane, we use a linear combination of two beta functions at each value of stellar mass (which we label 1 and 2). However, the proportion of galaxies which are drawn from each beta distribution at a given stellar mass is not fixed. We allow the "mixture probability" $p$ to vary smoothly as a function of stellar mass, which captures the well-known dependence of kinematic morphology and mass \citep[e.g.,][]{ emsellem2011,brough2017,veale2017b,vandesande2017b,green2018,graham2018}.

Note that the expected relation of both populations with stellar mass is also the primary reason for not using the \lre-\ee\ diagram to fit the data. Even though inclination has a significant impact on the observed \lre\ values that could be partially accounted for by using ellipticity instead of mass, we argue that without an inclination correction we only get an increase in the scatter and overlap of both \lre\ distributions. While we could attempt to correct for inclination, this parameter is poorly constrained for galaxies below $\lre<0.2$. Only correcting a subset of the data could lead to a bimodality by construction (see Section \ref{subsec:fast_slow_seeing}), which we want to avoid here. We further investigate the impact of inclination in Appendix \ref{app:lambdar_eo_random}.

\begin{figure*}
\includegraphics[width=\linewidth]{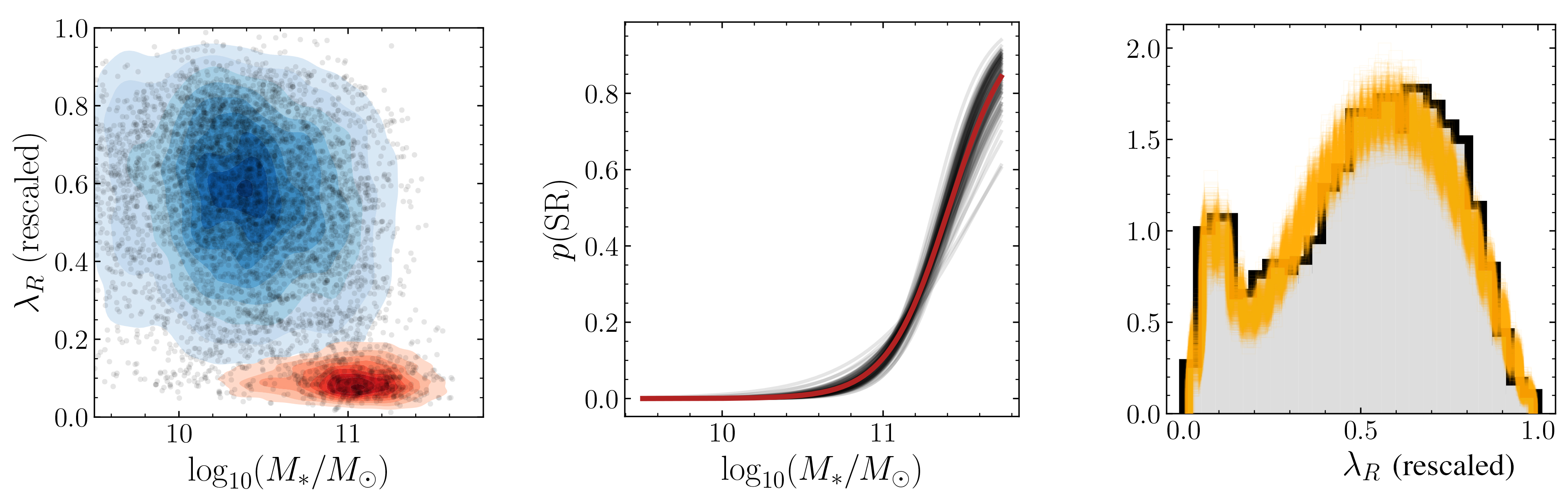}
\caption{Bayesian mixture model analysis to identify different kinematic populations. In panel (a), we show the seeing-corrected spin parameter proxy versus stellar mass, where the blue and red density contours show the amplitude of the beta distributions that we fit to the volume-corrected data. Note that we only show 5000 randomly drawn galaxies here. The "mixture probability" (i.e., the probability of being drawn from the second beta distribution describing the "slow" rotators) as a function of stellar mass is given in panel (b), where the black lines show 2000 realisations of the mixture model and the red line shows the average. Panel (c) shows the total distribution in \lre\ summed over all stellar masses for the data (black), together with 2000 realisations of the mixture model in orange. At low stellar mass ($\logm<10.5$) the probability of finding galaxies that belong to a second low-\lre\ population goes to zero, whereas at high stellar mass, the probability for a second low-\lre\ population is very high. We note that the high \lre\ Beta distribution shows a small deviation from the observed data, which is further explored in Fig.~\ref{fig:jvds_lambdar_mass_distribution}. Nonetheless, the Bayesian mixture model analysis provides the most principled separation of the two distributions.
\label{fig:spv_mass_lambdar_bayesian}}
\end{figure*}

At this point, we also apply a volume correction to our cluster sample. The complete volume correction analysis will be presented by van de Sande et al. (in preparation), but we provide a short description here. The SAMI targets are drawn from the volume-limited GAMA survey with high completeness ($\sim90$ percent). However, the GAMA regions lack high over-density regions with halo mass greater than $\loghm\sim14.5$. For that reason the SAMI Galaxy Survey targeted an additional 8 cluster regions to fill this density gap. Nonetheless, the probability of finding an extremely massive cluster such as Abell 85 (the most massive cluster in the SAMI cluster sample) within the GAMA volume is less than one. Hence, a volume correction needs to be applied. 

We first calculate the total survey volume, using the stepped series of stellar mass limits as a function of redshift from which the SAMI Galaxy Survey targets were selected \citep[see][]{bryant2015}. For each volume, we calculate the predicted halo mass function from \citet{angulo2012} using \textsc{HMFcalc}: An Online Tool for Calculating Dark Matter Halo Mass Functions \citep{murray2013}. With that halo mass function, we can then obtain a probability of finding a cluster galaxy within the SAMI-GAMA volume. For example, we find that the probability of observing a galaxy in the most massive cluster Abell 85 is $\sim1/38$. To take this over-abundance of cluster galaxies into account, we randomly draw each galaxy in the full survey - with replacement- using an oversampling of 38 multiplied by a galaxy's volume correction. In practise, a galaxy in the most massive cluster (Abell 85) will be drawn only once, whereas a galaxy in the GAMA region will be drawn 38 times. For each draw, we add a random number to each data point derived from the $1\sigma$ measurement uncertainty on \lre\ and a typical $1-\sigma$ stellar mass uncertainty of 0.1dex. The total volume-corrected dataset consists of 53,587 data points.

We then use this volume-corrected sample to fit the \lre\ distribution as a function of stellar mass. The shape parameters of both beta functions ($\alpha_1, \beta_1$ and $\alpha_2, \beta_2$ respectively) are defined to be linear functions of stellar mass. This allows the two beta distributions to vary their width and location in the \lr-\logm{} plane to match the observational data. Note that the model has the freedom to let one set of parameters have zero contribution if the data do not motivate two populations. A full mathematical description of the model including priors is given in Appendix \ref{app:bayesian_mm_priors}.

We fit this model using the \textsc{python} interface to the probabilistic programming language \textsc{Stan} \citep{stan2017}. \textsc{Stan} uses a modified version of the Hamiltonian Monte Carlo algorithm \citep{Duane1987, Hoffman2014} to sample the model's posterior probability distribution and perform full Bayesian inference of the parameters. During the fitting, we run 4 separate chains for 500 warm-up steps and 500 sampling steps each. The warm-up steps are then discarded. We ensure that there are no divergent transitions during the sampling and that the Gelman-Rubin convergence diagnostic $\hat{R}$ \citep{GelmanRubin1992} for each parameter is within normal values ($1<\hat{R}<1.1$). Note that no binning in \logm\ or \lre\ is applied in the fitting process; each data-point is treated independently.

\begin{figure*}
\includegraphics[width=\linewidth]{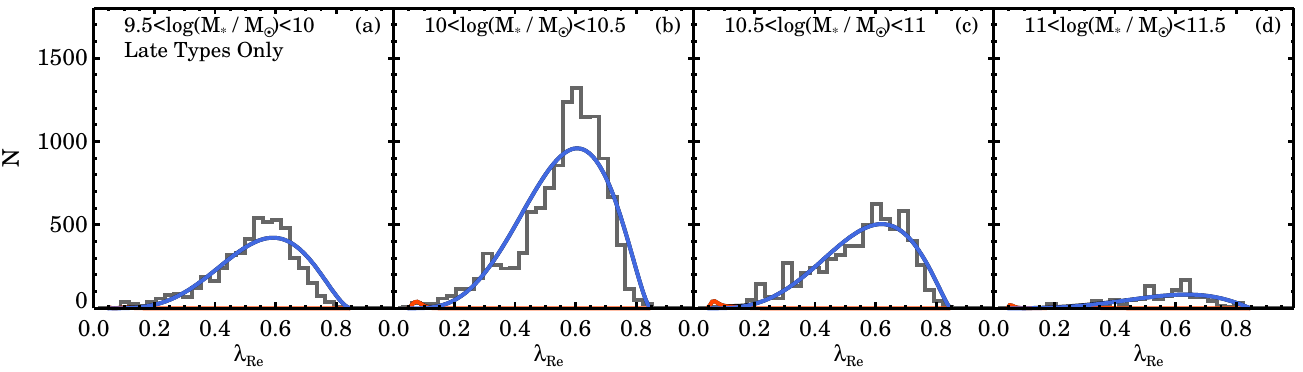}
\includegraphics[width=\linewidth]{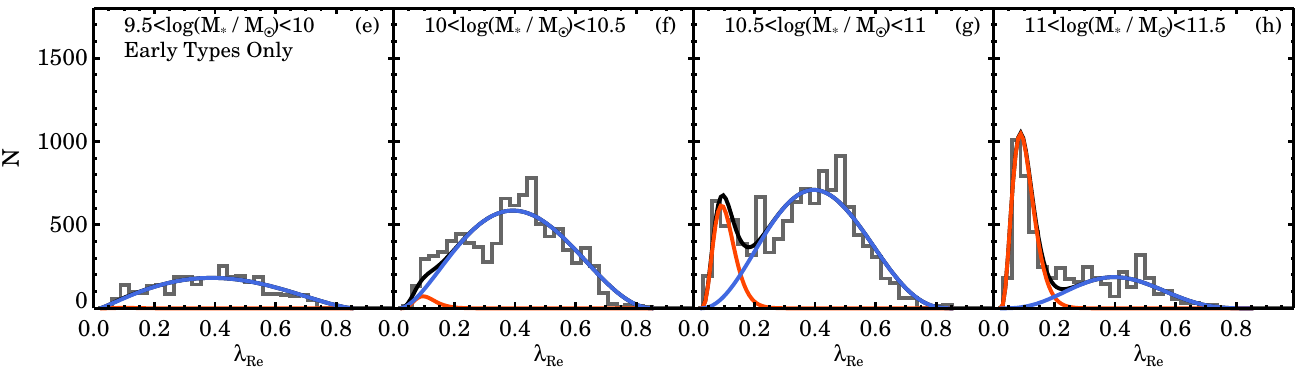}
\includegraphics[width=\linewidth]{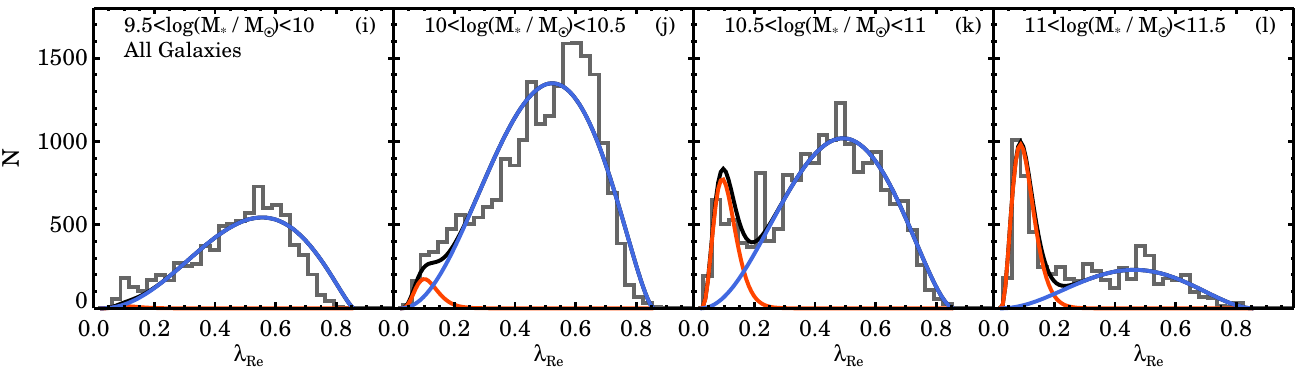}
\caption{Distribution of the seeing-corrected \lre\ from the volume-corrected sample in four stellar mass bins, split by visual morphology into late-type galaxies (top row), early-type galaxies (middle-row), and the full sample (bottom-row). The observed distribution is shown in grey, the best-fitting mixture model in black with the two beta distributions shown separately on top in blue and red. We emphasise that this mixture model has been fit to all galaxies in our sample simultaneously, and has \textit{not} been fit to the binned data shown here. Late-type galaxies are well described by a single beta distribution, with a near zero contribution from a second distribution. For early-type galaxies, above stellar mass  $\logm>10$, we find an increasingly dominant population of pSRs at low-\lre. The position and amplitude of this second distribution remains the same when we fit the entire population versus early-types only. In general we find a good fit to the data at low and high-stellar masses (column 1, 3, and 4), but between $10.0<\logm<10.5$ (second column) the shape of the high-\lre\ beta distribution does not match the data as well as for other stellar mass bins.
\label{fig:jvds_lambdar_mass_distribution}}
\end{figure*}

\subsubsection{Probabilistic Fast and Slow Rotators}

The key results from this analysis are shown in Fig.~\ref{fig:spv_mass_lambdar_bayesian}. We identify two clear distributions within the \lre-\logm\ diagram with moderate overlap. In Fig.~\ref{fig:spv_mass_lambdar_bayesian}(a), the blue high \lr\ distribution, which is consistent with the location of galaxies traditionally called fast rotators, dominates at low and intermediate stellar masses. Above $\logm>10.5$ the contribution from a second population at low \lre\ as shown in red, consistent with traditional slow rotators, becomes more and more dominant towards high stellar mass. While these two populations occupy the exact regions where we expect fast and slow rotators to reside, we want to avoid using the exact same terminology when the process of identifying these two populations is very different from previous studies. Instead, we will refer to these distributions as probabilistic fast and slow rotators (pFRs and pSRs).

In ~Fig.~\ref{fig:spv_mass_lambdar_bayesian}(b), we find that the probability of a galaxy being drawn from the pSR distribution rapidly increases as a function of stellar mass, particularly above $\logm>11$, in agreement with previous studies  \citep[e.g.,][]{ emsellem2011,brough2017,veale2017b,vandesande2017b,green2018,graham2018}. However, the model prediction becomes increasingly uncertain above \logm>11.4 where the number of observed SAMI Galaxy Survey galaxies rapidly drops. The \lre\ distribution summed over the entire mass range is shown in Fig.~\ref{fig:spv_mass_lambdar_bayesian}(c). There is a minor offset of the peak of the pFR distribution as compared to the peak of the data, but the peak of the pSR is well matched to the data. 

In Fig.~\ref{fig:jvds_lambdar_mass_distribution} we split the sample into four equal bins of stellar mass to investigate this offset further. In particular, we are interested in determining whether or not the main assumption that the \lre\ distribution can be described by a beta function is valid. Furthermore, we separate the late and early-type distributions because we expect the behaviour of these populations to be different. Indeed, for late-types only, we find that the \lre\ distribution can be described by a single beta function associated with the pFRs, with minimal contribution of a second beta distribution. However, for early-type galaxies, a second dominant peak appears approximately around $\logm>10.5$, which is also well fitted by a beta distribution.

In the combined sample (Fig.~\ref{fig:jvds_lambdar_mass_distribution} bottom row), we see how the relative contributions of early-types and late-types as a function of stellar mass impact the \lre\ distribution. Below $\logm<10.5$ the late-type population dominates, which is reflected by the strong peak at $\lre\sim0.6$, whereas towards higher stellar mass the contribution from early-type galaxies becomes more dominant. Between $10.5<\logm<11$, we find the combined late-type pFR and early-type pFR distribution, which have roughly equal numbers of galaxies, is also well described by a single beta distribution. This is perhaps surprising as the individual late-type and early-type pFR distributions are different in shape with peak values that are offset by $\sim0.2$ in \lre. While this does not exclude that the two populations are kinematically different, it validates the choice of a single beta distribution for the combined early-type and late-type pFR population. We also note that the peak and width of the pSR distributions are identical when analysed as part of the full sample or within the early-type sample. We emphasize that this is not by construction, but an outcome of our mixture model analysis. 

Nonetheless, while in three out of four stellar mass bins we find a relatively good fit of our model to the data, in the bin with mass interval $10.0<\logm<10.5$ we see a poorer fit to the data. The discrepancy between the model and the data could be caused by a relatively high broad peak in the distribution at $\lre\sim0.6$ for late-types, or because we enforce a smooth transition of the beta distributions as a function of stellar mass using a linear relation. Instead, we attribute the poor fit in this mass regime to the lower peak around $\lre \sim 0.2$. The larger abundance of galaxies at these low \lre\ values could be explained by a population of galaxies that we previously identified as NOR-WF or OR-WF (see Section \ref{subsec:or_nor}). These galaxies might have outer kinematic structures consistent with either the pSR or pFR population, but the inner kinematics offset the \lre\ measurements from their main distribution. Removing these galaxies from the sample indeed results in a better visual fit, but as our goal here is to use only the spin parameter proxy and stellar mass without secondary identifiers to clean or pre-select our sample, we did not attempt to improve this further.

To summarise, using a Bayesian mixture model analysis we have demonstrated that two beta distributions are required to describe the observed \lre\ distribution as a function of stellar mass. For early-type galaxies, the location of pFR peak has a lower \lre\ value as compared to pFR late-type galaxies, but the locations of the pFR and pSR peaks do not change with stellar mass. The amplitude of pSR distribution rapidly increases with stellar mass, but the peak and width remain constant. When we analyse the full SAMI Galaxy Survey sample, we find that the data are well-described by two beta distributions, but because the relative fraction of late and early-type galaxies changes as a function of stellar mass, we also find  that the width and peak of the pFR distributions change moderately. These results are consistent with the findings of \citet{guo2020}, who show that in the local Universe, above $\logm>10.5$, both late- and early-type populations become important in the total stellar mass budget, whereas below $\logm>10.5$ only one population is needed to reproduce the stellar mass function of galaxies.


\begin{table}
	\centering
	\caption{Confusion matrix for the condition of pSR versus pFR using the SR versus FR test.}
	\label{tbl:tbl3}
	\begin{tabular}{l c c} 
		\hline
		\hline
		  & pSR & pFR \\
		\hline
		SR  & True Positive & False Positive \\
		\hline
		FR  & False Negative & True Negative\\
		\hline
		\hline
	\end{tabular}
\end{table}

\begin{figure*}
\includegraphics[width=\linewidth]{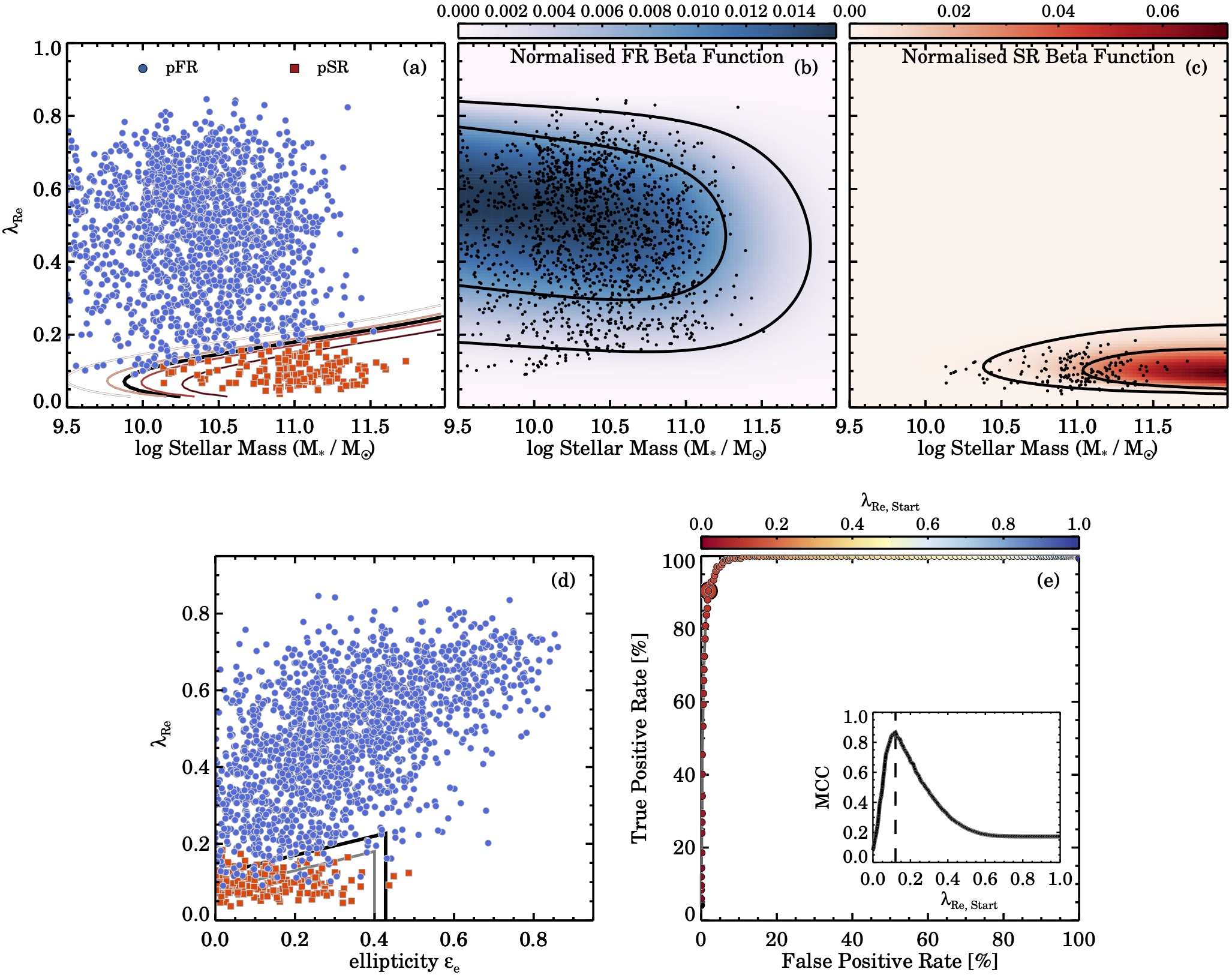}
\caption{Seeing-corrected spin parameter proxy versus stellar mass and ellipticity. We present the probabilistic Fast and Slow rotators (pFRs and pSRs) as blue and red coloured symbols in panel (a), with contours marking the probability for a galaxy to be a slow rotator (p(SR)=50\% black, and from light to dark red as 20\%, 40\%, 60\%, 80\%). Panels (b) and (c) show the fast and slow rotator PDFs normalised in each mass bin, with contours enclosing 68 and 95 percent of the PDF, and indicates the probability of finding a pSR or pFR if the mass-function is flat. The SAMI classified pFRs and pSRs are shown in the \lre-\ee\ space in panel (d) where the optimal selection region from our ROC analysis is shown in black together with the \citet{cappellari2016} SR selection box in grey. Because we select pFRs and pSRs from the p(SR)=50\% contour in the \lre-\logm, the small increase of the \lre\ limit with stellar mass results in some minor contamination in the \lre-\ee\ space. Nonetheless, panel (e) indicates an extremely high True Positive Rate with low False Positive Rate, but this is partly by construction.
\label{fig:jvds_mass_lambdar_psr_pfr}}
\end{figure*}

\subsubsection{Identifying Fast and Slow rotators using Bayesian Mixture Models as a Prior}

We now use the Bayesian mixture model to identify which galaxies are most likely pFRs and pSRs. In Fig.~\ref{fig:jvds_mass_lambdar_psr_pfr}(a) we show the SR probability contours, where $p({\rm{SR}}) = PDF_{\rm{\,SR}}~/~(PDF_{\rm{\,SR}}~ +~PDF_{\rm{\,FR}})$. We define a galaxy as a pSR when the $p({\rm{SR}})$ is higher than 50 percent. Note that this selection does not take into account ellipticity or visual morphology. For that reason, counter rotating disks that are often excluded using an ellipticity cutoff, or face-on spirals can still be selected as pSR when they are clearly different in structure and kinematics as compared to massive-triaxial ellipticals. We find that the fraction of pSRs strongly increases with stellar mass, which was already demonstrated in Fig.~\ref{fig:spv_mass_lambdar_bayesian}. But, the pSR contours are more tightly packed in the \lre\ direction, whereas the stellar mass range from the 20th to 80th probability covers nearly a dex in stellar mass. 

In Figs \ref{fig:jvds_mass_lambdar_psr_pfr}(b) and (c) we show the mass normalised FR and SR PDFs. The contours indicate the 68 and 95 percentiles or how likely we are to find a pFR or pSR in that region. We overlay the SAMI Galaxy Survey data to identify low probability pFR and pSR galaxies. For example, in Fig.~\ref{fig:jvds_mass_lambdar_psr_pfr}(b) below $\logm<10.5$ there are a number of pFR galaxies that lie below the 95 percentile contours. In our visual kinematic classification analysis (Fig.~\ref{fig:jvds_mass_lambdar_nor_or}) we already found that this region is predominately occupied by OR-WF, whereas the majority of OR have no such feature. Similarly, we find low-mass pSRs that are outside the 95 percentile. However, whereas the PDFs are normalised as a function of stellar mass, the SAMI observed mass function peaks around $\logm\sim10.5$. It is therefore no surprise that we find more pSR at $\logm<10.5$ than the PDF from a flat stellar mass distribution suggests.

Similar to the test we performed for \kinemetry\  and the visual kinematic classification, we will now treat pFR versus pSR as a condition that a galaxy can have, with  \lre-\ee\ diagram as the diagnostic to identify this condition. The confusion matrix is given in Table \ref{tbl:tbl3}. In Fig.~\ref{fig:jvds_mass_lambdar_psr_pfr}(d) we investigate where pFRs and pSRs reside in the \lre-\ee\ plane. Unsurprisingly, there is a clear separation between both classes because the pFR and pSR classifications come directly from the \lre-\logm\ probability cutoffs. Therefore, we are mainly gauging how much overlap of the pFR and pSR distribution there is when swapping \logm\ for \ee. While this may seem somewhat artificial, we note that both \kinemetry\ and the kinematic visual classification are also based on the kinematic data. Thus, all kinematic identifiers have some degree of interdependence. 

We find no dependence on the location of pSR with respect to the ellipticity (Fig.~\ref{fig:jvds_mass_lambdar_psr_pfr}d), and in particular towards low \ee, we do not detect a decline in the \lre\ values of pSR. This result is similar to the NOR category defined from visual classification. To quantify this trend, we explore different selection boxes with varying slopes, start and end positions in both \lre\ and \ee. Indeed, the optimal selection function has a nearly flat slope starting at \lre=0.14 and extends out \ee=0.5, with an MCC value that is higher than the MCC value from the default selection region from Eq.~\ref{eq:sr_vary} (0.890 vs. 0.865, respectively).

In Fig.~\ref{fig:jvds_mass_lambdar_psr_pfr}(e) we show the TPR versus FPR of our test as well as the MCC distribution for the default selection region region from Eq.~\ref{eq:sr_vary}. The optimal selection only has $\lr_{\rm{start}}$ value of 0.12, with a TPR of 90.4 percent with a small FPR of 1.9 percent, and a PPV of 83.0 percent. Overall, there is an excellent agreement between the selection of probabilistic fast and slow rotators using the \lre-\ee\ diagram, although we re-emphasise that this is mostly by construction.

\section{Different Kinematic Distributions in Cosmological Hydrodynamical Simulations}
\label{sec:simulations_approach}

Cosmological hydrodynamical simulations offer great insight into the formation and evolution of galaxies from high-redshift ($z\sim50$) to the present-day ($z=0$). By simultaneously comparing structural, dynamical, and stellar population measurements from simulations and observations,  \citet{vandesande2019} demonstrate that recent large cosmological simulations are now capable of reproducing many of the known galaxy relations. While recent comparisons with IFS measurements showed a qualitatively good agreement for several fundamental galaxy relations \citep[see e.g.,][]{penoyre2017,schulze2018,lagos2018b,choi2018,walomartin2020,pulsoni2020}, quantitatively some fundamental parameters are not well reproduced \citep{lange2016,vandesande2019,xu2019}; moreover, areas of discrepancy and agreement vary between the different simulations \citep{vandesande2019}. Nonetheless, these simulations are useful to interpret the kinematic properties of different galaxy populations across time and different environments \citep{teklu2015, dubois2016,welker2017,remus2017,penoyre2017,choi2017,kaviraj2017,choi2018,lagos2018a,lagos2018b, schulze2018, martin2018,pillepich2019, walomartin2020, pulsoni2020, schulze2020}. 

To assess whether observational selection criteria can be successfully applied to data from simulations to separate fast and slow rotators, we will now repeat the mixture model analysis on IFS mock-observations from cosmological hydrodynamical simulations. We use the data as presented by \citet{vandesande2019} where we used the \ea, \ha, and \textit{Magneticum Pathfinder} simulations. All simulations model key physical processes of galaxy formation, including gas cooling, star formation, feedback from stars and from supermassive black holes, although each simulation adopts different philosophies for calibrating to and reproducing observational results. The details of the simulations and mock-observations are summarised below.

\subsection{EAGLE and Hydrangea}
\label{subsec:eagle_sims}

From the publicly available \ea\ project \citep[Evolution and Assembly of GaLaxies and their Environments;][]{schaye2015,crain2015,mcalpine2016} data, we use the reference model Ref-L100N1504 that has a volume of (100 Mpc)$^3$ co-moving. We combine \ea\ with \hy\ that consists of 24 cosmological zoom-in simulations of galaxy clusters and their environments \citep{bahe2017} to provide a better environmental match to the observed SAMI Galaxy Survey.
\hy\ is part of the larger Cluster-\ea\ project \citep{barnes2017}. Cluster-\ea\ is similar to \ea\, but with different parameter values for the active galactic nuclei (AGN) feedback model, to make it more efficient. Both \ea\ and \hy\ adopt the Planck Collaboration XVI (\citeyear{planck2014}) cosmological parameters ($\Omega_\mathrm{m}$=0.307, $\Omega_{\Lambda}=0.693$, $H_{0}=67.77$\kms\ Mpc$^{-1}$). Each dark matter particle has a mass of $9.7\times10^6\msun$, and the initial gas particle mass is $1.81\times10^6\msun$. The typical mass of a stellar particle is similar to the gas particle mass. In what follows, we will refer to the joined \ea\ and \hy\ sample as \eap. 

\subsection{Horizon-AGN Simulations}
\label{subsec:horizon_agn_sims}
The second set of cosmological hydrodynamic simulations is \ha\, with the details presented by \citet{dubois2014}. Here, we use the simulation box with a volume of (142 Mpc)$^3$ co-moving with an adopted cosmology that is compatible with the Wilkinson Microwave Anisotropy Probe 7 cosmology \citep[$\Omega_\mathrm{m}$=0.272, $\Omega_{\Lambda}=0.728$, $H_{0}=70.4$\kms\ Mpc$^{-1}$;][]{komatsu2011}. \ha\ uses a grid to compute the hydrodynamics, employing adaptively refinement to the local density following a quasi Lagrangian scheme \citep{teyssier2002}, with cells that are 1kpc wide at maximal refinement level. The dark matter particle mass is $8\times10^7\msun$, and the adopted resolution is such that the typical mass of a stellar particle is $2\times10^6\msun$.

\subsection{MAGNETICUM Simulations}
\label{subsec:magneticum_sims}

The third set of cosmological hydrodynamical simulations that we will use are the \textit{Magneticum Pathfinder} simulations (www.magneticum.org), hereafter simply \ma\ (see Dolag et al., in preparation, \citealt{hirschman2014} and \citealt{teklu2015} for more details on the simulation). We use the data from the medium-sized cosmological box (Box 4) with a volume of (68 Mpc)$^3$ co-moving at the ultra high resolution level. \ma\ adopts a cosmology compatible with the Wilkinson Microwave Anisotropy Probe 7 cosmology \citep[$\Omega_\mathrm{m}$=0.272, $\Omega_{\Lambda}=0.728$, $H_{0}=70.4$\kms\ Mpc$^{-1}$;][]{komatsu2011}. The dark matter and gas particles have masses of respectively $5.1\times10^7\msun$ and $1.0\times10^7\msun$, and each gas particle can spawn up to four stellar particles.

\subsection{Mock Observations}
\label{subsec:mock_obs}

The method for extracting kinematic measurements from \eap\ are described in \citet{lagos2018b}, for \ha\ in \citet{welker2020}, and in \citet{schulze2018} for \ma, all corrected to $H_{0}=70.0$\kms\ Mpc$^{-1}$. For all simulations, we extract $r$-band luminosity-weighted effective radii, ellipticities, line-of-sight velocities and velocity dispersions, adopting techniques that closely match the observations. The \lr\ values for \ha\ and \ma\ are derived using Eq.~\ref{eq:lr}, whereas for \eap\ we use the definition as described in \citet{emsellem2007}. Note that these different \lr\ definitions do not impact our analysis as we are investigating the separation of two kinematic families within each distribution, without a direct quantitative comparison. Specifically, the different \lr\ definitions will only significantly impact galaxies with high values of ellipticity, well above the region where both kinematic distributions are expected to overlap.

A lower mass limit of $\mstar=5\times10^9\msun$ is used for \ea, \hy, and \ha, but a higher mass limit of $\mstar=1\times10^{10}\msun$ for \ma, to ensure that the simulated measurements from the mock-observations are well-converged. Nonetheless, we acknowledge that with the spatial resolution of these simulations, effects similar to observational beam-smearing might play a role in the kinematic measurements of mock-observed simulated galaxies. Lastly, a mass-matching technique is used to remove the difference between the observed and simulated stellar mass function for a clearer comparison of the results \citep[for more details see][]{vandesande2019}, but we note that we find consistent results when no mass-matching is enforced. 

\begin{figure*}
\includegraphics[width=0.99\linewidth]{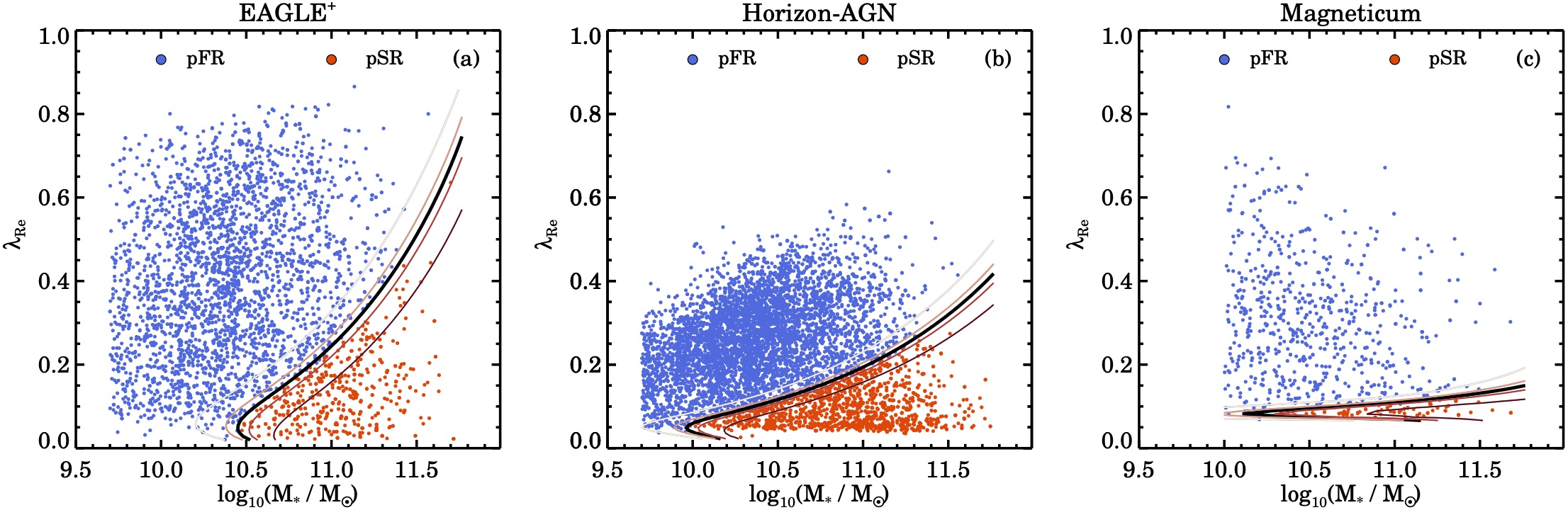}
\includegraphics[width=0.33\linewidth]{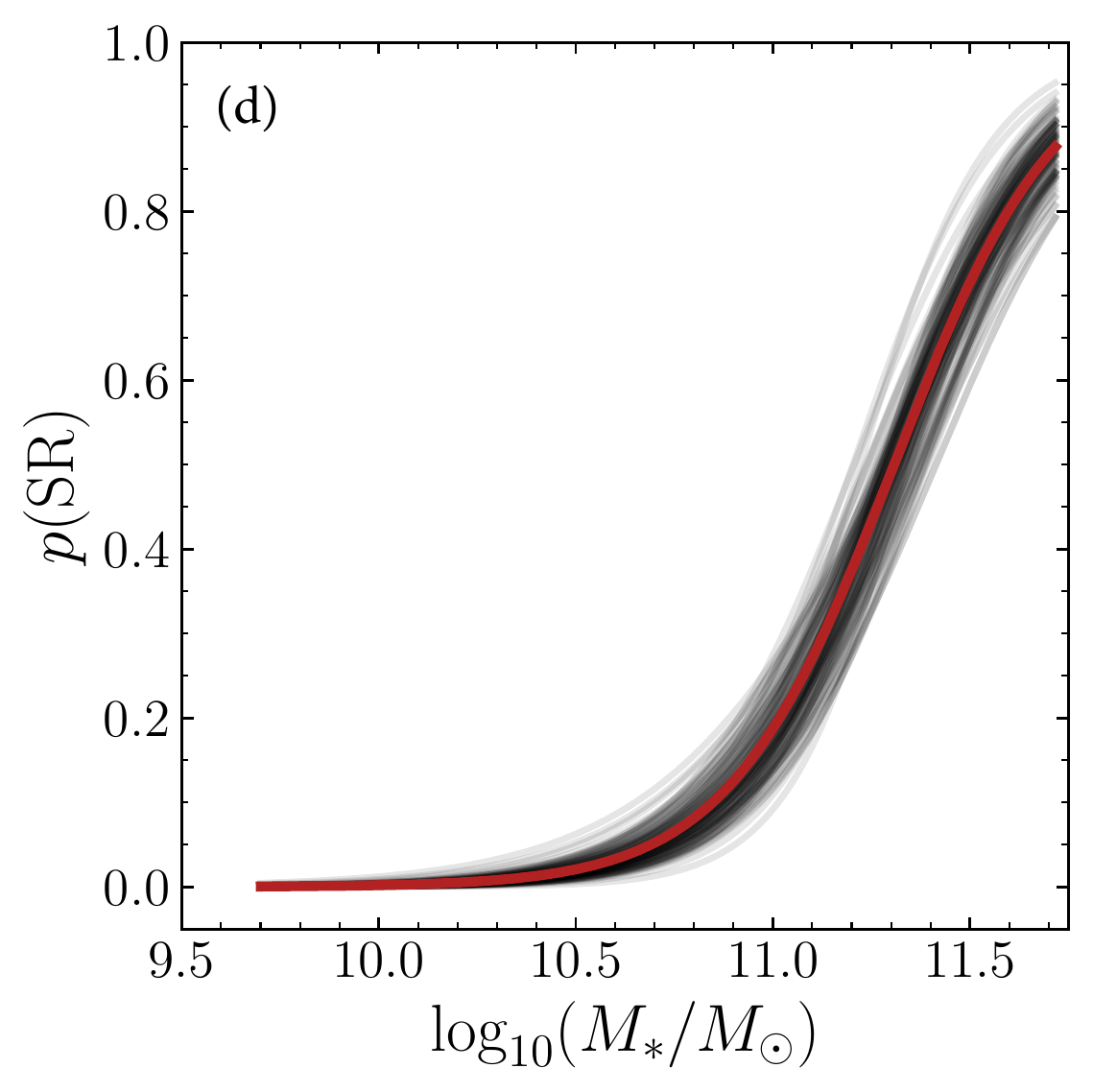}
\includegraphics[width=0.33\linewidth]{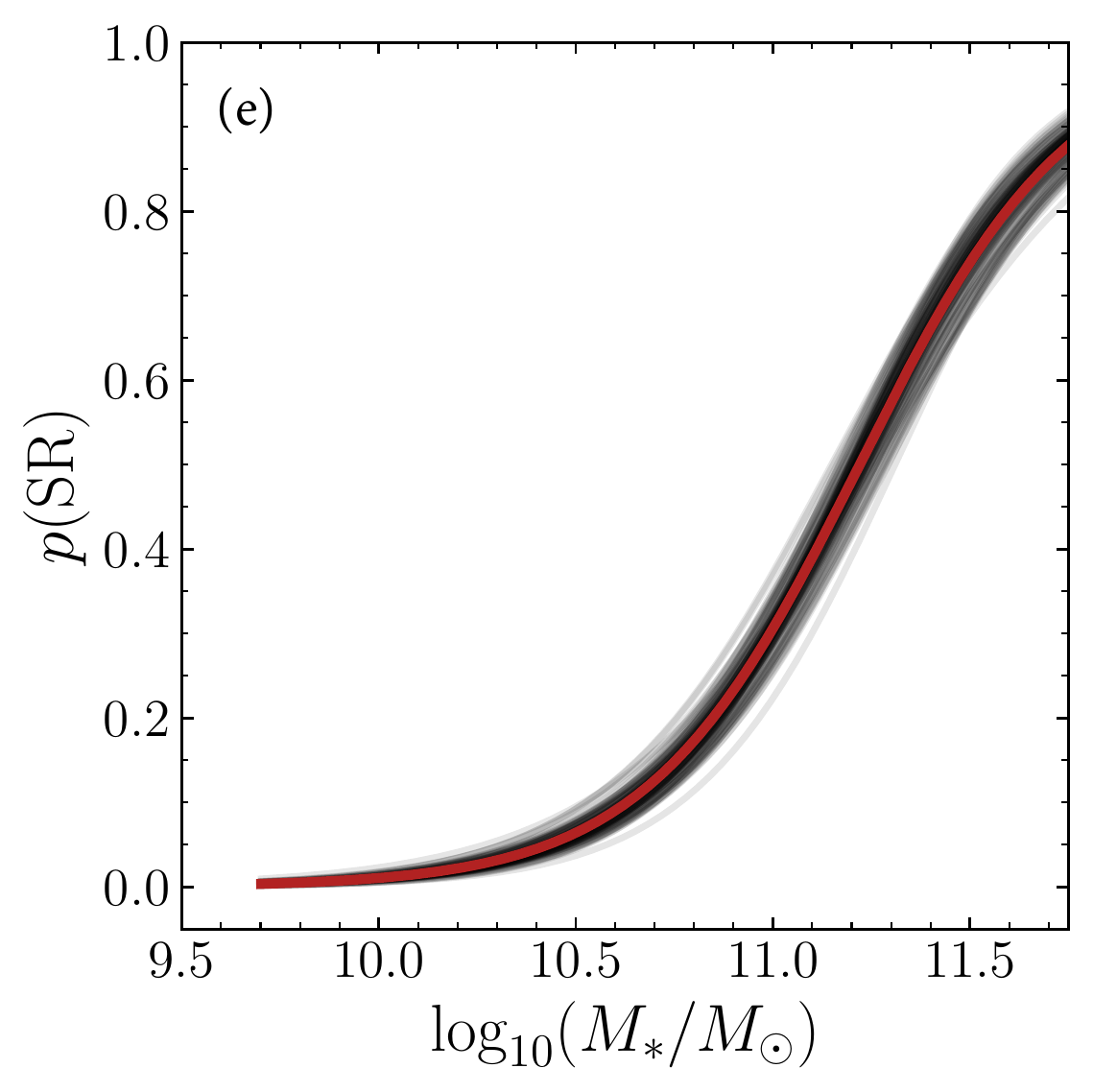}
\includegraphics[width=0.33\linewidth]{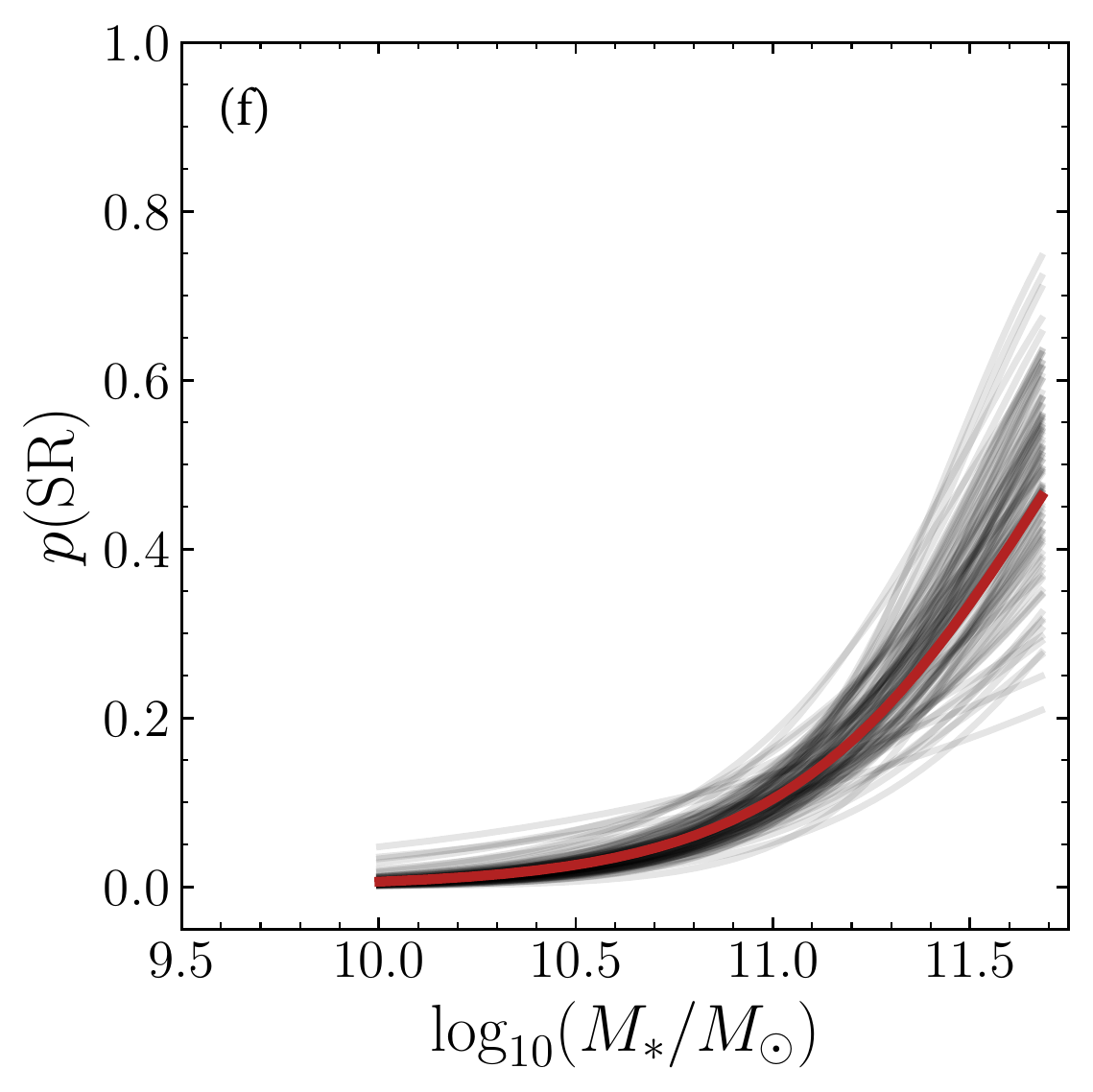}
\caption{Bayesian mixture model analysis to identify different kinematic populations in the \eap\ (left), \ha\ (middle), and \ma\ (right) simulation using the spin parameter proxy versus stellar mass. The probabilistic fast and slow rotators (pFRs and pSRs) are shown as blue and red coloured symbols. The contours mark the probability for a galaxy to be a slow rotator (p(SR)=50\% black, and from light to dark red as 20\%, 40\%, 60\%, 80\%). We find that the pSR/pFR divide in \ha\ closest matches the observations, whereas the \eap\ and \ma\ pSR/pFR cutoff reveal a respectively stronger and milder increase of the pSR distribution as a function of stellar mass.
\label{fig:jvds_mass_lambdar_psr_pfr_sim}}
\end{figure*}

\begin{figure*}
\includegraphics[width=1.0\linewidth]{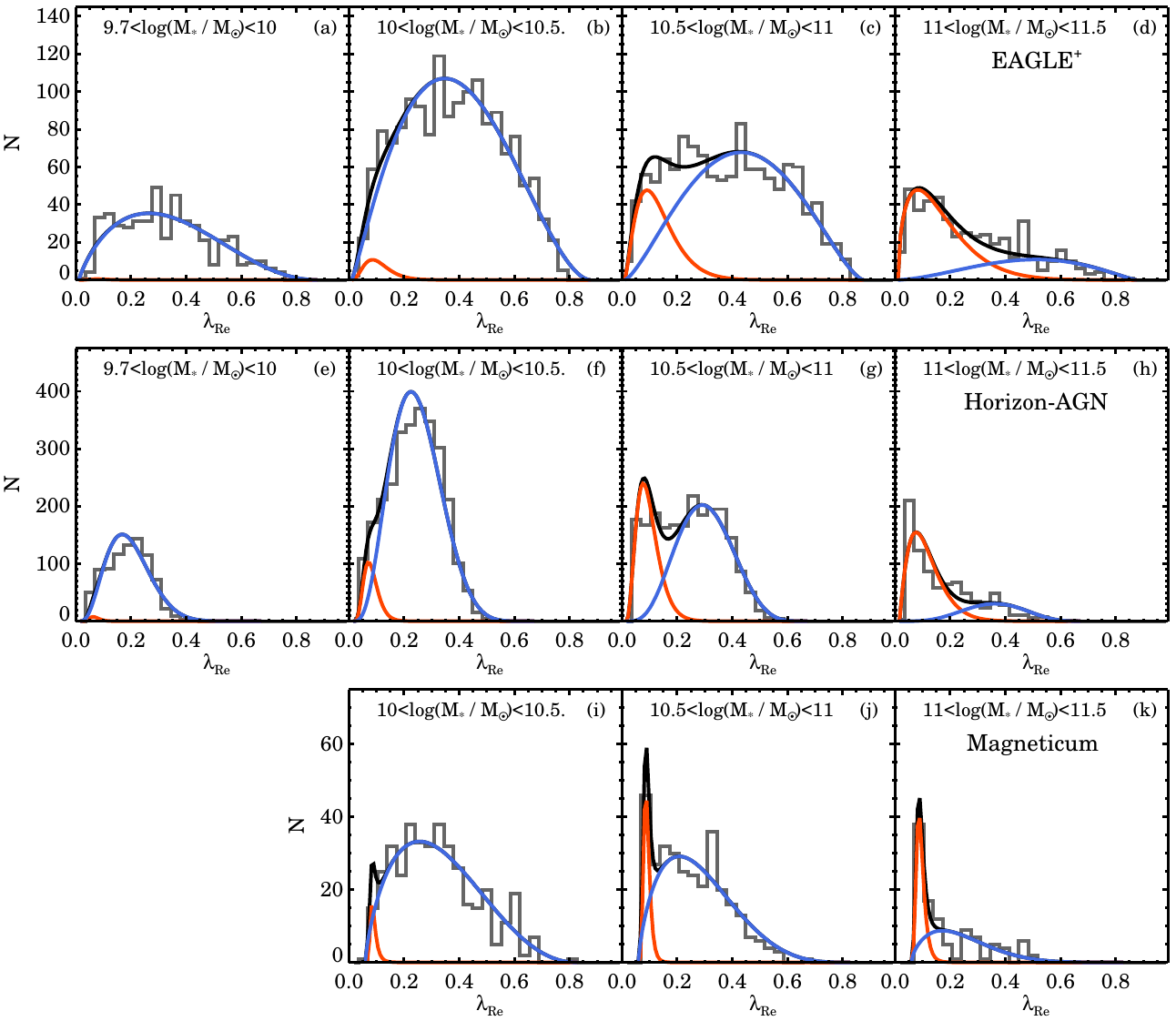}
\caption{Distribution of \lre\ from \eap\ (top), \ha\ (middle), and \ma\ (bottom) in four stellar mass bins. The observed distribution is shown in grey, the best-fitting mixture model in black with the two beta distributions shown separately on top in blue and red. Note that the mixture models have been fit to all mock-observed galaxies in the samples simultaneously, and has \textit{not} been fit to the binned data shown here. In all three simulations, the Bayesian mixture model indicates a bimodal distribution, but the differences between the \lre\ distributions from the three simulations are considerable with a large overlap of the pFR and pSR distributions.}
\label{fig:jvds_lr_distr_sims}
\end{figure*}

\subsection{Separating Fast and Slow Rotators in Simulations using Bayesian Mixture Models}
\label{subsec:bayesian_simulation}

We now repeat the Bayesian mixture model analysis from Section \ref{subsec:bayesian_approach}. Our goal is to see whether or not our mixture model recovers a meaningful separation of the two kinematic distributions within the simulated data, even though we have not demonstrated yet that two kinematic populations exist. The results for all three simulations are presented in Fig.~\ref{fig:jvds_mass_lambdar_psr_pfr_sim}, with the left column showing the \eap\ analysis, \ha\ in the middle column, and \ma\ on the right-hand side. We present the separation of pSRs (red) and pFRs (blue) using the 50 percent probability levels in the top row, whereas the bottom row shows the probability of being drawn from the pSR beta distribution.

The difference in the location of the pSR population is striking for all three simulations. As compared to pSR selection region from observations, we detect a steeper upturn in \lre\ towards high stellar masses for \ha\ and even steeper for  \eap. In contrast, the \ma\ pSR contours increase much slower as a function of stellar mass, with a narrow range in permitted \lre\ values, although the upper limit of the pSR selection is similar to observations. The other striking difference between the observations and simulations is the location and the shape of the pFR distribution. Below $\logm<10.5$ for \eap\ and $\logm<11$ for \ma, the pFR distribution covers the full \lre\ range, whereas for \ha\ the peak of the pFR distribution is very low from $\lre\sim0.2$ to $\sim0.4$.

From the probability of the pSR beta distribution as a function of stellar mass (Fig.~\ref{fig:jvds_mass_lambdar_psr_pfr_sim}f) it is clear that the Bayesian mixture model for the \ma\ simulation data is not as well constrained as compared to the other two simulations and the observed SAMI Galaxy Survey data (Fig. \ref{fig:spv_mass_lambdar_bayesian}). The numerous model realisations indicate that there is a considerable range of possible solutions. The probability of finding the pSRs distribution at the highest stellar masses in \ma\ is also lower as compared to the observations (respectively, $\sim$0.5 versus $\sim$0.75), whereas \eap\ and \ha\ predict consistent values. Furthermore, in Fig.~\ref{fig:jvds_mass_lambdar_psr_pfr_sim}d)-e) we find that for \eap\ and \ha\ the p(SR) as a function of stellar mass are similar. It is not obvious that this should be the case, especially given the large differences in the ranges of \lre\ from the two simulations (Fig.~\ref{fig:jvds_mass_lambdar_psr_pfr_sim}a-b).

In order to see how well the mixture model separates the pSR and pFR distributions in the simulations data, we present the \lre\ distributions in different stellar mass bins in Fig.\ref{fig:jvds_lr_distr_sims}. We find a wide variety of \lre\ distributions, both in terms of shape and maximum \lre\ extent. Most noticeably for \eap, and to a lesser extend in \ha, we find that the width of the pSR distribution increases with increasing stellar mass, with a tail towards higher and higher \lre\, even though the peak of the pSR distribution remains at the same location. However, for \ma\ the pSR distribution is extremely narrow and does not change considerably as a function of stellar mass.  

In all three simulations the mixture model suggests a bimodal distribution, even though two distinct peaks are not evident for each simulation in Fig.\ref{fig:jvds_lr_distr_sims}. While this does not imply that multiple kinematic populations do not exist, it does demonstrate the value of investigating the kinematic distributions beyond the work as presented in  \citet{vandesande2019}. Here, we find that the overlap between the pSR and pFR distributions is more considerable in the simulations as compared to observations, and that the dividing line for pSR and pFR is at different \lre\ values as a function of stellar mass. Thus, a good agreement between the observed and simulated distributions does not automatically imply that the ratio of sub populations matches as well. These results also show that the observational selection criteria that are used to classify galaxies into fast and slow rotators are not suitable to study the fractions of the simulated populations as a function of stellar mass or environment. Within the same SR selection region, between observations and simulations it is unlikely that a comparable population of galaxies will be selected without considerable contamination.

\section{Discussion} 
\label{sec:discussion}

The taxonomy of galaxies determined from their visual morphological properties has been a powerful tool to advance our knowledge of the processes that shape galaxies, with the Hubble Sequence \citep{hubble1926} and the De Vaucouleurs system \citep{devaucouleurs1959} still in active use today. However, like any other area where taxonomy is used, an introduction of hard boundaries between classes can create artificial dichotomies when in reality the transition between these classes could be continuous \footnote{See \citet{agraham2019} for a detailed discussion on the "artificial division of the early-type galaxy population" from size measurements.}. 

With increasingly large samples of galaxies with resolved kinematic measurements, various kinematic classifications have now been proposed. Some of these naming conventions have perhaps led to an oversimplification of the way we view the kinematic galaxy population with an assumption that the previously proposed classes are distinct and independent. In this paper, we have investigated how well we can separate a bimodal kinematic distribution in the galaxy population, specifically when the data quality is more severely impacted by seeing and spatial sampling. In the second half of this analysis, we convincingly show that we can separate two kinematic populations, yet when relying on secondary classifiers such as kinematic visual classification or \kinemetry\ we find that the overlap of these different classes can be considerable. Because of the mixing of the different distributions, we are cautious to assign individual galaxies to a certain class. Instead, we advocate using probabilities to assess how likely it is that galaxies share the same properties. Nonetheless, historically various kinematic tracers have been used to promote the existence of a dichotomy. These will be reviewed in Section \ref{subsec:dich_vis}-\ref{subsec:dich_sims}, whereas the implications of our work are discussed in Section \ref{subsec:best_separate}-\ref{subsec:implications}.

\subsection{Separating Fast and Slow Rotators based on Visual Kinematic Classification}
\label{subsec:dich_vis}

We will start with a historical context on the visual kinematic identification of the first resolved kinematic maps that formed the foundation of the work that we present in this paper. Kinematic visual classification only became advantageous with the introduction of the SAURON IFS \citep{bacon2001}, followed by several IFS surveys. But, as we will argue in this section, the lack of a clear and well-defined classification scheme and the limited field of view has made kinematic visual classification overly subjective with some key results left open to alternative interpretations. 

The SAURON survey \citep{dezeeuw2002} yielded kinematic maps for a significant sample of 48 nearby early-type galaxies. A visual analysis revealed that most early-type galaxies show a significant amount of rotation, whereas others have complex dynamical structures inconsistent with being simple rotating oblate spheroids \citep{emsellem2004}. Visual classification of the kinematic maps was further explored in \citet{emsellem2007} and \citet{cappellari2007} who introduced  the fast and slow rotator classes. However, a detailed look at some of the early results suggests that even with good quality data, the classification is not always obvious. For example, we would argue that from fig.~1 in \cite{emsellem2007}, it is not clear that elliptical NGC 5982 (2nd row, 6th column) is a slow-rotator galaxy as the outskirts show rapid rotation. In our revised kinematic classification scheme from Section \ref{subsec:or_nor} this galaxy would be classified as an obvious rotator with features (OR-WF). 

Similar ambiguities can be found in the kinematic maps from the \at\ Survey using the SAURON IFS, as presented in fig.~1 from \citet{krajnovic2011}. For example, galaxies NGC 4472 and NGC 4382 are classified as NRR-CRC (Counter-Rotating Core) and RR-2M (Double Maxima), respectively. From \kinemetry\ the classification into NRR and RR is clear: $\mk=0.197\pm0.075$ for NGC 4472 and $\mk=0.025\pm0.009$ for NGC 4382. However, when attempting a visual classification of the velocity fields, we would argue that these velocity fields in the outskirts do not look that different, where there are clear signs of rapid ordered rotation. Both galaxies are round (\ee=0.17, 0.2) and have respective \lre\ values of 0.08 and 0.16, which puts them well-below and relatively close to the fast/slow rotator dividing line (SRs must have \lre<0.14 at \ee=0.20). Combined with the fact that both velocity fields do not extend beyond 0.26-0.36\re, it is hard to argue that one galaxy is a clear slow rotator whilst the other is not. 

Our revised kinematic classification scheme was purposely designed to take into account such ambiguity by adopting a new terminology of "obvious" and "non-obvious" rotation (ORs and NORs). Even though the extent of the kinematic maps is still important, outer versus inner rotation is more clearly defined in this revised scheme (see also Section \ref{subsec:disc_scale}). Additionally, our SAMI kinematic sample has at least one \re\ kinematic coverage for $\sim80$ percent of the galaxies, and only a relatively small fraction of galaxies do not extend beyond 0.5\re\ ($\sim5$ percent). 

The new visual classification scheme also allows each user to come up with their own interpretation of what ORs and NORs could look like, although we offer some examples of what the classes might look like. "Self-calibration" is important in this classification scheme, and to facilitate this, each classifier was shown their collection of galaxies assigned to the same class after each subset. By being allowed to swap galaxies between classes, the most optimal selection could be made. Given this ambiguity and flexibility in the classification scheme, the bimodal distribution of ORs and NORs in the \lre-\mstar\ space (Fig.~\ref{fig:jvds_mass_lambdar_nor_or}) is surprisingly clear, and confirms that two classes indeed exist.

Unlike some previous classifications \citep{graham2018}, we advocate for the aggregation of classifications from many independent classifiers. A comparison of visual classifications from three different authors on the SAMI maps, using the classifying scheme from \citet{krajnovic2011}, resulted in a large range in classification, with poor overall agreement. Results based off single classifiers may thus be biased and artificially skew the resulting distributions. A supervised machine learning approach (e.g., boosting), or a citizen science project \citep[e.g., Galaxy Zoo;][]{lintott2008}, could provide a viable solution for the near future when the number of galaxies with 2D kinematic maps is expected to grow beyond 10,000. We further emphasise that a more quantitative approach guided by these visual classifications should always be sought to connect to other studies and simulations

\subsection{Separating Regular and Non-Regular Rotators using Kinemetry}
\label{subsec:dich_RR_NRR}

\kinemetry\ offers a quantification of the irregularity of the velocity field \citep{krajnovic2006,krajnovic2008,krajnovic2011} and has been exploited to classify galaxies into Regular and Non-Regular classes. This classification scheme formed the basis for the revised \lr-\ee\ separation line of fast and slow rotators in \citet{emsellem2011} and \citet{cappellari2016}. We re-analyse the separation of \at\ fast and slow rotators using our ROC analysis (see Appendix \ref{app:kinemetry_a3d}) and find a clean separation of RR and NRR with a high Positive Prediction Value (89.7), but with an optimal selection region that has a higher \lre\ limit as compared to \cite{emsellem2011} or \citet{cappellari2016}.

Nonetheless, using a sub-set of high-quality SAMI Galaxy Survey data, \citet{vandesande2017a} showed that the \smk\ distribution from both \at\ and SAMI was peaked around $\smk\sim0.02-0.03$ but with a continuous tail towards higher \smk\ values. Yet, the shape of this distribution does not suggest that the distribution in \smk\ is bimodal. Furthermore, in Section \ref{subsec:kinemetry} we show that with lower quality data, the overlap of regular and non-regular galaxies in the \lr-\mstar\ and \lr-\ee\ space is considerable. As \smk\ is intrinsically correlated with \vs\ and \lr\ through the rotational component, galaxies with high \vs\ or \lr\ values will always have lower \smk\ if the $k_5$ component remains constant. Thus, while \kinemetry\ provides a useful and quantifiable measure of the kinematic asymmetry of the velocity field, we argue that this method does not provide strong evidence for a kinematic dichotomy.

\subsection{Separating two kinematic families using JAM Modelling}
\label{subsec:dich_jam}

Using Jeans Anisotropic MGE (JAM) modelling of \at\ galaxies \citep{cappellari2013b}, \citet{cappellari2016} shows that the distribution of $\kappa$ is bimodal, where $\kappa$ is the ratio of the observed velocity $V_{\rm{obs}}$ and the modelled velocity $V(\sigma_{\phi}=\sigma_{R})$ from JAM using an oblate velocity ellipsoid. Regular rotators are Gaussian distributed around $\kappa=1$, whereas non-regular rotators conglomerate towards zero, with minor overlap of both distributions (see also \ref{fig:jvds_a3d_distributions}c).

In Appendix \ref{app:kappa_a3d} we repeat our ROC analysis with \at\ data using $\kappa<0.65$ as the identifier for slow rotators. We only find a marginal improvement of the MCC parameter if we use JAM modelling as compared to using \kinemetry. Further, as noted by \citet{cappellari2016}, for non-regular rotators without a disk "the shape of the predicted $V(\sigma_{\phi}=\sigma_{R})$ is, even qualitatively, very different from the observed velocity field." JAM models assume axisymmetry, so perhaps it is unsurprising that triaxial ETGs have low values of $\kappa$. Nonetheless, dynamical modelling clearly shows that there is a family of galaxies consistent with being axisymmetric oblate rotating spheroids, and a class of galaxies with more complex dynamical properties that are not well-fitted by Jeans models. Schwarzschild modelling  \citep{schwarzschild1979}, has great potential for understanding the orbital structure of all types of galaxies 
\citep[see for example][using SAURON and CALIFA IFS data]{vandenbosch2008,vandeven2008,zhu2018a,zhu2018b}. Yet, the technique remains computationally expensive and requires high-quality data for the orbital decompositions to be non-degenerate.

\subsection{The impact of a Radial Extent on the Slow Rotator Classification}
\label{subsec:disc_scale}

The radial extent out to which the kinematic measurements are analysed also has a significant impact on the fast and slow rotator classification. When visually classifying kinematic maps, more attention may subconsciously be given to larger radii, with the eye being drawn to the larger number of spaxels in the outskirts. This is one of the reasons why our revised classification scheme distinguishes between obvious versus non-obvious rotation, which can be more easily picked up in the outskirts, with a refinement option for kinematic features towards the centre. Nonetheless, if the radial coverage of the kinematic maps do not extend beyond the central region (e.g., $<0.5\re$), the classification will be inherently biased.

For quantitative measurements, a scale of one \re\ is typically adopted because of observational constraints that are necessary to obtain the required S/N ratio to extract the LOSVD beyond this radius. However, there is no physical reason to restrict our kinematic measurements to within this radius and this approach might have hampered our understanding of galaxies \citep[e.g., see][]{agraham2019}. The radial coverage varies considerably between different IFS surveys, but more importantly, it typically changes as a function of stellar mass within surveys as well. The necessity for aperture-correcting \lr\ and \vs\ measurements is demonstrated in \citet{vandesande2017b} \citep[see also][]{deugenio2013} who show that there is a strong bias in the largest measurable kinematic radius as a function of stellar mass, which significantly impacts the fraction of slow rotators.

Studies measuring the kinematic parameters out to larger radius have demonstrated that the rotational properties of galaxies can change when measured at a different radius \citep[e.g.,][]{weijmans2009, proctor2009, arnold2011}. More recently, radial tracks within the $\lr-\varepsilon$ space have been utilised to study how the radial kinematic behaviour changes using increasingly large samples \citep{graham2017,bellstedt2017,foster2018,rawlings2020}. Kinematic full-spectral bulge–disk decomposition now also offer the possibility to explore \lr\ for bulges and disks separately \citep{tabor2017,mendezabreu2018,tabor2019,oh2020}. For a single kinematic galaxy classification that is based on an average quantity such as \lr, a larger aperture will always be preferred to include the largest possible fraction of stellar mass. But for classifying internal sub-components the aforementioned methods will be essential. While several past (e.g., SLUGGS, \citealt{brodie2014}; CALIFA, \citealt{sanchez2012}) and upcoming (e.g., Hector \citealt{bryant2016}; MAGPI, \citealt{foster2020}) IFS surveys are aimed at providing larger \re\ coverage, for the coming years the largest samples of galaxies will still be restricted to 1-2\re. 

Therefore, well-calibrated large cosmological simulations will be crucial to offer insight into the build-up of mass and angular momentum at large radius \citep[e.g.,][]{schulze2020,pulsoni2020} which can be tested observationally with smaller samples that have large \re\ coverage \citep[e.g.,][]{sarzi2018,gadotti2019}.

\subsection{Do Simulations Predict Two Kinematic Families?}
\label{subsec:dich_sims}

Many theoretical studies have tried to explain the origin of the different kinematic classes of galaxies, in particular in relation to the impact of mergers \citep[for a review on the topic, see ][]{naab2014}. While several early-type formation models managed to create galaxies with little rotation, the detailed properties of those simulated galaxies still differ significantly from observations \citep[e.g.,][]{bendo2000,jesseit2009,bois2011}.

Binary galaxy merger simulations demonstrate that the majority of merger remnants are consistent with being fast rotating galaxies \citep{bois2010,bois2011}, with the mass ratio of the progenitors being a crucial parameter for creating slow rotators, although the orbit-spin orientation of the merger might be as important \citep{moody2014}. 
A clear bimodality in \lre-\ee\ is seen in both \citet{jesseit2009} and \citet{bois2011}, but \citet{bois2011} caution that this bimodality \textit{"could likely result from the specific choices of simulated mass ratios and the limited number of simulated incoming orbits"}. The relative importance of a dissipational component in the formation of a bimodal populations is still unclear, with contrasting results from \citet{cox2006} and \citet{taranu2013}.

Cosmological simulations offer a more realistic insight into the kinematic distribution of modelled galaxies, 
although the fast/slow selection nearly always follows the observational criteria. In this paper, we demonstrate that considerable differences in the location of the different distributions of fast and slow rotators exist as compared to observations and between simulations. Quantitative offsets in galaxy structural, kinematic, and stellar population parameters were already demonstrated to exist as shown by \citet{vandesande2019}, so it is not surprising that the selection criteria for fast and slow rotators should be adapted for the different simulations. 

Nonetheless, the large differences of the pSR populations towards higher stellar mass in the simulations is perhaps surprising. While lower mass galaxies assemble their stellar material primarily though star formation \citep{robotham2014}, mergers dominate the addition of stellar material in galaxies above $\mathcal{M}^{*}$ ($\logm\sim10.75$) and also are key in lowering angular momentum in galaxies. Large differences between the frequency and mass-ratio of mergers are not expected between different cosmological simulations, which points towards a different problem within the simulations. The key might lie in the fast-rotating population. None of the simulations showed a close match to the observed pFR distribution, and if the progenitors of slow rotators do not match the observed distribution, perhaps we should not expect the pSR distribution to match either. 

\begin{figure*}
\includegraphics[width=\linewidth]{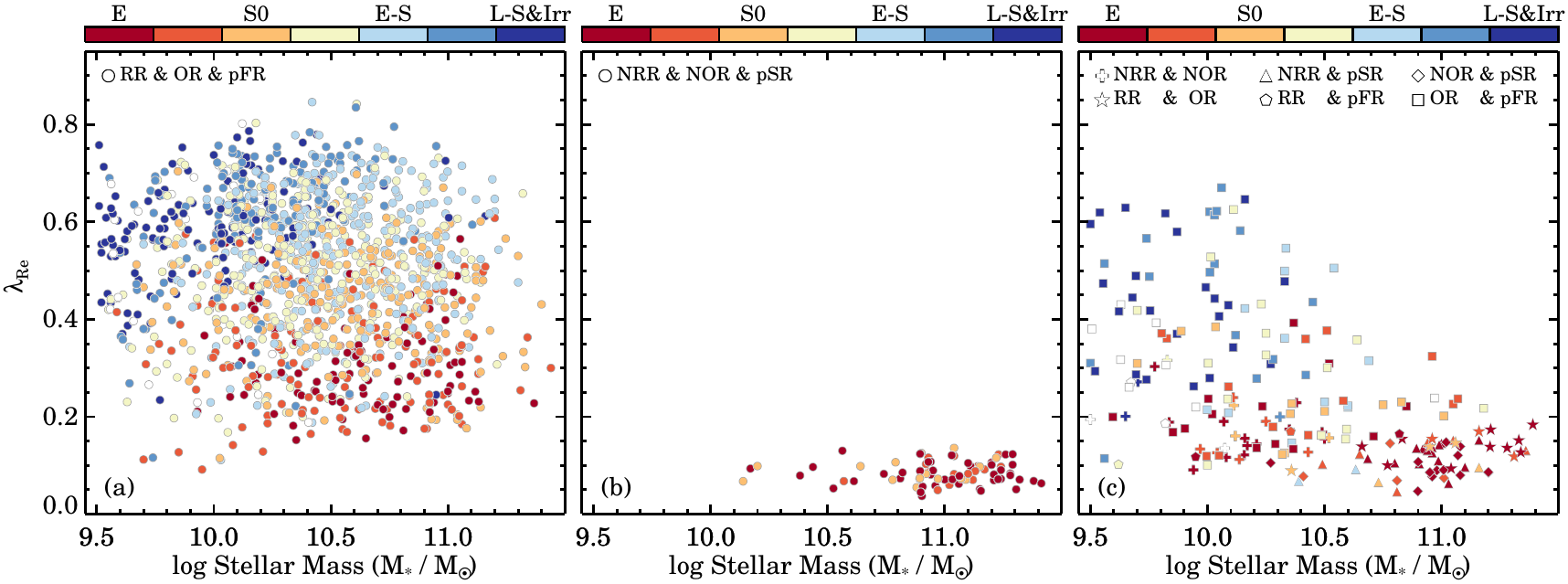}
\caption{Seeing-corrected spin parameter proxy versus stellar mass colour coded by visual morphology. The three panels show combined classes of FRs (panel a), SRs (panel b), and galaxies with mixed classifications (panel c). The fraction of galaxies with mixed classification is considerable (12.6 percent, 205 / 1625) and the mixed sample spans a large in \lre.
\label{fig:jvds_mass_lambdar_method_compare}}
\end{figure*}

Alternatively, the large differences in the \lr\ distributions could indicate that the presence and treatment of gas, star-formation, as well the feedback subgrid models in these simulations are more critical for the kinematic properties of \emph{all} galaxies than previously assumed, although the effects could be indirect. The exact prescription for how feedback is modelled will change the abundance of gas in galaxies and will therefore impact the frequency in which dry versus wet mergers happen as well as their typical mass ratios \citep[e.g., see][]{dubois2013,dubois2016,lagos2018a}. Both of these factors are important in the formation of slow rotators. This idea is in contradiction with the finding from \citet{penoyre2017} who conclude that no major difference is found due to presence of gas in mergers in \ill, whereas \citet{naab2014} using cosmological zoom-in simulations, and \citet{lagos2018a,lagos2018b} using \ea, found a clear impact of the gas content on galaxy spin. The latter results are also confirmed by \citet{martin2018} who find that the morphology of merger remnants strongly depends on the gas fraction of the merger and that re-grown disks are common in gas-rich mergers. However, as no clear picture into the formation of slow rotators has yet emerged from large-volume cosmological simulations, it will be paramount to accurately compare distributions of observations and simulations in consistent ways.

\subsection{How To Best Separate Two Kinematic Families}
\label{subsec:best_separate}

The main aim of this paper is to investigate how we can best separate different kinematic populations and to what extent these different kinematic populations overlap. In our analysis, we have investigated three kinematic classifications in detail (\kinemetry, visual kinematic morphology, and Bayesian mixture model classification), but have not yet directly compared the different classifications to each other. We address this here by looking at the agreement and disagreement between these methods. We select galaxies which have all three classifications, which reduces the total sample from 1765 to 1625 galaxies, caused by the fact that not all galaxies have \smk\ measurements out to  one \re. Galaxies are then grouped according to where their classifications agree: 1) a galaxy is an RR, an OR, and a pFR (Fig.~\ref{fig:jvds_mass_lambdar_method_compare}a), or 2) a galaxy is an NRR, an NOR, and a pSR ( Fig.~\ref{fig:jvds_mass_lambdar_method_compare}b), or 3) a galaxy only has two out of three matching classifications (Fig.~\ref{fig:jvds_mass_lambdar_method_compare}c). It should be noted that none of the kinematic identifiers used here are truly independent as they all rely on the velocity and velocity dispersion maps. 

The overlapping FR classifications form the biggest group with 82.6 percent (1342/1625), the SR classifiers are the smallest (4.8 percent; 78/1625), whereas galaxies with mixed classifications make up 12.6 percent of the total sample (205/1625). Within the mixed classified sample, the largest subgroup is where galaxies are classified as ORs and pFRs, but where \kinemetry\ suggests the galaxies are NRRs (114/205). As most of these galaxies reside towards lower stellar masses and higher \lre\, we argued before that the high \mk\ values that lead to this classification are more likely caused by observational effects rather than their intrinsic properties. This suggests that when using \kinemetry\ with SAMI-like data quality, the NRR population has the highest probability to be contaminated with ORs and pFRs.

We conclude that the visual kinematic morphology and the Bayesian mixture model analysis provide the most consistent classification. Using a modified version of the \cite{cappellari2016} slow rotator selection region, we also find a high positive predictive value for separating the NOR/OR and pSR/pFR classes. Therefore, a combination of the two methods with the associated selection box is recommended for selecting SRs from the SAMI Galaxy Survey or data with similar quality when aiming to compare to previous studies that separate fast and slow rotators.

\subsection{Implications for Galaxy Formation Scenarios}
\label{subsec:implications}

A critical stellar mass limit of $\logm\sim11.3$ has been proposed as the limit above which passive slow rotators with cores dominate \citep[see discussion in][]{cappellari2016}, to the point where galaxies below this mass limit are no longer classified as slow rotators \citep{graham2019b}. Although our data quality does not allow us to address the question of whether slow rotators in our sample have core or power-law inner light profiles, we do not find evidence for a limit below which there are no NNRs, NORs, or pSRs, as based on \kinemetry, visual kinematic classification, or the Bayesian mixture models. Depending on the kinematic galaxy identifier (e.g., \kinemetry, visual classification, Bayesian mixture models) we find the fraction of these different populations to start rising at different stellar masses. The analysis from \citet[][fig. 5]{falconbarroso2019} using CALIFA data also clearly shows a group of slow-rotator ellipticals with stellar masses below $\logm\sim11.3$. Furthermore, some of the cosmological simulation data analysed here also reveal a small, but non-negligible fraction of slow rotators towards low stellar mass. Hence, we are cautious to use mass as a selection criterion for slow rotators, in particular when their formation process is still not well understood.

One of the striking results from our Bayesian mixture model analysis is that the \lr\ peaks of the pFR and pSR distributions are nearly constant as a function of stellar mass. While this appears to be in conflict with the general notion that galaxies have lower \lr\ with increasing stellar mass, our results indicate that this kinematic trend is caused by a changing fraction of galaxies in the pFR and pSR distribution. Towards higher stellar mass, the fraction of galaxies in the pFR decreases while it increases for the pSR, leading to a lower \lr\ of the entire population.

The constant \lr\ peak for pSRs as a function of stellar mass implies that the formation process of SRs, i.e., the near complete removal of a galaxy's angular momentum, is likely similar across all stellar masses. If the mechanism  transforming pFRs into pSRs was different as a function of stellar mass, we would either observe a different pSR \lr\ peak, or a change in the width of the pSR distribution. We detect neither. Nonetheless, the strong fractional increase of the pSRs distribution as a function of stellar mass may indicate that the SRs formation process is more efficient towards higher stellar mass, or else that the processes which cause a galaxy to evolve into a slow rotator also tend to lead to it becoming very massive. This picture is consistent with predictions from cosmological simulations that show that approximately 30-50 percent of the SRs are produced in a single massive merger \citep[e.g.,][]{schulze2018,lagos2020}, a process that can happen at all stellar masses, albeit it is most efficient in transforming galaxies into slow rotators if the merger is dry, which is more likely at higher masses.

These predictions can be tested with enhanced number statistics combined with better kinematic data quality \citep[e.g., Hector, ][]{bryant2016} or by going to higher redshift (e.g., MAGPI at $z\sim0.3$, \citealt{foster2020}; LEGA-C at $z\sim0.8$, \citealt{vanderwel2016}). Early results from the LEGA-C survey  indicate that there is increased rotational support in $z\sim0.8$ quiescent galaxies \citep{bezanson2018}, although it's not clear yet whether this is due to a change in the \vs\ or \lr\ peak of the pFR and pSR distributions or a change in the fractions of pFRs and pSRs distributions.

\section{Conclusion} 
\label{sec:conclusion}

The dynamics of galaxies offers great insight into the assembly and redistribution of stellar mass within galaxies over time. The prevailing physical explanation for drastically altering the dynamical properties of galaxies is undoubtedly merging and accretion. Yet many questions still remain on the importance of gas as a dissipational component as well as the frequency of major and minor mergers and their impact on the inner and outer stellar distributions. Key to answering these questions is identifying the different kinematic populations that exist and link these to the various proposed formation scenarios.

Using data from the SAMI Galaxy Survey we investigate whether or not we can detect a bimodality in the kinematic properties of the entire galaxy population using \lre\ versus stellar mass and ellipticity. The main goal of the paper is to use different techniques to identify whether we can accurately separate a bimodal kinematic distribution in relatively low-signal-to-noise, seeing-impacted data, and to what extent different kinematic populations overlap. By doing so, we aim to consolidate results from ongoing multi-object IFS surveys with the conclusions from previous IFS surveys that had better S/N and spatial resolution, but where the sample size did not allow a statistical analysis. We also provide a framework for comparing these results to mock-observations from cosmological simulations.

We find the following results:

(i) \textbf{Partially applied seeing corrections can lead to an artificially enhanced bimodality.}
Using 1765 SAMI galaxies with \lre\ measurements we investigate the impact of the seeing corrections. No clear bimodal distribution in \lre\ is detected in the SAMI seeing-dominated data or when the \citet{harborne2020b} correction is applied to all galaxies (Figs \ref{fig:jvds_lambdar_eps}a and c). However, when only regular rotators are seeing-corrected, as was done in \citet{graham2018}, we detect a clear bimodal distribution in \lre, but we argue this is an artificial construct as the correction is applied to a subset of the sample (Fig.~\ref{fig:jvds_lambdar_eps}b). Thus, from the \lre-\ee\ diagram alone using SAMI Galaxy Survey data, we do not find strong evidence for distinct kinematic populations of galaxies. 

(ii) \textbf{There is considerable overlap of regular and non-regular rotator distributions with SAMI.}
With galaxies classified as regular and non-regular rotators from the kinematic asymmetry of the rotational velocity fields using the \kinemetry\ method, we investigate the amount of overlap of the RR and NRR distributions within the \lre-\logm\ and \lre-\ee\ diagrams. At low stellar mass, NRR (\mk>0.07) have higher values of \lre\ than at high stellar mass (Figs \ref{fig:jvds_mass_lambdar_kinemetry}a and c). We find considerable mixing of the regular and non-regular rotator populations in our SAMI sample, in particular below $\logm<10.5$. The trend of decreasing \mk\ values with increasing stellar mass also leads to considerable overlap of RRs and NRRs within the \lre-\ee\ diagram. We use a "Receiver Operating Characteristic Curve" (ROC) and "Matthews correlation coefficient" ($MCC$) to determine the best possible selection box to separate RR and NRR within the \lre-\ee\ diagram. For SAMI Galaxy Survey data the optimal selection region has a higher \lre\ threshold as compared to the selection box from \citet{cappellari2016}, but overall does not provide a clean separation of regular and non-regular rotator classes as the Positive Prediction Value is only 65.7 percent.

(iii) \textbf{Visual kinematic classification of SAMI data leads to a cleaner separation of two kinematic populations.}
We devise a new visual classification scheme that first separates galaxies with obvious rotation (ORs) from galaxies with no obvious rotation (NORs), combined with a second layer of refinement to find galaxies with inner kinematic features (with-features versus no-features). There is a well-defined separation of ORs and NORs within the \lre-\logm\ and \lre-\ee\ planes. Similar to \kinemetry\ we find that the NORs have higher values of \lre\ towards lower stellar mass. The optimal selection region for selecting NORs using \lre\ and \ee\ is close to the selection region from \citet{cappellari2016}. Furthermore, the ROC analysis reveals a significantly higher success rate as compared to using \kinemetry\ for SAMI, which suggests that our visual classification scheme is more suitable for data that has a large range of S/N and the typical spatial resolution of SAMI or MaNGA.

(iv) \textbf{Bayesian mixture models provide the cleanest separation of two kinematic families.}
Rather than looking for two populations using \kinemetry\ or visual classification, we use a Bayesian mixture model analysis to determine whether multiple populations can be identified as a function of stellar mass. At all stellar masses, the \lre\ distribution for late-type galaxies is well described by a single beta distribution that peaks at $\lre\sim 0.6$. However, for early-type galaxies above $\logm>10.5$ a second beta distribution is required with a lower peak at $\lre\sim 0.1$. These results demonstrate clearly that we can separate two stellar kinematic populations from the \lr\ distribution, even when these distributions have non-negligible overlap. Based on these results, we then refer to galaxies as probabilistic fast or slow rotators (respectively, pFRs and pSRs). In contrast to NRRs and NORs, pSRs have lower \lre\ values at lower stellar mass. This could indicate that the NRRs and NORs found at low stellar mass are simply the lower tail of a broad \lre\ distribution, but not a separate class. Even though pSR have slightly higher \lre\ values towards high stellar mass, the pSRs and pFRs are extremely well-separated within the \lre-\ee, but with a caveat that pSRs were selected primarily using \lre.

(v) \textbf{Mixed results from Cosmological Hydrodynamical Simulations.}
We apply the same Bayesian mixture model analysis to mock-observations from the \ea\ and \hy, \ha, and \ma\ cosmological hydrodynamical simulations. Although the mixture model predicts two populations of stellar rotators in all three simulations, the \lre\ peak of the two beta distributions is significantly offset from observations, and the fraction of both beta distributions as a function of stellar mass also change considerably. The overlap of the two beta distributions also differs significantly between the simulations. Our results indicate that the treatment of the ISM and feedback subgrid models within these simulation have a considerable impact on the distribution of fast and slow rotators. More importantly, observational selection criteria for fast and slow rotators should not be applied to data from simulations to derive the fraction of different kinematic populations unless the distributions in \lre\ and \ee\ are well-matched to the observations they are compared to.

(vi) \textbf{The optimal classification of galaxy stellar kinematics.}
By comparing three kinematic classification methods (\kinemetry, visual kinematic morphology, and a Bayesian mixture model) we find the best agreement between the visual kinematic morphology and Bayesian mixture model classification. Nonetheless, we  argue that visual classification provides a unique spatial kinematic insight, and maintains its usefulness for future work. For comparing to previous studies that adopted the \citet{cappellari2016} selection box to separate fast and slow rotators, we advise using Eq.~\ref{eq:sr_vary} with $\lr_{\rm{start}}=0.12$ when the data quality is similar to that of the SAMI Galaxy Survey. However, we stress that our analysis revealed a significant amount of overlap between the different kinematic distributions which should be acknowledged.

\textbf{Going forward.}
Many claims about a kinematic bimodality have been made in the past with an ongoing unquenchable drive to separate galaxies into binary classification. In this paper, we confirm key findings from previous and ongoing IFS studies that the vast majority of galaxies are consistent with being a family of oblate rotating systems viewed at random orientation. At the same time, there is a group of mainly massive early-type galaxies that show complex dynamical structures, irregular velocity fields, 2-sigma peaks, or kinematic misalignment, with indications that some fraction of these galaxies are triaxial systems. 

The rapid increase in the number of galaxy IFS observations has been achieved by a compromise between multiplexing, spatial resolution and S/N. Nevertheless, we have demonstrated that we can extract different kinematic populations in seeing-impacted data, but only when the analysis techniques are matched to the data quality. However, even when using higher-quality data (e.g., see Appendix \ref{app:fast_slow_a3d}), with different kinematic identifiers the same galaxy can be simultaneously classified into opposite groups (e.g., a RR-SR or NRR-FR). Our results show that it has become essential to take into consideration that the distributions of various kinematic populations overlap. When the overlap and mixing of classes is ignored, and naming conventions for various kinematic classifications slowly morph into a singular class (e.g., all non-regular rotators are slow rotators and vice versa), the complexity of galaxy evolution is disregarded. Cosmological simulations have shown that galaxy evolution is a highly stochastic process, hence we do not expect distinct, cleanly separated classes. Instead, we promote the analysis of stellar kinematic data using probability distribution functions instead of bimodal classes. With more than 10,000 IFS galaxy observations becoming publicly available soon (e.g., SAMI Galaxy Survey, SDSS-IV MaNGA), now is the perfect time to further pursue such an endeavour.

\section*{Acknowledgements}
We thank the anonymous referee for their thorough and thoughtful comments that have improved the clarity of this paper. We also thank Tim de Zeeuw and Michele Cappellari for insightful and helpful comments on the paper, and thank Alister Graham for useful discussions on this topic. The SAMI Galaxy Survey is based on observations made at the Anglo-Australian Telescope. The Sydney-AAO Multi-object Integral field spectrograph (SAMI) was developed jointly by the University of Sydney and the Australian Astronomical Observatory, and funded by ARC grants FF0776384 (Bland-Hawthorn) and LE130100198. The SAMI input catalogue is based on data taken from the Sloan Digital Sky Survey, the GAMA Survey and the VST ATLAS Survey. The SAMI Galaxy Survey is supported by the Australian Research Council Centre of Excellence for All Sky Astrophysics in 3 Dimensions (ASTRO 3D), through project number CE170100013, the Australian Research Council Centre of Excellence for All-sky Astrophysics (CAASTRO), through project number CE110001020, and other participating institutions.

JvdS acknowledges support of an Australian Research Council Discovery Early Career Research Award (project number DE200100461) funded by the Australian Government. LC is the recipient of an Australian Research Council Future Fellowship (FT180100066) funded by the Australian Government. NS acknowledges support of an Australian Research Council Discovery Early Career Research Award (project number DE190100375) funded by the Australian Government and a University of Sydney Postdoctoral Research Fellowship. The research of JD is supported by the Beecroft Trust and STFC. JBH is supported by an ARC Laureate Fellowship (FL140100278) that funded the SAMI prototype. JJB acknowledges support of an Australian Research Council Future Fellowship (FT180100231). M.S.O. acknowledges the funding support from the Australian Research Council through a Future Fellowship (FT140100255). FDE acknowledges funding through the H2020 ERC Consolidator Grant 683184. Parts of this research were conducted by the Australian Research Council Centre of Excellence for All Sky Astrophysics in 3 Dimensions (ASTRO 3D), through project number CE170100013. The Magneticum Pathfinder simulations were partially performed at the Leibniz-Rechenzentrum with CPU time assigned to the Project 'pr86re'. This work was supported by the DFG Cluster of Excellence 'Origin and Structure of the Universe'. 

This paper made use of the \texttt{mpfit} \textsc{IDL} package \citep{mpfit}, as well as the \texttt{astropy} \textsc{python} package \citep{astropy}, \texttt{IPython} \citep{ipython}, the \texttt{matplotlib} plotting software \citep{matplotlib}, the scientific libraries \texttt{numpy} \citep{numpy}, \texttt{seaborn} \citep{seaborn}, and \texttt{scipy} \citep{scipy}. 


\section*{Data availability}

All observational data presented in this paper are available from Astronomical Optics' Data Central service at https://datacentral.org.au/ as part of the SAMI Galaxy Survey Data Release 3. The \ea\ simulations are available at http://icc.dur.ac.uk/Eagle/database.php, and the \ma\ simulations are  available at http://www.magneticum.org. Data from the \at\ Survey are available at http://www-astro.physics.ox.ac.uk/atlas3d/.

\bibliographystyle{mnras}
\bibliography{jvds_sami_mnras_lowres}{}

~\\
{\it \footnotesize \noindent$^{1}${Sydney Institute for Astronomy, School of Physics, A28, The University of Sydney, NSW, 2006, Australia}\\
$^{2}${ARC Centre of Excellence for All Sky Astrophysics in 3 Dimensions (ASTRO 3D), Australia}\\
$^{3}${International Centre for Radio Astronomy Research, The University of Western Australia, 35 Stirling Highway, Crawley WA 6009, Australia}\\
$^{4}${School of Physics, University of New South Wales, NSW 2052, Australia}\\
$^{5}${Australian Astronomical Optics, AAO-USydney, School of Physics, University of Sydney, NSW 2006, Australia}\\
$^{6}${University of Oxford, Astrophysics, Keble Road, Oxford OX1 3RH, UK}\\
$^{7}${Institut d’Astrophysique de Paris, UMR 7095, CNRS, UPMC Univ. Paris VI, 98 bis boulevard Arago, 75014 Paris, France} \\
$^{8}${Sterrenkundig Observatorium, Universiteit Gent, Krijgslaan 281 S9, B-9000 Gent, Belgium} \\
$^{9}${Australian Astronomical Optics, Faculty of Science and Engineering, Macquarie University, 105 Delhi Rd, North Ryde, NSW 2113, Australia}\\
$^{10}${Research School of Astronomy and Astrophysics, Australian National University, Canberra ACT 2611, Australia}\\
$^{11}${Department of Physics and Astronomy, Macquarie University, NSW 2109, Australia } \\
$^{12}${Astronomy, Astrophysics and Astrophotonics Research Centre, Macquarie University, Sydney, NSW 2109, Australia} \\
$^{13}$ Universit\"ats-Sternwarte M\"unchen, Scheinerstr. 1, D-81679 M\"unchen, Germany \\
$^{14}${SOFIA Science Center, USRA, NASA Ames Research Center, Building N232, M/S 232-12, P.O. Box 1, Moffett Field, CA 94035-0001, USA}\\
$^{15}$ Max Planck Institute for Extraterrestrial Physics, Giessenbachstra{\ss}e 1, D-85748 Garching, Germany \\
$^{16}${School of Mathematics and Physics, University of Queensland, Brisbane, QLD 4072, Australia} \\
$^{17}${Department of Physics and Astronomy, The Johns Hopkins University, Baltimore,MD 21210, USA} \\}


\appendix

\section{Testing Seeing Corrections on Repeat Observations}
\label{sec:repeat_obs}

\subsection{Analytic correction to account for atmospheric seeing}
\label{subsec:seeing_corr_analytic}

In this paper we use an analytic seeing correction for \lr\ as presented by \citet{harborne2020b} using the public code \textsc{SIMSPIN} \citep{harborne2020a}, optimised for SAMI Galaxy Survey data. Specifically, the corrections in \citet{harborne2020b} cover the full range of $0 < \sigpsfre < 0.9$, whereas the median \sigpsfre=0.22 for SAMI data with a defined maximum limit of $\sigpsfre<0.6$. The updated equation for \lr\ is (where $R$ is the semi-major axis of the ellipse on which each spaxel lies):


\begin{equation} 
    \Delta \lambda_{\varepsilon-R}^{\text{corr}} = f\left(\frac{\sigma_{\text{PSF}}}{R_{\text{maj}}}\right) + \left(\frac{\sigma_{\text{PSF}}}{R_{\text{maj}}}\right) \times f(\varepsilon, n, R_{\text{eff}}^{\text{fac}}), \label{eq:elR_corr}
\end{equation}

\noindent where,
\begin{equation} 
    f\left(\frac{\sigma_{\text{PSF}}}{R_{\text{maj}}}\right)^{\Delta \lambda_{R}^{\varepsilon}} = \frac{7.44}{1 + \text{exp}{[4.87 \left(\frac{\sigma_{\text{PSF}}}{R_{\text{maj}}}\right)^{1.68} + 3.03]}} - 0.34,
\end{equation} 

\begin{align} 
    f(\varepsilon, n, R_{\text{eff}}^{\text{fac}})^{\Delta \lambda_{R}^{\varepsilon}} =\,  
    & [0.011 \times \text{log}_{10}(\varepsilon)] 
    - [0.278 \times \text{log}_{10}(n)] \nonumber \\
    & + 0.098
\end{align}

\noindent Here, $n$ is the \ser\ index, and $R_{\text{eff}}^{\text{fac}}$ is the radius (in units of \re) at which \lr\ is measured ($R_{\text{eff}}^{\text{fac}} = 1 $ in our case). Similarly, for \vs:



\begin{equation}
    \Delta V/\sigma^{\text{corr}} = f\left(\frac{\sigma_{\text{PSF}}}{R_{\text{maj}}}\right) + 3 \left(\frac{\sigma_{\text{PSF}}}{R_{\text{maj}}}\right) \times f(\varepsilon, n, R_{\text{eff}}^{\text{fac}}), \label{eq:vsig_corr}
\end{equation}

\noindent where,

\begin{equation}
    f\left(\frac{\sigma_{\text{PSF}}}{R_{\text{maj}}}\right)^{\Delta V/\sigma} = \frac{7.47}{1 + \text{exp}[5.31 \left(\frac{\sigma_{\text{PSF}}}{R_{\text{maj}}}\right)^{1.68} + 2.89]} - 0.39,
\end{equation} 

\noindent and

\begin{align}
     f(\varepsilon, n, R_{\text{eff}}^{\text{fac}})^{\Delta V/\sigma} =\, 
     & [-0.078 \times \varepsilon] 
     + [0.0038 \times \text{log}_{10}(n)] \nonumber \\
     & + 0.029.
\end{align}


\noindent Using these equations, we can then calculate  \lrei\ (the intrinsic or true value of the spin parameter proxy) from the observed \lreo.

\begin{align}
    \lambda_{\,R_{\rm{e}}}^{\rm{\,intr}} &= 10^{\left[\text{log}_{10}(\lreo) - \Delta \lambda_{R}^{\text{corr}}\right]},
    \label{eq:lrinvert}
\end{align}

\noindent and similarly for \vs:

\begin{align}
    (V / \sigma)_{{\rm{e}}}^{\rm{\,intr}} &= 10^{\left[\text{log}_{10}(\vseo) - \Delta V/\sigma^{\text{corr}}\right]}.
    \label{eq:vsinvert}
\end{align}

\noindent The SAMI stellar kinematic sample has a median \sigpsfre=0.22 that results in a median \lr\ correction factor of 0.14 dex, or an average absolute increase in \lre\ of +0.11. Above $\sigpsfre>0.6$, the correction factor increases rapidly, which is the main motivation for not using data above this limit for the main analysis.

\subsection{Testing Seeing correction on Repeat Observations}
\label{subsec:seeing_corr_sc}

We now use SAMI Galaxy Survey repeat observations to test the analytic seeing correction as described in the previous section. Repeat observations are ideal for estimating uncertainties due to weather conditions, such as seeing and transmission, but also in the use of different hexabundles. Due to the SAMI Galaxy Survey's optimal field tiling and plate configuration, there are a total of 210 galaxies that have repeat observations. For this analysis, we only use galaxies that meet our selection criteria from Section \ref{sec:data}, with full stellar kinematic \re\ coverage. The full \re\ coverage selection is applied to avoid confusing uncertainties from our \lr\ aperture correction with those due to the impact of seeing. This selection reduces the number of galaxies with repeat observations to 169. 

\begin{figure*}
	\includegraphics[width=0.95\linewidth]{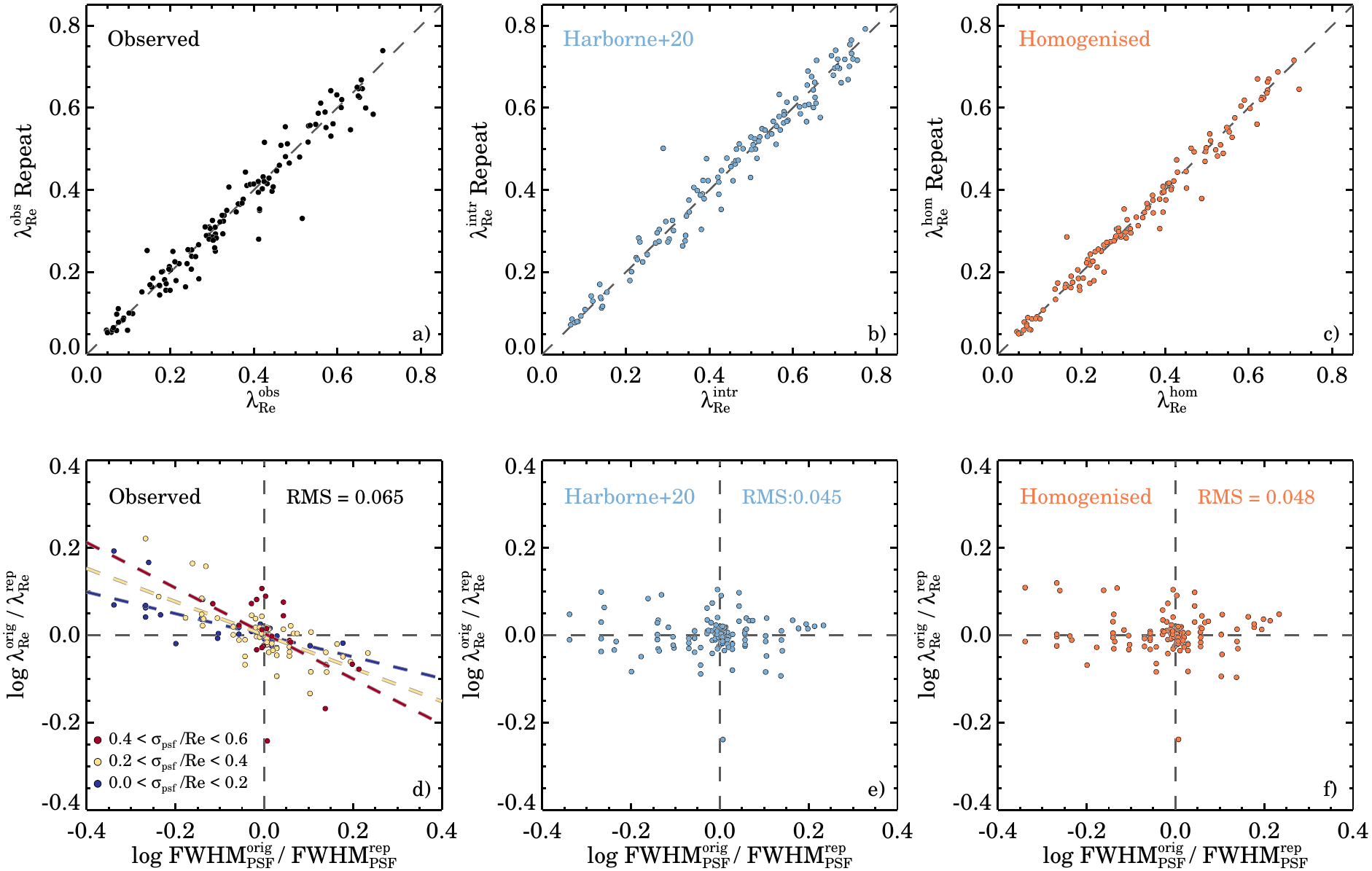}
    \caption{Comparison of \lre\ measurements from repeat observations. In the top row we show \lre\ for the original data (panel a), with the seeing correction from \citet{harborne2020b} applied (panel b), and after homogenising the sample to a common seeing of 2\farcs0 (panel c). In the bottom row we show the fractional difference in \lre\ versus the difference in seeing of the original and repeat observations. The dashed line in panel d) shows the best-fit relations as given by Eq.~\ref{eq:lr_see_hom} that were used to homogenise the data (panels c and f).}
    \label{fig:app_lr_repeat_obs}
\end{figure*}

In the top row of Fig.~\ref{fig:app_lr_repeat_obs} we present the \lre\ measurement for the original and repeat observations, with and without different methods to correct for the seeing. The bottom row of Fig.~\ref{fig:app_lr_repeat_obs} shows the fractional difference of the PSF's FWHM versus the fractional difference in \lre. Galaxies from observations with the best combination of seeing and S/N are called "original", whereas the other secondary observations are named "repeats". We note that our sample of repeat observations is a representative sub-sample of the total stellar kinematic sample, with similar stellar mass and morphological type, and is observed under similar seeing conditions. The median FWHM$_{\rm{PSF}}$ of the repeat observations is 2\farcs06, whereas the average seeing of the entire stellar kinematic sample is 2\farcs04. The best-seeing repeat observation has FWHM=1\farcs37, whereas the worst repeat has FWHM=2\farcs85. 

The seeing correction from \citet{harborne2020b} applied to the \lre\ repeat measurements are shown in Fig.~\ref{fig:app_lr_repeat_obs}(c). This figure demonstrates that the seeing correction works well across the large range in \lre\ measurements. For low values of \lre<0.35, where we expect most galaxies with complex kinematic features, the agreement between the seeing-corrected original and the repeat measurement is excellent. Thus, the analytic correction works for all types of galaxies, including slow rotators. Figs \ref{fig:app_lr_repeat_obs}(b) and (d) show that the RMS of the fractional differences goes down from 0.065 to 0.045 when the seeing correction is included.


\subsection{Empirical correction to account for atmospheric seeing}
\label{subsec:seeing_corr_emp}

In this section, we test an alternative method to correct for the impact of seeing on our kinematic data. The idea here will be to use an empirical relation derived from the repeat observations to homogenise the sample to a common average seeing. Such a data driven approach has the advantage that it is not biased to the choice of simulations to derive the analytic seeing correction. For example, the analytic correction has not been designed for galaxies with complex dynamics (although the correction works relatively well with little scatter and the absolute \lr\ correction for these galaxies is small). An empirical seeing homogenisation can be applied to the whole sample because it is based on the exact same observational setup and data quality as the sample that it will be applied to. The large range in morphology, stellar mass, and \lr\ values for the repeat observations also remove any morphological bias. A homogenised sample might also be more appropriate to use for comparing to simulations. The philosophy here is that it is more reliable to convolve mock-observations from simulations to the seeing of an observational survey then it is to deconvolve observational results.

For these reasons, we will now derive an empirical seeing correction or homogenisation, and test how well it removes the scatter in \lr\ and \vs\ using repeat observations. Figs \ref{fig:app_lr_repeat_obs}(a) and (d) show the results from the repeat observations without applying any corrections. We find that there is a linear trend between $\Delta$FWHM and $\Delta\lre$, such that with a larger difference in seeing between the original and repeat, the larger the difference in \lr. By fitting a linear relation to our data from Fig.~\ref{fig:app_lr_repeat_obs}(d), our goal is to remove the trend between FWHM and \lr. Motivated by the results from the analytic correction we separate the sample into different bins of \sigpsf\ where the impact of seeing will be different. We find that:

\begin{align}
\label{eq:lr_see_hom}
\log_{10} ( {\lreorig}/{\lrerep}) &= -0.247 \times \log_{10} ( {\rm{FWHM_{\rm{PSF}}^{\rm{\,orig}}}}/{\rm{FWHM_{\rm{PSF}}^{\rm{\,rep}}}}) \nonumber \\
& \hspace{1.5cm} \rm{for~} 0.0<\sigpsf<0.2 
\end{align}

\begin{align}
\log_{10} ( {\lreorig}/{\lrerep}) &= -0.380 \times \log_{10} ( {\rm{FWHM_{\rm{PSF}}^{\rm{\,orig}}}}/{\rm{FWHM_{\rm{PSF}}^{\rm{\,rep}}}}) \nonumber \\
& \hspace{1.5cm} \rm{for~} 0.2<\sigpsf<0.4 
\end{align}

\begin{align}
\log_{10} ( {\lreorig}/{\lrerep}) &= -0.520 \times \log_{10} ( {\rm{FWHM_{\rm{PSF}}^{\rm{\,orig}}}}/{\rm{FWHM_{\rm{PSF}}^{\rm{\,rep}}}}) \nonumber \\
& \hspace{1.5cm} \rm{for~} 0.4<\sigpsf<0.6 
\end{align}

\noindent The best-fit relation are shown in Fig.~\ref{fig:app_lr_repeat_obs}(b). As expected, with increasing fractions of \sigpsf\ the slope of the relation between \lr\ and FWHM also increases. The analysis is repeated on the \vs\ measurements (not shown here), from which we find the following relations:

\begin{align}
\label{eq:vs_see_hom}
\log_{10} ( {\vseorig}/{\vserep}) &= -0.380 \times \nonumber \\
&  ~~\log_{10} ( {\rm{FWHM_{\rm{PSF}}^{\rm{\,orig}}}}/{\rm{FWHM_{\rm{PSF}}^{\rm{\,rep}}}}) 
\nonumber \\
& ~~~\rm{for~} 0.0<\sigpsf<0.2
\end{align}

\begin{align}
\log_{10} ( {\vseorig}/{\vserep}) &= -0.495 \times \nonumber \\
&  ~~\log_{10} ( {\rm{FWHM_{\rm{PSF}}^{\rm{\,orig}}}}/{\rm{FWHM_{\rm{PSF}}^{\rm{\,rep}}}}) 
\nonumber \\
& ~~~\rm{for~} 0.2<\sigpsf<0.4 
\end{align}

\begin{align}
\log_{10} ( {\vseorig}/{\vserep}) &= -0.580 \times \nonumber \\
&  ~~\log_{10} ( {\rm{FWHM_{\rm{PSF}}^{\rm{\,orig}}}}/{\rm{FWHM_{\rm{PSF}}^{\rm{\,rep}}}}) 
\nonumber \\
& ~~~\rm{for~} 0.4<\sigpsf<0.6 
\end{align}

We use Eq.~\ref{eq:lr_see_hom} to homogenise our \lre\ measurements to a single seeing value of 2\farcs0, which are shown in Fig.~\ref{fig:app_lr_repeat_obs}c,f. The difference between the original and repeat observations has become smaller, with a clear reduction in the RMS scatter. The method works surprisingly well given the low number of free parameters in the fit. Even more so, if we use a single relation to fit all data between $0.0<\sigpsf<0.6$, the RMS scatter only increases marginally to 0.048. In summary, homogenising the data reduces the scatter similarly as the analytic correction from \citet{harborne2020b}. In this paper, we adopt the analytic corrected to derive the intrinsic \lr\ such that we can compare to previous surveys where seeing was not a limitation. However, for a comparison to mock observations from large cosmological simulations, which might be impacted by numerical resolution effects, a seeing homogenisation method might be more suitable.


\section{Fast and Slow rotators in \at}
\label{app:fast_slow_a3d}

\begin{figure*}
\includegraphics[width=0.95\linewidth]{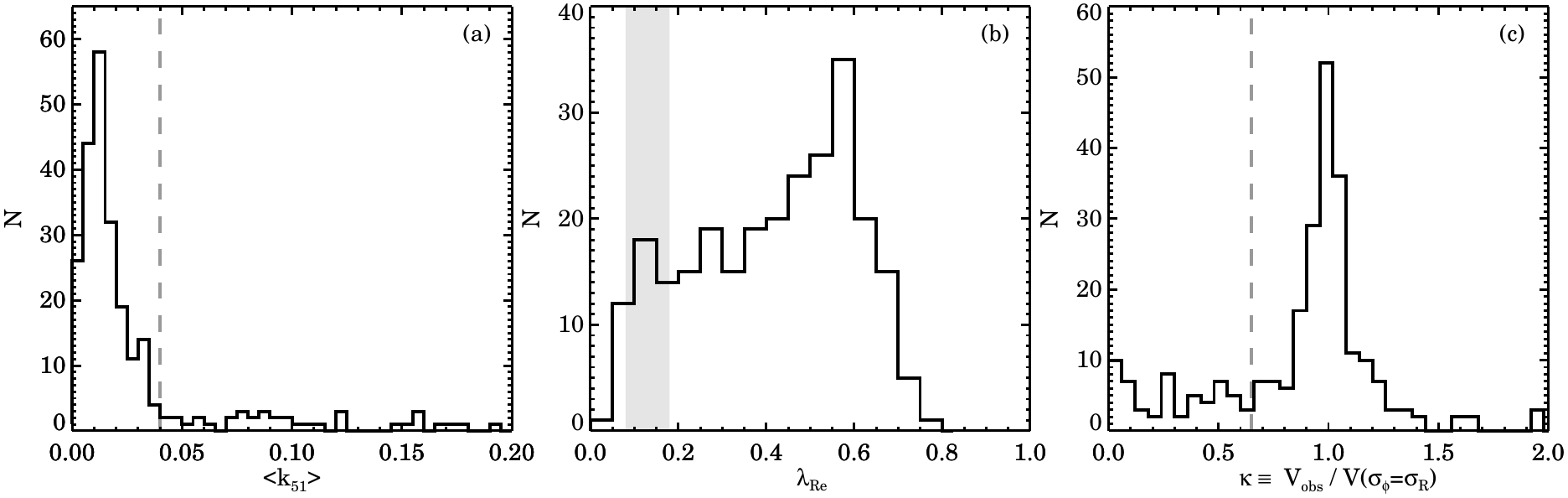}
\caption{Distributions of \mk, \lre\, and $\kappa$ from the \at\ survey. The vertical lines in panel (a) and (c) and indicate the proposed division for regular and non-regular rotators (panel a) and slow and fast rotators using JAM modelling panel (c), whereas the wider gray region in panel (b) indicates the \citet{cappellari2016} FR/SR selection region that includes an ellipticity term.
\label{fig:jvds_a3d_distributions}}
\end{figure*}

One of the goals of this paper is to investigate whether or not we can identify multiple kinematic populations in data where impact of seeing and data quality cannot be ignored. We present a framework to quantify how well kinematic identifiers are separated in the \lre-\ee\ diagram based upon the ratio of the True Positive Rate and the True Negative Rate. Here, we re-analyse results from the \at\ survey, using data as presented by \citet{emsellem2004}, \citet{cappellari2011a}, \citet{krajnovic2011}, \citet{emsellem2011}, and \citet{cappellari2013a,cappellari2013b}, adapted to definitions used in this paper. For more details on these measurements we refer to \citet{vandesande2019}. 

In Fig.~\ref{fig:jvds_a3d_distributions} we first present the distributions of \mk, \lre\, and $\kappa$ (see Section \ref{subsec:dich_jam}) which is the ratio of the observed velocity $V_{\rm{obs}}$ and the modelled velocity $V(\sigma_{\phi}=\sigma_{R})$ from JAM models. Note that for the $\kappa$ values, we did not imply a JAM quality cut, as non-regular rotators are not expected to be well described by JAM models. The vertical grey lines in Fig.~\ref{fig:jvds_a3d_distributions} show the commonly used selection regions for each parameter. However, we do not find clear evidence for a bimodal distribution from these three parameters, although this does not exclude the existence of a bimodality. Instead, it shows that a larger sample of galaxies is required to detect a multimodal distribution if these parameters are used independently.

\begin{figure*}
\includegraphics[width=0.95\linewidth]{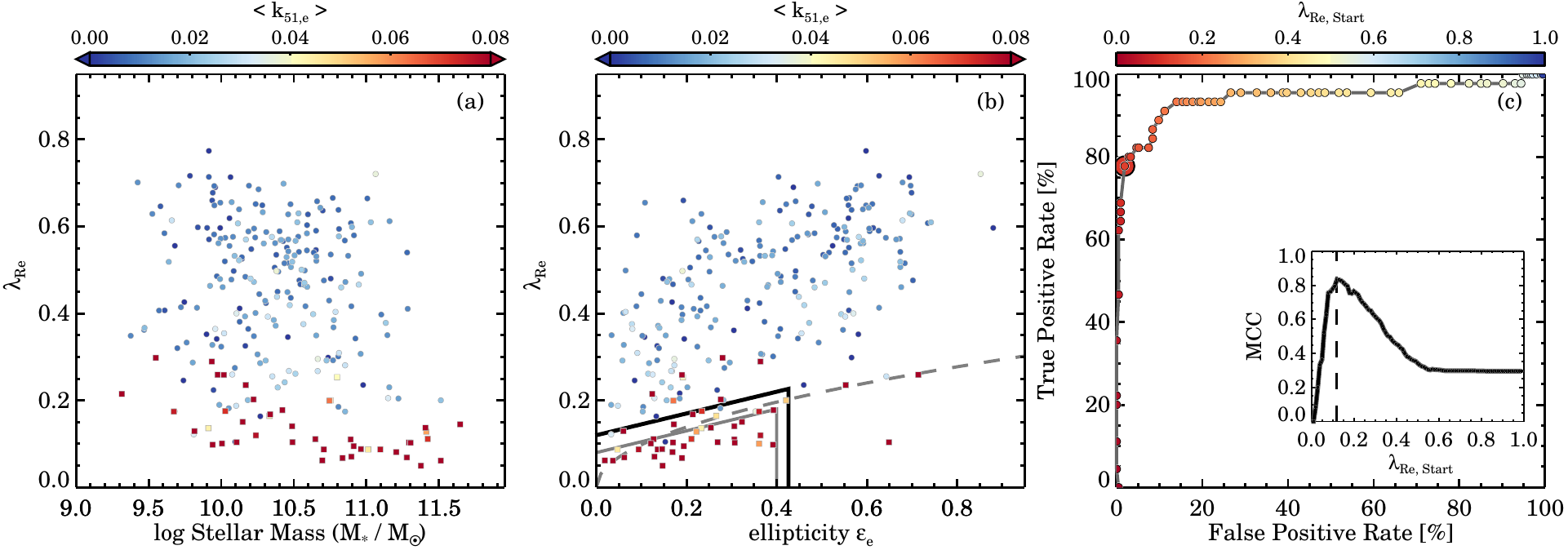}
\caption{Spin parameter proxy versus stellar mass and ellipticity using \at\ data. Data are colour coded by \mk\ (panels a and b). Round symbols show Regular Rotators, and square symbols show Non Regular Rotators. We show the RRs and NRRs in the \lre-\ee\ space in panel (b) with the optimal selection region (black), the SR selection box from \citet{cappellari2016} in grey, and from \citet{emsellem2011} as the dashed line. Panel (c) suggests that the \lre-\ee\ space is an effective way to distinguishing between regular and non-regular rotators derived from \at\ data.
\label{fig:jvds_mass_lambdar_kinemetry_a3d}}
\end{figure*}

\begin{figure*}
\includegraphics[width=0.95\linewidth]{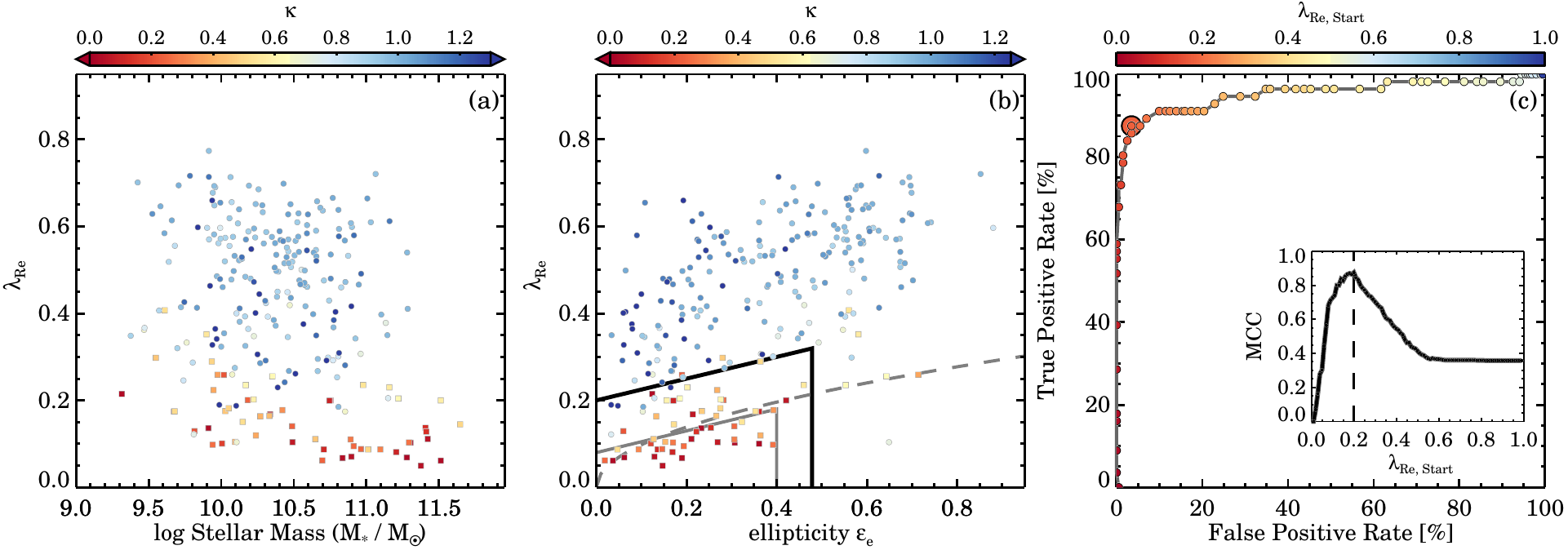}
\caption{Spin parameter proxy versus stellar mass and ellipticity using \at\ data. Data are colour coded by $\kappa$ (panels a and b). Round symbols show the $\kappa$FR, and square symbols show the $\kappa$SR. We show the $\kappa$FR and $\kappa$SR in the \lre-\ee\ space in panel (b) with the optimal selection region (black), the SR selection box from \citet{cappellari2016} in grey, and from \citet{emsellem2011} as the dashed line. Panel (c) suggests that the \lre-\ee\ space is a powerful method to distinguishing between the two types of rotators as identified by $\kappa$ from JAM modelling of \at\ data. 
\label{fig:jvds_mass_lambdar_kappa_a3d}}
\end{figure*}

\subsection{Regular and non-regular rotators from \kinemetry\ using \at\ data}
\label{app:kinemetry_a3d}

Following \citet{krajnovic2011} we use \kinemetry\ to mark the condition a galaxy can have, using $\smke < 0.04$ to select fast rotators. We present the measurements in Fig.~\ref{fig:jvds_mass_lambdar_kinemetry_a3d}. Similar to SAMI data, we find that galaxies with high \mk\ also have higher values of \lre\  towards low stellar mass. When using the ROC analysis with the confusing matrix from Table \ref{tbl:tbl1}, we find that the optimal selection region that starts at $\lr=0.12$ only has a True Positive Rate of 77.8 percent with a False Positive Rate of 1.9 percent. However, the Positive Prediction Value for \at\ data is significantly higher than for SAMI data, with respectively 89.7 versus 70.6  percent, with a similar result for the MCC with \at=0.83 and SAMI=0.66. The TPR for the \citet{emsellem2011} and \citet{cappellari2016} selection criteria are similar but relatively low at 62.2 percent. The MCC values are 0.747 and 0.757, respectively. Thus, from a statistical point of view, we do not find a significant difference between the selection criteria from \citet{emsellem2011} and \citet{cappellari2016}, but we note that for example the $\ee<0.4$ criteria was introduced to exclude counter-rotating disks from the SR class.

\subsection{Fast and Slow rotators from JAM modelling using \at\ data}
\label{app:kappa_a3d}

We will now use the $\kappa$ parameter to classify galaxies as $\kappa$SR ($\kappa<0.65$) and $\kappa$FR ($\kappa>0.65$). We use the JAM model parameters from \citet{cappellari2013b}, without applying a quality cut or limit on the inclination, but we note that with "$Quality$"$>0$ or inclination $i>60^{\circ}$ the results are qualitatively the same. The data and the ROC analysis are presented in Fig.~\ref{fig:jvds_mass_lambdar_kappa_a3d}. Similar to \kinemetry, we find that galaxies with low $\kappa$ also have higher values of \lre\  towards low stellar mass. The optimal selection region is now considerably higher than the selection box from \citet{cappellari2016}, with a starting value of $\lr=0.20$ and a TPR = 87.5, an FPR = 3.5, a PPV = 87.5, and an MCC = 0.875. In contrast, the TPR for the \citet{emsellem2011} and \citet{cappellari2016} selection regions are both 51.8 with MCC = 0.674 and 0.682, respectively. 

While the $\kappa$ parameter presents the cleanest separation within the \lre-\ee\ diagram of all identifiers that we have used, the optimal selection is significantly higher as compared to the \kinemetry\ optimal selection box. However, a closer look at the MCC distribution in Fig.~\ref{fig:jvds_mass_lambdar_kappa_a3d}(c) reveals that the peak of the MCC values is quite broad, and the relatively small \at\ sample size could offset the \lre\ starting value to higher values. Nevertheless, the fact the \kinemetry\ and $\kappa$ selection boxes differ quite considerably should be a warning that even with high-quality IFS data, the question of how to select fast and slow rotators from the \lre-\ee\ diagram is sensitive to the kinematic identifier used.

\section{The impact of inclination on the \lre\ distributions}
\label{app:lambdar_eo_random}

To test the impact of inclination on the \lre\ distributions, we will compare our default \lre\ measurement to \lre\ values corrected to an edge-on projection. These \lre\ edge-on estimates are derived from the observed \lre\ and \ee\ measurements following the method described in \citet{vandesande2018}. The method combines the observed properties with theoretical predictions from the tensor Virial theorem \citep{binney2005} and builds on the assumption that galaxies are simple rotating oblate axisymmetric spheroids with varying intrinsic shape and mild anisotropy \citep{cappellari2007}. As this is an oversimplification of the known complexities of galaxy structure and dynamics, in particular for massive early-type galaxies, we therefore only use late-type galaxies for the analysis here.

\begin{figure}
\includegraphics[width=\linewidth]{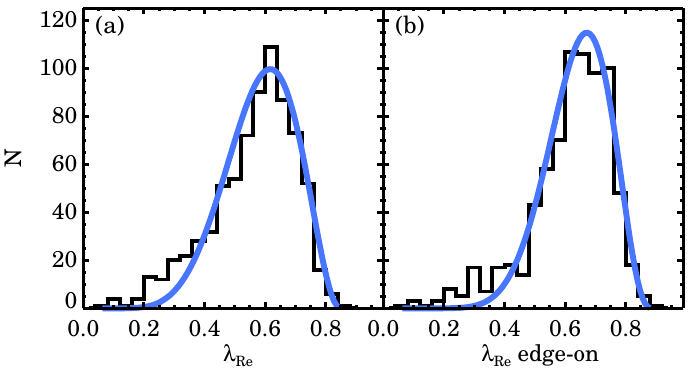}
\caption{Observed and edge-on projected \lre\ distributions for SAMI Galaxy Survey late-type galaxies. The data are shown in black, whereas the blue line shows the best-fitting Beta function. The $1-\sigma$ width of the distribution changes from 0.275 to 0.234 between \lre\ and \lreeo, and we also detect an offset of $\Delta \lre=0.055$ towards higher \lre\ when the measurements are projected to edge-on.
\label{fig:jvds_lambdar_edge_on}}
\end{figure}

We present the \lre\ and \lreeo\ distributions for SAMI late-type galaxies in Fig. \ref{fig:jvds_lambdar_edge_on}. The low-\lre\ galaxies that are not well-fitted by the Beta function, do not disappear after applying our inclination correction. As many of these galaxies are observed close to face-on, uncertainties on the ellipticity measurements play an increasingly negative role on the inclination correction. Moreover, morphological features such as bars and spiral arms can make the galaxies' ellipticity to appear flatter than they really are, inhibiting an accurate deprojection. Alternative methods to determine edge-on \lre\ measurements will be explored in future work.

Nonetheless, the edge-on projected distribution becomes only mildly narrower ($\Delta \lre = 0.041$) and shifts slightly towards higher $\lre$ from 0.586 to 0.641. As these changes are relatively small, we do not expect our results on separating a bimodal \lre\ distribution to change significantly due to the effects of inclination.

\section{The Bayesian Mixture Model in detail}
\label{app:bayesian_mm_priors}

\subsection{Model description}

For each galaxy in our sample, we have measurements of its stellar mass ($M_*$) and a proxy for the spin parameter measured within one effective radius ($\lambda_R$). We aim to model a galaxy's spin parameter, which we label $y$ to keep with standard notation in the literature, in terms of its stellar mass. We do this by building a probabilistic mixture model. In this appendix, we describe the Bayesian Mixture model used in Section \ref{subsec:bayesian_approach}  in detail. 

A mixture model uses a number (in this case two) of different probability distributions to model a set of observed data. The probability of a single data point being drawn from one distribution is denoted $\pi$. This implies that the likelihood function is of the form:

\begin{equation}
p(y | \boldsymbol{\theta}) \propto \prod_{n=1}^{N} \big( \pi  p_1(y_n | \boldsymbol{\theta_1}) + (1 - \pi) p_2(y_n | \boldsymbol{\theta_2}) \big)
\end{equation}

\noindent where $p_1$ and $p_2$ refer to the different probability distributions and $\boldsymbol{\theta}$ is a vector of model parameters. 

In this case, we assign $p_1$ and $p_2$ to be two distinct beta distributions, each with shape parameters $\alpha$ and $\beta$ (i.e., $\boldsymbol{\theta_1} = (\alpha_1, \beta_1)$ and $\boldsymbol{\theta_2} = (\alpha_2, \beta_2)$).  We allow these shape parameters to vary as a function of stellar mass via a first order polynomial:

\begin{align}
i &= 1,2\\
\log(\alpha_i) &= c_i + d_i M_*\\
\log(\beta_i) &= e_i + f_i M_*
\end{align}

\noindent These correspond to eight free parameters. Note that the beta distribution's shape parameters must be constrained to be positive, and as such we vary them on the \textit{logarithmic} scale, such that $\alpha_i$ and $\beta_i$ are always greater than zero.

Furthermore, we allow the mixture probability, $\pi$, to vary with stellar mass. This represents the well-known dependence of kinematic morphology with stellar mass, with massive galaxies much more likely to be slow rotators \citep[e.g.,][]{ emsellem2011,brough2017,veale2017b,vandesande2017b,green2018,graham2018}. Since $\pi$ is a probability, it must lie between 0 and 1. To ensure this is always the case, we use the sigmoid function to map any real number to the interval [0, 1], which also introduces a further two parameters to the model ($\mu$ and $\sigma$):

\begin{align}
\pi(M_*) = \frac{1}{1 + \exp(-(M_* - \mu)/\sigma)}
\end{align}

To summarise, our model has 10 free parameters; eight corresponding to the change in beta distribution shape parameters with stellar mass and two corresponding to how the probability of being drawn from either beta distribution varies with stellar mass. The final likelihood function is therefore:

\begin{align}
p(y | M_*,\boldsymbol{\alpha}, \boldsymbol{\beta}, \mu, \sigma) \propto & \prod_{n=1}^{N} \Big( \,\pi(M_*, \mu, \sigma) \, p_1(y_n | \alpha_1(M_*), \beta_1(M_*)) + \nonumber \\
 & (1 -  \pi(M_*, \mu, \sigma))\,  p_2(y_n | \alpha_1(M_*), \beta_1(M_*)) \Big)
\end{align}

\subsection{Priors}

\begin{table*}
 \caption{A summary of our prior choices for the Bayesian mixture model presented in Section \ref{subsec:bayesian_approach}. Each parameter (or transformation of a parameter) below is assigned a Gaussian prior with the given location (mean) and scale (standard deviation). The exception is the $\sigma$ parameter, which has a half-Gaussian prior (i.e a Gaussian probability distribution for positive values and zero probability for negative values).}
\begin{tabular}{crrrrrrrr} 
 \hline
 & \multicolumn{2}{c}{SAMI Observations} & \multicolumn{2}{c}{EAGLE} & \multicolumn{2}{c}{HorizonAGN} & \multicolumn{2}{c}{Magneticum} \\
 Parameter & Location & Scale & Location & Scale & Location & Scale & Location & Scale\\ 
 $\log(c_1)$ & $\log_e(4.5)$ & 0.5 & $\log_e(3.3)$ & 0.3 & 0 & 3 & $\log_e(4.3)$ & 1 \\ 
 $\log(d_1)$ & 0 & 0.5 & 0 & 0.5 & 0 & 3 & 0 & 3\\ 
 $\log(c_2)$ & $\log_e(6.9)$ & 0.5 & $\log_e(4.5)$ & 0.3& $\log_e(9.75)$ & 3 &$\log_e(25)$ &1\\
 $\log(d_2)$ & 0 & 0.5 & 0& 0.5 &$\log_e(1.49)$ & 3 & 0 &3\\ 
 $\log(e_1)$ & $\log_e(4.4)$ & 0.5& $\log_e(5)$ & 1 & $\log_e(15)$& 3 & $\log_e(9.4)$ & 1\\ 
 $\log(f_1)$ & 0 & 0.5 & 0& 0.5& 0 & 3 & 0  & 3 \\ 
 $\log(e_2)$ & $\log_e(100)$ & 0.5 & $\log_e(45)$ & 0.3 & $\log_e(100)$ & 0.5 & $\log_e(200)$ & 1\\ 
 $\log(f_2)$ & 0 & 0.5 & 0& 0.5& 0& 3& 0& 3\\ 
 $\mu$ & 0 & 1 & 1 & 0.3 & 0 & 1 & 0 & 1\\
 $\sigma$ & 0 & 2 & -1 & 0.3 & 0 & 2 & 0& 2\\
 \hline
\end{tabular}
\label{tbl:priors}
\end{table*}

As with any Bayesian analysis, each free parameter must be assigned a prior. Our prior choices are described in Table \ref{tbl:priors}. We conduct simulations to see the effect of our prior choices (known as ``prior predictive checks") and ensure that our prior choices do not give rise to unphysical distributions of simulated data. Reasonable changes to these priors do not change our conclusions.


\bsp	
\label{lastpage}
\end{document}